\documentclass[prd,aps,twocolumn,a4paper,floatfix]{revtex4-1}

\usepackage{graphicx}
\usepackage{amsmath}
\usepackage{color,soul}

\newcommand{\pda}[2]{\frac{\partial {#1}}{\partial {#2}}}

\begin{document}


\author{Carsten Gundlach, Ian Hawke and Stephanie J Erickson}
\affiliation
       {School of Mathematics, 
         University of Southampton,
         Southampton SO17 1BJ, UK}

\title{A conservation law formulation of nonlinear elasticity in
  general relativity}

\begin{abstract}

  We present a practical framework for ideal hyperelasticity in
  numerical relativity. For this purpose, we recast the formalism of
  Carter and Quintana as a set of Eulerian conservation laws in an
  arbitrary 3+1 split of spacetime. The resulting equations are
  presented as an extension of the standard Valencia formalism
  for a perfect fluid, with additional terms in the stress-energy
  tensor, plus a set of kinematic conservation laws that evolve a
  configuration gradient $\psi^A{}_i$. We prove that the equations can
  be made symmetric hyperbolic by suitable constraint additions, at
  least in a neighbourhood of the unsheared state. We discuss the
  Newtonian limit of our formalism and its relation to a second
  formalism also used in Newtonian elasticity.  We validate our framework
  by numerically solving a set of Riemann problems in Minkowski
  spacetime, as well as Newtonian ones from the literature.

\end{abstract}


\maketitle

\tableofcontents


\section{Introduction}

Neutron stars are believed to form a crystalline outer crust as they
age and cool, but retain a fluid (probably superfluid) core
\cite{ChamelHaensel}. A mathematical framework for weak solutions of
general relativistic elasticity is likely to be indispensable for the
modelling of neutron star crusts in at least two scenarios: starquakes
and binary mergers. 

Pulsars are observed to spin down at a regular rate, losing angular
momentum through gravitational and/or electromagnetic
radiation. Occasionally the rotation spins up suddenly. One model
suggests that such a ``glitch'' occurs when the elastic crust breaks
and the inertial moment of the star decreases suddenly as a
consequence (e.g.~\cite{AlparEtAl96}). It has also been suggested
\cite{Duncan98} that starquakes are the cause of soft gamma
repeaters (SGRs). Quasi-periodic oscillations in the tails of giant
flares in SGRs have been suggested \cite{SamuelssonAndersson07} to
provide direct observational evidence for crust oscillation modes,
although the modelling of neutron star oscillations even in
perturbation theory is complicated by the coupling between the crust,
the fluid core and a strong magnetic field (see
e.g.~\cite{Gabler11}). A correct model would of course have to be
nonlinear. Finally we note that strong shocks also arise when two old
neutron stars in a binary system merge. The detailed dynamics and
features, such as the breaking strain (see \cite{HorowitzKadau09}), of
the crust, will determine when and where the crust melts and
breaks. This will in turn impact on the post-merger dynamics, such as
the time taken by the remnant to collapse to a black hole (see
e.g.~\cite{Baiotti08}). 

For all these scenarios, models must therefore comprise an elastic
crust, a fluid core, and a magnetic field permeating both.  As a step
towards such models, we present here a formulation of
(hyper)elastic matter in general relativity in the form of
conservation laws amenable to solution by high-resolution
shock-capturing (HRSC) numerical methods. These conservation laws are
the union of the usual stress-energy conservation (dynamics), and a
set of conservation laws for a deformation tensor (kinematics).

The kinematic equations are essentially the same in Newtonian and
relativistic physics, but the literature on weak solutions of
Newtonian elasticity uses Cartesian tensor notation, which obscures
the geometric nature of the theory. In Sec.~\ref{section:kinematics},
we derive these equations carefully, using the language of
differential geometry. Following Carter and Quintana \cite{CQ}, we
begin with a map from spacetime to a 3-dimensional matter space. The
main object we calculate with is its derivative $\psi^A{}_i$. As a
partial derivative, this is subject to integrability conditions. Under
a 3+1 split these become evolution equations and constraints, of a
purely kinematic nature, both of which can be written as conservation
laws. We show that their physical significance is to allow
discontinuities in the density and kinks but forbid discontinuities in
the crystal axes and particle world lines.

The other, dynamical, half of the problem consists in finding the
stress-energy tensor from $\psi^A{}_i$ and an equation of state. We do
this in Sec.~\ref{section:dynamics}, following Karlovini and
Samuelsson \cite{KS}. In particular, demanding covariance on both
spacetime and matter space restricts the possible dynamics. For
isotropic matter, the equation of state can relate only two
deformation scalars, besides the number density, internal energy, and
entropy. Similarly, the stress-energy tensor depends on the equation
of state through two generalised forces.

In Sec.~\ref{section:hyperbolicity}, we prove that our kinematic and
dynamical equations together form a first-order system of evolution
equations that, by constraint addition, can be made hyperbolic {\em if
  the constraints (\ref{CAij}) are obeyed or not.} This property is
crucial for the stability of numerical solutions in which the
constraints are left to evolve freely, and hence finite difference
error generically generates constraint violations.  We use the methods
of Beig and Schmidt \cite{BS}, who proved symmetric hyperbolicity of
an inequivalent first-order system (one which would not be appropriate
for modelling weak solutions).

In order to make contact with existing work on ideal fluid dynamics
and magnetohydrodynamics in general relativity, in
Sec.~\ref{section:valencia} we present our dynamical equations as a
generalisation of the Valencia \cite{Font} formulation of
hydrodynamics. We give an algorithm for the conversion between
conserved and primitive variables.

As a first test of our formalism, we present numerical time evolutions of
Riemann problems in Sec~\ref{section:numerics}. The variables are
three-dimensional, and the grid is either one-dimensional, or
two-dimensional with the Riemann problem at an angle to the grid. We
compare the relativistic code in the Newtonian limit with an
explicitly Newtonian code, and both with published Newtonian Riemann
problems \cite{TRT,BDRT}. We also compare against exact Riemann
solutions in the relativistic regime in Minkowski spacetime. We
compare the Eulerian and mixed formalisms, and evolutions where the
number density is either read off from the deformation tensor, or
evolved separately.

In Sec.~\ref{section:conclusions} we summarize the results of the
paper, and discuss the work remaining to apply these methods to full
3+1 nonlinear simulations.

We collect relevant formulas from the standard 3+1 split of spacetime
in Appendix~\ref{appendix:3+1}, and relevant definitions of
hyperbolicity in Appendix~\ref{appendix:hyperbolicity}.  One of two
existing Newtonian formalisms
\cite{TrangensteinColella,MillerColella2001} is essentially the
Newtonian limit of our formalism. We derive the Newtonian limit in
Appendix~\ref{appendix:newtonianlimit}. In
Appendix~\ref{appendix:mixed} we derive the equations of an
alternative Newtonian formalism
\cite{GodunovRomenski,GodunovPeshkov,PS}, and prove that the two have
the same weak solutions.

The remaining Appendixes contain auxiliary material on our numerical
method and our numerical tests:
Appendix~\ref{section:discreteconstraints} proposes a general
framework for discrete constraint preservation (similar to
``constrained transport'' for MHD), and Appendix~\ref{appendix:2d}
presents our implementation of Riemann tests on a 2-dimensional grid.
Appendix~\ref{appendix:eos} describes the equations of state we
use. Appendix~\ref{sec:exactsolns} summarizes how we construct exact
solutions for specific Riemann problems and
Appendix~\ref{appendix:initialdata} the initial data for our Riemann
tests used here.

We have attempted as far as possible compatibility with the notation
of \cite{KS}, \cite{BS} and \cite{Font}. Throughout this paper, tensor
indices are assumed to be in a generic local coordinate basis. Partial
derivatives in this basis are indicated by commas. Indices
$a,b,c,\dots=0,1,2,3$ are spacetime indices, $i,j,k,\dots=1,2,3$ are
spatial indices on $x^0=t=\rm const.$ hypersurfaces, and
$A,B,C,\dots=1,2,3$ are matter space indices on a 3-dimensional matter
space $X_3$. In Appendix~\ref{appendix:mixed}, the indices
$\alpha,\beta,\gamma,\dots=0,1,2,3$ are matter space indices on an
extended matter space $X_4$. In Secs.~\ref{section:dynamics} and
\ref{section:hyperbolicity}, $\alpha,\beta=1,2$ are used to label
elastic forces. In Appendix~\ref{appendix:hyperbolicity},
$\alpha,\beta$ label the variables of a generic hyperbolic system. For
all these indices a summation convention applies.

In order to take determinants of 2-index objects which are not (1,1)
tensors, we introduce the non-tensorial totally antisymmetric symbol
$\delta$, which is defined to be $\delta_{0123}=1$, etc. With the
exception of the objects $\delta$, throughout this paper, all objects
transform as tensors of the type indicated by their free indices,
unless we indicate otherwise by a suffix: for example, the determinant
of the spacetime metric in coordinates $x^a$ will be denoted by
$-g_x$.


\section{Kinematics}
\label{section:kinematics}


\subsection{The configuration gradient and its 3+1 split}

In the relativistic framework of \cite{CQ,KS}, the matter configuration
is encoded in a map from 4-dimensional spacetime to 3-dimensional
matter space
\begin{equation}
\chi:\quad M_4\to X_3,
\end{equation}
or in local coordinates $x^a$ on spacetime and $\xi^A$ on matter space,
\begin{equation}
x^a\mapsto \xi^A=\chi^A(x^a).
\end{equation}
For simplicity of notation we denote the derivative $d\chi$ of $\chi$
by a new symbol $\psi$,
\begin{equation}
\psi:\quad M_4\to TX_3\otimes T^*M_4, 
\end{equation}
\begin{equation}
x^a\mapsto \psi^A{}_a:={\partial \xi^A\over \partial x^a}.
\end{equation}

For time evolutions, we introduce a time-foliation of the
spacetime, so that we have
\begin{eqnarray}
\chi:\quad R\times M_3&\to& X_3, \\
(t,x^i)&\mapsto& \xi^A
\end{eqnarray}
with derivatives
\begin{equation}
{\psi^A}_i:={\partial \xi^A\over \partial x^i}, 
\qquad \psi^A{}_t:={\partial \xi^A\over \partial t}.
\end{equation}
Following \cite{Wernig-Pichler}, we shall call $\chi^A$ the
\emph{configuration} and both $\psi^A{}_a$ and $\psi^A{}_i$ the
\emph{configuration gradient}.

The matter space coordinates $\xi^A$ label particles and must
therefore be constant along particle world lines, so that
\begin{equation}
u^a\psi^A{}_a=0,
\end{equation}
where the 4-velocity $u^a$ is tangential to the matter world
lines. Parameterising the 4-velocity in the standard way as
\begin{equation}
\label{uaup}
u^a=(u^t,u^i)=\alpha^{-1}W(1,\hat v^i)
\end{equation}
(see Appendix~\ref{appendix:3+1} for more details), we have
\begin{equation}
\label{psiAt}
\psi^A{}_t=-\hat v^i\psi^A{}_i.
\end{equation}

The configuration gradient $\psi^A{}_i$ is also used as the primary
variable in the Newtonian framework of \cite{TrangensteinColella,
  MillerColella2001, MillerColella2002} (denoted there by $g$). This
framework is the Newtonian limit of our relativistic one. In
Appendix~\ref{appendix:newtonianlimit} we derive the Newtonian limit
of our framework. Other Newtonian papers \cite{PS,GodunovPeshkov} use
the the $3\times 3$ matrix inverse of $\psi^A{}_i$, which we shall
denote by $F^i{}_A$, as the primary variable (denoted there by
$F$). We review this alternative framework in
Appendix~\ref{appendix:mixed}. In the Newtonian literature, $F^i{}_A$
is commonly called the (Lagrangian) deformation gradient, and
$\psi^A{}_i$ the inverse deformation gradient. From a geometric point
of view, however, these objects on their own carry no information
about what one might intuitively call a deformation.


\subsection{Kinematic equations and hyperbolicity fix}

From the definition of $\psi^A{}_a$ as a partial derivative, we have
the integrability conditions
\begin{equation}
\label{CAab}
C^A{}_{ab} :=\psi^A{}_{[a,b]}=0.
\end{equation}
In a 3+1 split, these become 
\begin{eqnarray}
\label{CAij}
{C^A}_{ij} &:=& {\psi^A}_{[i,j]} = 0, \\
\label{EAi}
{E^A}_i &:=& 2C^A{}_{it} = {\psi^A}_{i,t}+ \left(\hat v^j{\psi^A}_j\right)_{,i}=0.
\end{eqnarray}
The constraints (\ref{CAij}) are conserved by the evolution equations
(\ref{EAi}). Note that these equations are already in
conservation law form: more explicitly, 
\begin{eqnarray}
\label{CAijbis}
{C^A}_{ij} &=& \left({\psi^A}_k\delta^k_{[i}\delta^l_{j]}\right)_{,l} = 0, \\
\label{EAibis}
{E^A}_i &=& {\psi^A}_{i,t}+ \left(\hat v^j{\psi^A}_j\delta^k_i\right)_{,k}=0.
\end{eqnarray}

Instead of $E^A{}_i=0$ as an evolution equation for $\psi^A{}_i$, we
shall in fact use
\begin{equation}
\label{Ebari}
\bar E^A{}_i:=2\alpha W^{-1}u^aC^A{}_{ia}=
2\psi^A{}_{[i,t]}+2\hat v^j\psi^A{}_{[i,j]}=0.
\end{equation}
This can be written as a balance law obtained from the
conservation law (\ref{EAi}) by adding a source term that is
proportional to the constraint (\ref{CAij}), namely
\begin{equation}
\label{EAitres}
\psi^A{}_{i,t}+\left(\hat v^j \psi^A{}_j\right)_{,i}=2\hat v^j \psi^A{}_{[j,i]}.
\end{equation}
Note that this cannot be written in pure conservation law form. 

In handwaving anticipation of the hyperbolicity analysis presented in
Sec.~\ref{section:hyperbolicity}, we point out in passing that
(\ref{EAitres}) can be written as an advection equation for
$\psi^A{}_i$ with a source term that is of lower order in
$\psi^A{}_i$, namely
\begin{equation}
\label{EAi4}
\psi^A{}_{i,t}+\hat v^j \psi^A{}_{i,j}=-\psi^A{}_j\hat v^j{}_{,i}.
\end{equation}
For {\em given} $\hat v^i$, this is strongly hyperbolic in
$\psi^A{}_i$, whereas (\ref{EAi}) is only weakly hyperbolic.


\subsection{Kinematic jump conditions}
\label{section:kinematicjump}

The geometric meaning of the integrability conditions (\ref{CAab}) is
that the particle world lines and the instantaneous crystal lines
(i.e.\ lines of constant $\xi^A$) all mesh up into a four-dimensional
grid. In particular, the world lines and crystal lines are
continuous. The weak form of these equations must therefore keep them
continuous while allowing them to kink, thus forbidding dislocations
and fractures. To stress their purely kinematic nature, we shall
discuss them without invoking a metric on spacetime or matter
space. We ignore the source term in the evolution equations
(\ref{EAitres}) in deriving the Rankine-Hugoniot conditions, because
it has no effect on physical solutions, which obey the constraints.

Consider a surface of discontinuity in space (from now on called a
shock for briefness). Let $n_i$ be a covector normal to the shock
(uniquely defined up to an overall factor). Let $s^i$ be the shock
velocity vector (defined, in the absence of a metric, only up to the
addition of a vector tangential to the shock), and let $s:=s^i n_i$
(which inherits the arbitrary factor in $n_i$ but not the arbitary
vector in $s^i$) be the normal shock speed. The jump
(Rankine-Hugoniot) conditions arising from 
(\ref{CAijbis}) and (\ref{EAibis}) are then
\begin{eqnarray}
\label{CAijjump}
\left[{\psi^A}_k\delta^k_{[i}\delta^l_{j]}\right]n_l = 0, \\
\label{EAijump}
-s\left[{\psi^A}_i\right]+ \left[\hat v^j{\psi^A}_j\delta^k_i\right]n_k=0.
\end{eqnarray}

\begin{figure}
\includegraphics[width=6cm]{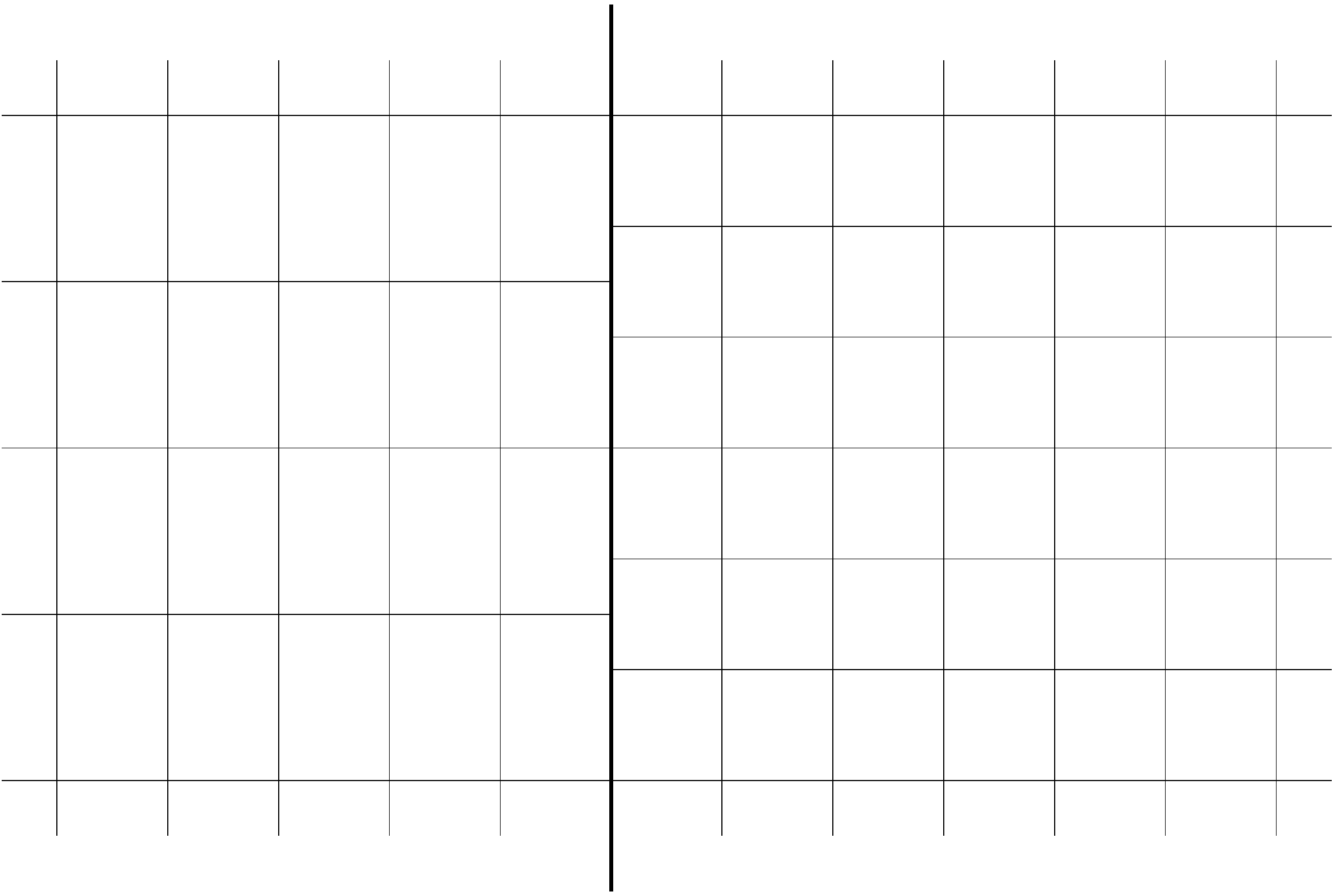}
\caption{
\label{fig:nosurgery1} 
A discontinuity of the type illustrated here is not allowed by the
jump conditions (it would require ``surgery'' on the material). For
simplicity and without loss of generality, we choose space and matter
space coordinates in this and the next three figures so that the shock
is along the $y$ axis and $\psi^A{}_i=\delta^A_i$ in the left
state. This type of surgery is then forbidden by $[\psi^Y{}_y]=0$.}
\end{figure}

\begin{figure}
\includegraphics[width=6cm]{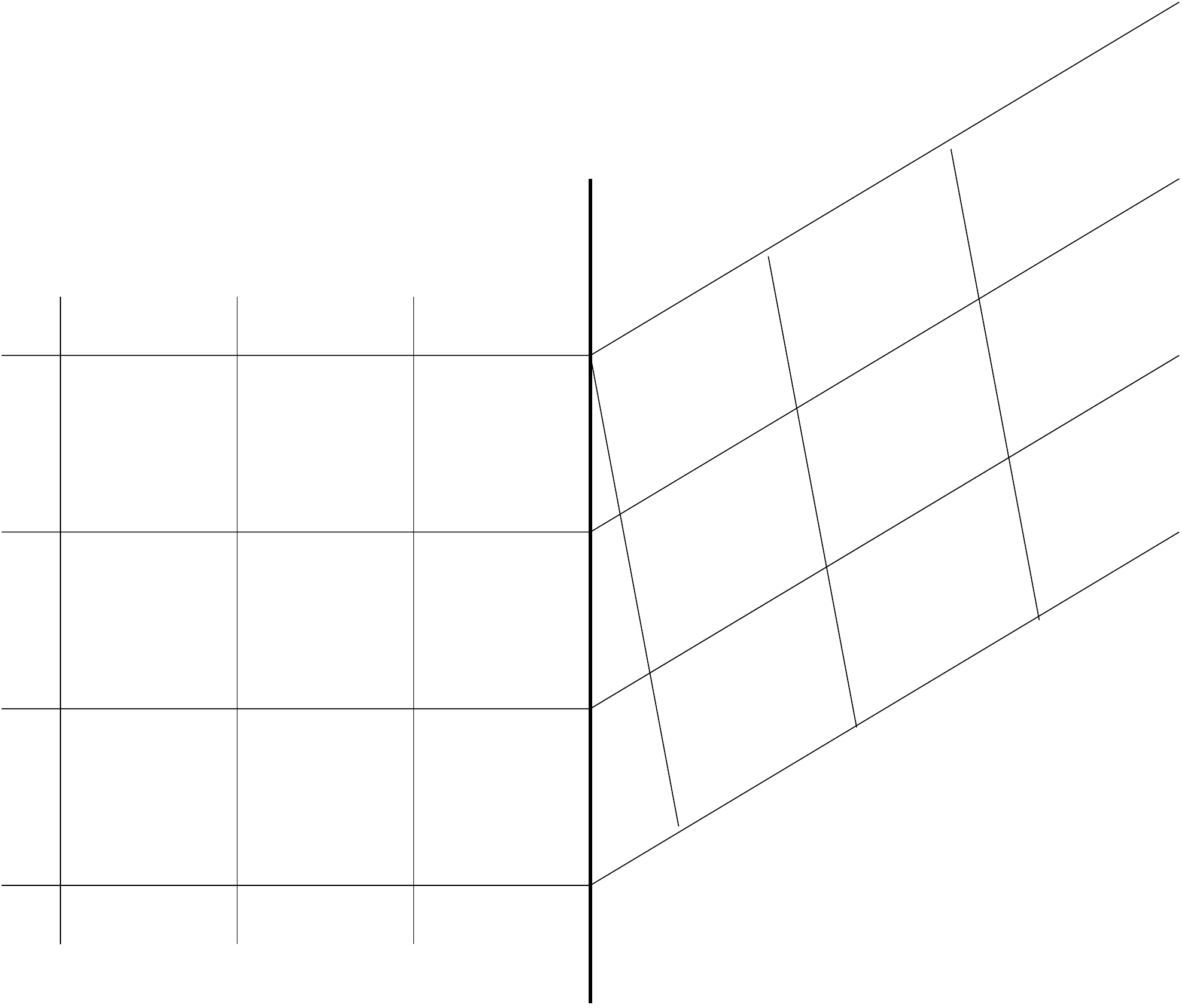}
\caption{
\label{fig:nosurgery2} 
This type of surgery is forbidden by $[\psi^X{}_y]=0$. The coordinate
choices are as in Fig.~\ref{fig:nosurgery1}.} 
\end{figure}

\begin{figure}
\includegraphics[width=6cm]{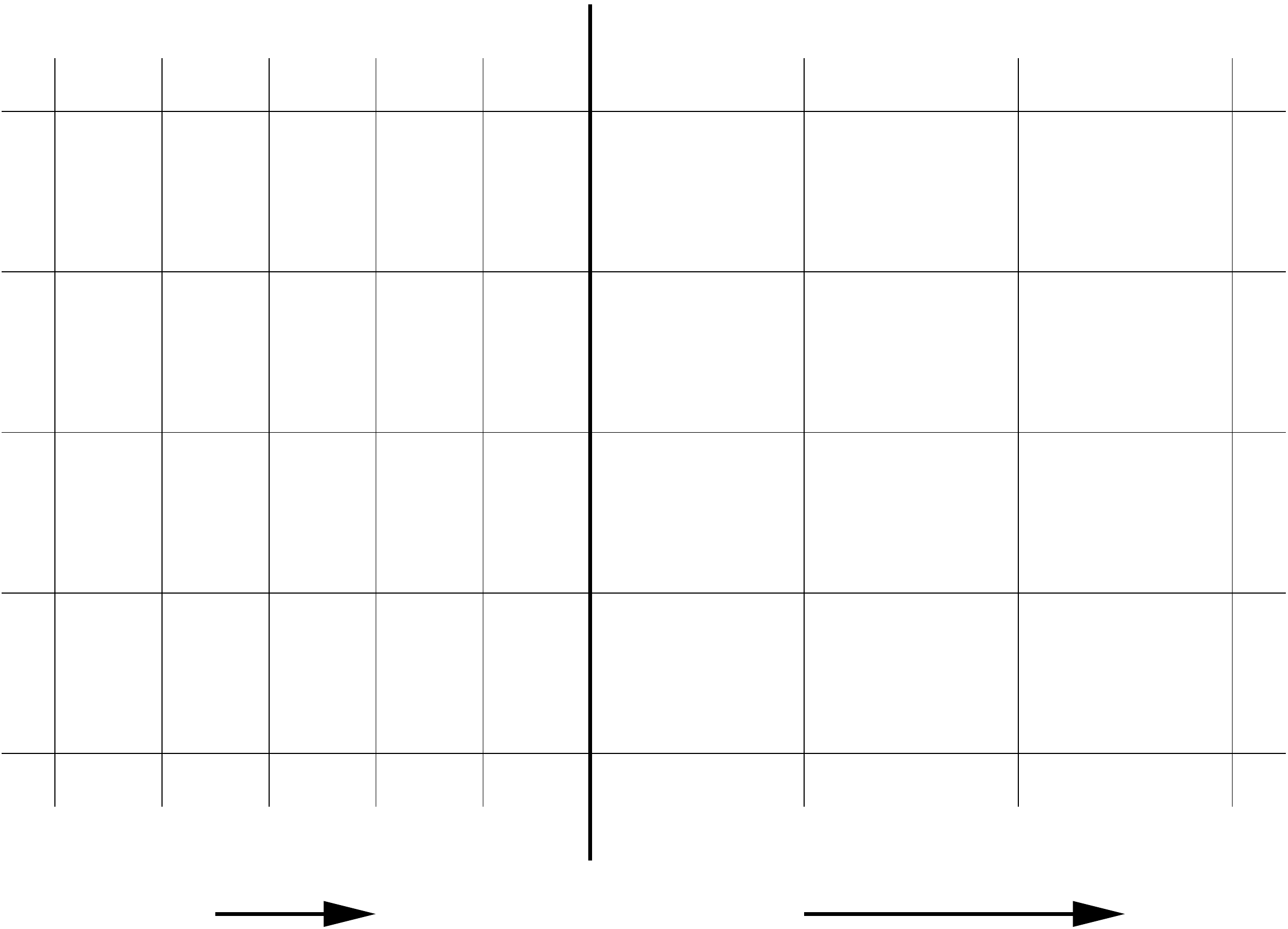}
\caption{
\label{fig:pressureshock}
A pure density shock, with $v^y$ continuous, shown in the rest frame
of the shock. This is essentially a one-dimensional
phenomenon. ``Density'' and velocity on the left and right (shown as
arrows) are related (in these coordinates) by $[\psi^X{}_xv^x]=0$.  }
\end{figure}

\begin{figure}
\includegraphics[width=6cm]{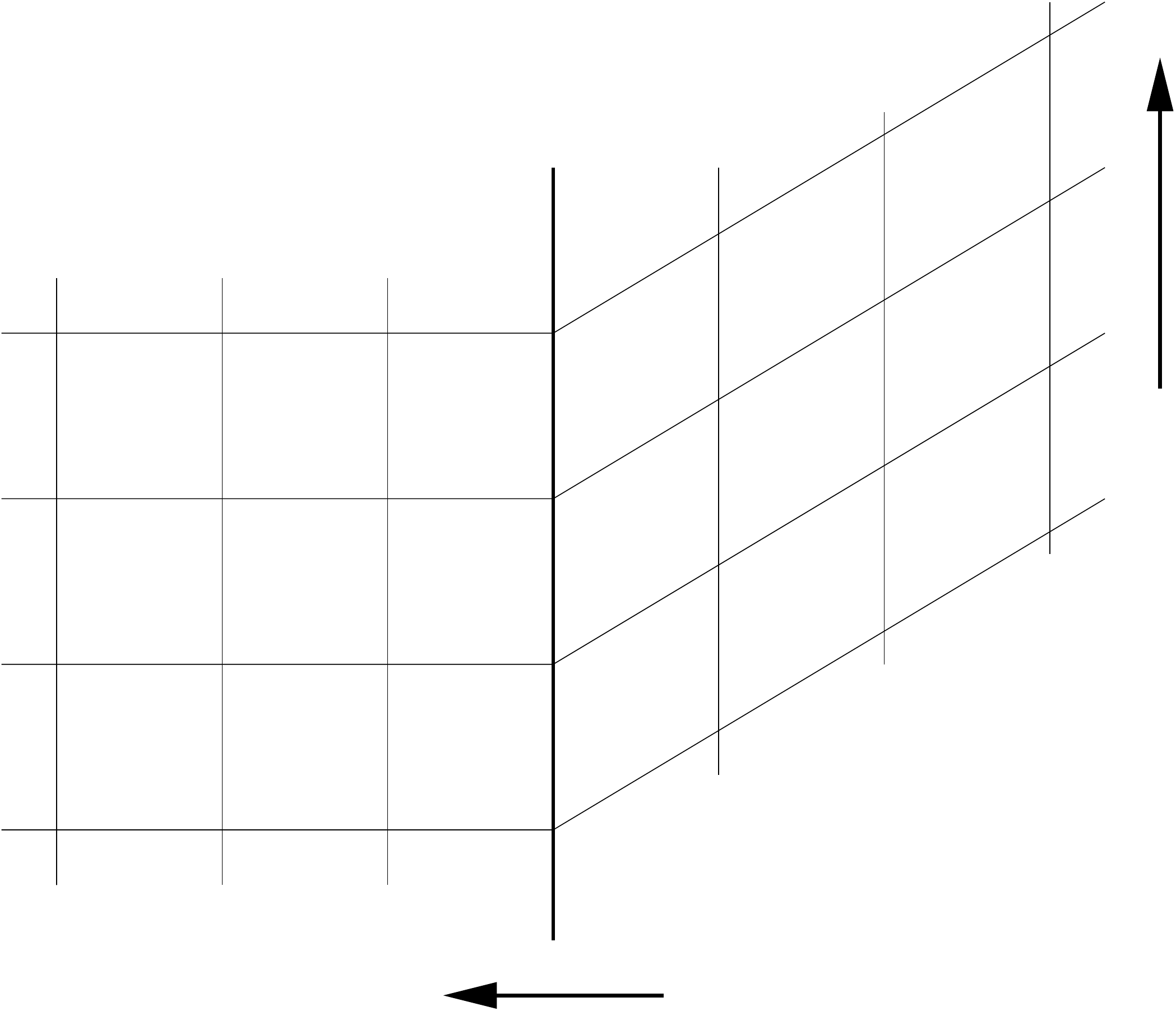}
\caption{
\label{fig:shearshock}
A pure travelling kink, shown in the rest frame of the left state. For
simplicity, we have assumed $\psi^X{}_x=1$ to be continuous, which
implies that $v^x=0$ is continuous and the ``volume density'' is
continous. However, the ``line density'' along the $Y$ crystal axis is
discontinuous. The shock speed and shear speed (shown as arrows) in
these coordinates are related by $s(\psi^Y{}_x)_R=(v^y)_R$.}
\end{figure}

We want to decompose these conditions into parts normal and parallel
to the shock.
Let $n^i$ be a vector that obeys $n^in_i=1$. $n^i$ is therefore
uniquely defined up to the factor in $n_i$, and the addition of an
arbitrary vector tangent to the shock. Define the tensor
\begin{equation}
\parallel^i{}_j:=\delta^i{}_j-n^in_j.
\end{equation}
It is the projection operator into the tangent plane of the shock in the
sense that
\begin{equation}
\parallel^i{}_j\,n_i=0, \quad \parallel^i{}_j\,n^j=0, \quad
\parallel^i{}_j\parallel^j{}_k=\parallel^i{}_k.
\end{equation}
Split into normal and tangential components defined by $\hat
v^n:=v^in_i$ and $\hat v^{\parallel i}:=\parallel^i{}_j\hat v^j$, the
jump conditions can now be compactly written as
\begin{eqnarray}
\label{psijump1}
[\psi^A{}_{\parallel i}]&=&0, \\
\label{psijump2}
{[\psi^A{}_n(\hat v^n-s)]}+\psi^A{}_{\parallel i} [\hat v^{\parallel i}]&=&0.
\end{eqnarray}
The first of these guarantees the continuity of crystal lines ($\xi^A$
lines) across the shock, or the absence of ``surgery across the
shock'', as illustrated in Figs.~\ref{fig:nosurgery1} and
\ref{fig:nosurgery2}. The second guarantees the conservation of
particles as they cross the shock. Consider the special case where
$\hat v^{\parallel i}$ is continuous. Then, in the rest frame of the
shock, $[\psi^A{}_n\hat v^n]=0$. This is a pure ``density'' shock of
the type familiar from fluid dynamics, and is illustrated in
Fig.~\ref{fig:pressureshock}. Conversely, consider the case where
$\psi^A{}_n$ is continuous. Then, again in the rest frame of the
shock, $\psi^A{}_n[\hat v^n] +\psi^A{}_{\parallel i}[\hat v^{\parallel
  i}]=0$. This is a pure travelling kink, set up by a discontinuity in
the tangential velocity, as illustrated in Fig.~\ref{fig:shearshock}.

Fluids allow for a contact discontinuity where the tangential velocity
jumps. This is replaced by the travelling kink in elastic matter. (The
only contact discontinuity that survives is the one where the entropy
jumps.) This holds even in the limit where the dynamics goes to the
fluid limit (the stiffness goes to zero and the stress-energy tensor
becomes that of a fluid), and so the fluid limit is singular.


\subsection{Matter space metric and particle number current}
\label{section:particlecurrent}

The minimal geometric structure on matter space is a volume form
$n_{ABC}$ whose integration over a volume in matter space gives the
number of particles in that part of matter space. In addition, at
least a conformal metric is required to define angles on matter space,
which can then be compared with angles on spacetime to define
deformations. But together these two structures define a full
Riemannian metric $k_{AB}$. (``Distances'' are measured in particles,
not meters). Therefore we now assume that $k_{AB}$ is defined and
$n_{ABC}$ is compatible with it. In matter coordinates $\xi^A$ this
means that
\begin{equation}
n_{ABC}=\sqrt{k_\xi}\,\delta_{ABC}, 
\end{equation}
where
\begin{equation}
k_\xi:={1\over 3!}\delta^{ABC}\delta^{DEF}
k_{AD}\,k_{BE}\,k_{CF}
\end{equation}
is the usual determinant. The suffix $\xi$ is a reminder that it is
not a scalar on matter space but depends on the $\xi^A$ coordinates.

We use $k_{AB}$ as an example to discuss the ``evolution'' of tensors
on matter space. Matter space itself has no time, but as we are using
a Eulerian framework, we effectively consider $k_{AB}(\chi^C(x^d))$ as
a function on spacetime. The push-forward of $k_{AB}$ to
a tensor $k_{ab}$ on spacetime obeys
\begin{equation}
{\cal L}_u k_{ab}=0, \quad u^a k_{ab}=u^b k_{ab}=0.
\end{equation}
Formally, tensor fields on matter space could be defined as tensors on
spacetime whose Lie derivative along $u^a$ and contractions with $u^a$
all vanish, and this is indeed the approach of \cite{CQ}, and partly
of \cite{KS}. However, equivalently the components
$k_{AB}(\chi^C(x^d))$ can be considered as scalars on spacetime that
are constant along particle world lines, so that
\begin{equation}
\label{matteradvection}
u^a k_{AB,a}=0,
\end{equation}
or in coordinates
\begin{equation}
\label{kadvection}
k_{AB,t}+\hat v^i k_{AB,i}=0.
\end{equation}
Numerically, we prefer to work with $k_{AB}$, which has fewer
components and a simpler evolution equation than $k_{ab}$. 

Following \cite{CQ,KS}, we consider the push-forward of $n_{ABC}$ to a
3-form $n_{abc}$ on spacetime
\begin{equation}
\label{nabcdef}
n_{abc}:={\psi^A}_a{\psi^B}_b{\psi^C}_c n_{ABC}.
\end{equation}
Spacetime also has a volume form $\epsilon_{abcd}$, compatible with
a Lorentzian metric $g_{ab}$. In arbitrary coordinates,
\begin{equation}
\epsilon_{abcd}=\sqrt{g_x}\,\delta_{abcd}, \quad 
g_x:=-{1\over 4!}\delta^{abcd}\delta^{efgh}
g_{ae}\,g_{bf}\,g_{cg}\,g_{dh}.
\end{equation}
(We have defined $g_x$ as positive for ease of notation).  We then
define the particle number current
\begin{equation}
\label{jdef}
j^a:={1\over 3!}\epsilon^{abcd}n_{bcd}.
\end{equation}
This is timelike, and conserved,
\begin{equation}
\nabla_a j^a=\epsilon^{abcd}\nabla_a n_{bcd}=0,
\end{equation}
where $\nabla_a$ is the covariant derivative compatible with
$g_{ab}$. The right-hand side vanishes because it is the push-forward of
$n_{[BCD,A]}$, which must vanish as it is a 4-form on a 3-dimensional
space. 
We split $j^a$ into a matter
4-velocity and a particle density
\begin{equation}
\label{ufromj}
j^a=: n u^a,
\end{equation}
where $u^a$ is normalised as
\begin{equation}
u^au_a=-1
\end{equation}
(and hence $n=-j_aj^a$).
In coordinates, using (\ref{gtogamma}) and (\ref{Valenciau}),
$\nabla_aj^a=0$ becomes
\begin{equation}
\label{restmassconservationGR}
\left(\sqrt{\gamma_x}Wn\right)_{,t}+\left(\sqrt{\gamma_x}Wn \hat
v^i \right)_{,i}=0.
\end{equation}

Conversely, we can relate the particle density and current via
$n=-u_a j^a$,
and substituting (\ref{Valenciaudown}) into this and using
(\ref{3eps4eps}), we obtain
\begin{equation}
\label{rhoGR}
n={1\over 3!}W^{-1}\epsilon^{ijk}n_{ijk}={\sqrt{k_\xi}\psi_{x\xi}\over W
  \sqrt{\gamma_x}}.
\end{equation}
Here $n_{ijk}$ are the space components of the 4-dimensional 3-form
$n_{abc}$ in the adapted coordinates $(t,x^i)$, and $\psi_{x\xi}$ is
the determinant
\begin{equation}
\psi_{x\xi}:={1\over
  3!}\delta^{ijk}\delta_{ABC}\psi^A{}_i\psi^B{}_j\psi^C{}_k.
\end{equation}

We now show explicitly that $\nabla_a j^a=0$ is a linear combination
of the evolution equations (\ref{EAi4}) for $\psi^A{}_i$, that is, the
kinematic evolution equations {\em with} the hyperbolicity
fix. Contracting (\ref{EAi4}) with $F^i{}_A$, the matrix inverse of
$\psi^A{}_i$, and using the matrix identity
$\delta(\ln\psi_{x\xi})=F^i{}_A\delta\psi^A{}_i$, we obtain
\begin{equation}
\label{detpsihyperbolic}
(\ln\psi_{x\xi})_{,t}+\hat v^i(\ln\psi_{x\xi})_{,i}+\hat
  v^i{}_{,i}=0. 
\end{equation}
Working from the other end, we insert (\ref{rhoGR}) and (\ref{uaup})
into (\ref{ufromj}) and use (\ref{gtogamma}) to obtain 
\begin{equation}
j^a={\sqrt{k_\xi}\psi_{x\xi}\over
  \sqrt{g_x}}(1,\hat v^i).
\end{equation}
Hence $\nabla_aj^a=0$ is equivalent to 
\begin{equation}
(\sqrt{k_\xi}\psi_{x\xi})_{,t}+(\sqrt{k_\xi}\psi_{x\xi}\hat
  v^i)_{,i}=0.
\end{equation}
But with the advection equation 
\begin{equation}
(k_\xi)_{,t}+\hat v^i(k_{\xi})_{,i}=0,
\end{equation}
which follows from (\ref{kadvection}), this is equivalent to
(\ref{detpsihyperbolic}).


\section{Relativistic dynamics}
\label{section:dynamics}


\subsection{Action and stress-energy tensor}

We begin with the matter action
\begin{equation}
\label{action}
S:=\int e(g^{ab},\psi^A{}_a,k_{AB},\dots,s)
\, g_x^{1/2}\, d^4x,
\end{equation}
where the dots stand for any other tensors on matter space and $s$ is
the entropy per rest mass (a scalar on matter space). Varying for
now only the metric, the standard definition of the stress-energy
tensor $T_{ab}$,
\begin{equation}
\label{Tabdef}
\delta S=:{1\over 2}\int T_{ab} \ \delta g^{ab} \  g_x^{1/2}\, d^4x,
\end{equation}
evaluates to
\begin{equation}
\label{Tabfrome}
T_{ab}=2{\partial e\over\partial g^{ab}}-e g_{ab}.
\end{equation}

We define a projector into the tangent space normal to the 4-velocity,
\begin{equation}
h_{ab}:=u_au_b+g_{ab}.
\end{equation}
$h_{ab}$ should not be confused with the projector $\gamma_{ab}$ into the
$t=\rm const$ hypersurfaces defined in (\ref{gammaab}).

We can now write
\begin{equation}
\label{Tabsplit}
T_{ab}=e u_au_b + p_{ab},
\end{equation}
where
\begin{equation}
p_{ab}: =2{\partial e\over\partial g^{ab}}-e h_{ab},
\end{equation}
which is by definition symmetric. 

We define the pull-back of the spacetime metric to matter space,
\begin{equation}
\label{gABdef}
g^{AB}:={\psi^A}_a{\psi^B}_b g^{ab}.
\end{equation}
We define $g_{AB}$ as its matrix inverse. We therefore now have two
Riemannian metrics on matter space, namely $g_{AB}$ and $k_{AB}$. As a
matter of convention and terminology, we will refer to $k_{AB}$ (only)
as {\em the} matter space metric, but we will later implicitly move
matter space indices (only) with $g_{AB}$ and $g^{AB}$. Note that in
this convention $k^{AB}:=g^{AC}g^{BD}k_{BD}$, and that this is not the
matrix inverse of $k_{AB}$. (We note in passing that the Newtonian
limit ${\psi^A}_i{\psi^B}_j\gamma^{ij}$ of $g^{AB}$ is commonly called
the Finger tensor in the Newtonian literature. The Newtonian
literature implicitly assumes that $k_{AB}$ and $\gamma_{ij}$ are flat
and given in Cartesian coordinates and moves indices
implicitly. Moreover, some expressions can only be made sense of if
$F^i{}_A$ and $\psi^A{}_i$ are also used implicitly to convert between
space and matter space indices.)

As a further illustration of these conventions, the quantity
\begin{equation}
\psi_A{}^a:=\psi^B{}_b g^{ab}g_{AB} 
\end{equation}
is the inverse of $\psi^A{}_a$ (which is not a square matrix, and so
has no matrix inverse) in the sense that
\begin{eqnarray}
\label{invPsi1}
\psi^A{}_a\psi_B{}^a &=&\delta^A{}_B, \\
\label{invPsi2}
\psi^A{}_a\psi_A{}^b&=& h_a{}^b.
\end{eqnarray}
(The first of these follows directly from the definition of $g_{AB}$ as
the matrix inverse of $g^{AB}$. The second can be shown by verifying
that the right-hand side is normal to $u^a$ and $u_b$, and obeys
$h_a{}^b h_b{}^c=h_a{}^c$.) 

From covariance in both spacetime and matter space, we must have
\begin{equation}
\label{gABonly}
e(\psi^A{}_a,g^{ab})=e(g^{AB}),
\end{equation}
as this is the only way the spacetime indices on $\psi^A{}_a$ and
$g^{ab}$ can be contracted. (A more formal proof is given in
\cite{BS}.) Hence
\begin{equation}
\label{dedgab}
{\partial e\over \partial g^{ab}}
={\partial e\over \partial g^{AB}}
{\partial g^{AB}\over \partial g^{ab}}
={\partial e\over \partial g^{AB}}{\psi^A}_a {\psi^B}_b.
\end{equation}
Hence $p_{ab}u^a=0$, and so $u_ah_{bc}T^{ab}=0$. This means that there
is no energy flux relative to the matter. In this sense we are dealing
with ideal (non-dissipative) elastic matter. $p_{ab}$ is called the
pressure tensor (for a perfect fluid, $p_{ab}=p h_{ab}$, where $p$ is
the pressure), and we now see that the Lagrangian $e$ in the action
(\ref{action}) evaluates (for solutions to the Euler-Lagrange
equations) to the total energy density (in the rest frame of the
matter).

We next note that
\begin{equation}
\label{n2fromnabc}
n^2={1\over 3!}n^{abc}n_{abc}={1\over
  3!}g^{ad}g^{be}g^{cf}n_{abc}n_{def}.
\end{equation}
From its relation to the matter space volume form (\ref{nabcdef}),
$n_{abc}$ is independent of $g^{ab}$ in the sense that it is
constructed only from $n_{ABC}$ and ${\psi^A}_a$. Hence, taking a
derivative of (\ref{n2fromnabc}),
\begin{equation}
{\partial n\over \partial g^{ab}}={1\over 2}n h_{ab},
\end{equation}
where in the partial derivative $n$ is considered as a function of
$g^{ab}$, ${\psi^A}_a$ and the matter tensors, as well as $s$.  Then,
defining $\epsilon$ by
\begin{equation}
\label{epsilondef}
e=:n(1+\epsilon),
\end{equation}
we have
\begin{equation}
\label{pabfromepsilon}
p_{ab}=2n{\partial\epsilon\over\partial g^{ab}},
\end{equation}
with the same definition of the partial derivative. (\cite{KS} and
\cite{BS} define $e=n\epsilon$. Here we take the rest mass out of the
energy density to agree with the usual definition of $\epsilon$ in
relativistic hydrodynamics as the {\em internal} energy per rest
mass.) Similarly to (\ref{dedgab}), we can write
(\ref{pabfromepsilon}) as
\begin{equation}
\label{pabfromtauAB}
p_{ab}=n\tau_{AB}\psi^A{}_a\psi^B{}_b,
\end{equation}
where we have defined
\begin{equation}
\label{tauABdef}
\tau_{AB}:=2{\partial\epsilon\over\partial g^{AB}}.
\end{equation}
(The Newtonian limit of $\tau_{AB}$ is commonly called the second
Piaola-Kirchhoff tensor in the Newtonian literature, modulo the
implicit assumptions mentioned above.)


\subsection{Isotropic matter}

We now specialise to the case that the specific internal energy
$\epsilon$ depends on $g^{ab}$, ${\psi^A}_a$, $s$ and a single matter
tensor, the metric $k_{AB}$.  (Modelling matter with an anisotropic
crystal structure would require $e$ to depend on additional tensor
fields on matter space, such as a preferred frame.) $e$ and hence
$\epsilon$ should transform as a scalar both on spacetime and on
matter space. We therefore need to find all double scalars that can be
made from $g^{AB}$ and $k_{AB}$.

From (\ref{gABdef}), we see that $g^{AB}$ transforms as a (2,0)-tensor
on matter space and as a scalar on spacetime. With this in mind we
define
\begin{equation}
{k^A}_B:=g^{AC}k_{BC}=g^{ac}\psi^A{}_a\psi^C{}_c k_{BC}.
\end{equation}
This transforms as a scalar on spacetime and as a $(1,1)$ tensor on
matter space. Hence its eigenvalues transform as scalars on matter
space. They are the required double scalars. (We note that \cite{KS}
work with the (1,1)-tensor on spacetime
${k^a}_b=g^{ac}\psi^B{}_b\psi^C{}_c k_{BC}$ instead. This has the same
eigenvalues as $k^A{}_B$ plus one zero eigenvalue.)

We split the matrix ${k^A}_B$ into its determinant
${k}$ and a unit determinant matrix ${\eta^A}_B$, 
\begin{equation}
{\eta^A}_B:={k}^{-1/3}{k^A}_B,
\end{equation}
and note that the determinant is related to the particle density by 
\begin{eqnarray}
{k}&:=&{1\over 3!}\delta_{ABC}\delta^{DEF}{k^A}_D\,{k^B}_E\,{k^C}_F \,
 \nonumber \\
&=&\delta_{[A}^D\delta_B^E\delta_{C]}^F{k^A}_D\,{k^B}_E\,{k^C}_F \,
 \nonumber \\
&=&g^{[A|D} k_{AD} \ g^{|B|E} k_{BE} \ g^{|C]F} k_{CF}
\nonumber \\
&=&{1\over 3!} g^{AD} g^{BE} g^{CF} n_{ABC} n_{DEF} \nonumber \\
&=& n^2,
\end{eqnarray}
where the first equality is the usual definition of the determinant of
a matrix, the second reminds us that for a (1,1)-tensor this is
actually a scalar, the third is the definition of ${k^A}_B$, the
fourth follows from the fact that $n_{ABC}$ is the volume form of
$k_{AB}$, and the last one is (\ref{n2fromnabc}) pulled back to
matter space.

We can now consider the specific internal energy $\epsilon$ as a
function of $n$, $\eta^A{}_B$ and $s$. In fact, it can depend on
$\eta^A{}_B$ only through its scalar invariants, of which there are
precisely two independent ones. Hence
\begin{equation}
\epsilon(k^A{}_B,s)=\epsilon(k,{\eta^A}_B,s)=\epsilon(n,I^1,I^2,s),
\end{equation}
where $n=k^{1/2}$ as just shown and we have defined
\begin{eqnarray}
\label{I1def}
I^1&:=&{\eta^A}_A ={k}^{-1/3}g^{AB}k_{AB}, \\
\label{I2def}
I^2&:=&{\eta^A}_B{\eta^B}_A={k}^{-2/3}g^{AB}g^{CD}k_{AC}k_{BD}.
\end{eqnarray}
With $g_{AB}$ defined as the matrix inverse of $g^{AB}$ we have
\begin{equation}
{\partial{k}\over\partial g^{AB}}={k}\,g_{AB},
\end{equation}
and hence
\begin{equation}
\label{dndgAB}
{\partial n\over\partial g^{AB}}={1\over 2}n\,g_{AB}.
\end{equation}
We find
\begin{equation}
\label{tauABexpression}
\tau_{AB}={p\over n} g_{AB}+2(f_1\pi^1_{AB}+f_2\pi^2_{AB}),
\end{equation}
where
\begin{eqnarray}
p&:=&n^2{\partial\epsilon\over\partial n}, \\
f_{1,2}&:=&{\partial\epsilon\over\partial I^{1,2}}, \\
\label{pi1def}
\pi^1_{AB}&:=&{\partial I^1\over \partial g^{AB}}=\eta_{AB}-{1\over
    3}g_{AB}I^1, \\
\label{pi2def}
\pi^2_{AB}&:=&{\partial I^2\over \partial g^{AB}}=2(\eta_{AC}\eta^C{}_B-{1\over
    3}g_{AB}I^2).
\end{eqnarray}

Substituting (\ref{tauABexpression}) into (\ref{pabfromtauAB}), we see that
\begin{equation}
\label{pabsplit}
p_{ab}=p h_{ab}+\pi_{ab}, 
\end{equation}
with the first term the stress tensor of a perfect fluid and the
second term representing the anisotropic stress,
\begin{equation}
\label{piab}
\pi_{ab}=\psi^A{}_a\psi^B{}_b\pi_{AB},
\end{equation}
where 
\begin{equation}
\label{piAB}
\pi_{AB}:=2n(f_1\pi^1_{AB}+f_2\pi^2_{AB}).
\end{equation}
Hence $\pi_{ab}$ is a tracefree spatial tensor in the sense that
\begin{equation}
\label{piabconstraints}
\pi_{ab}u^a=0, \qquad h^{ab}\pi_{ab}=0.
\end{equation}
Moreover, $\pi_{ab}$ vanishes if  $\epsilon$ depends only on $n$ and $s$,
which is the fluid limit.

We also note that with the temperature defined by 
\begin{equation}
T:={\partial\epsilon\over\partial s},
\end{equation}
the first law of thermodynamics on a per particle basis can be
written as
\begin{equation}
\label{firstlaw}
d\epsilon=T\,ds-p\,d\left({1\over n}\right) + f_1\,dI^1 + f_2\,dI^2,
\end{equation}
so $f_{1,2}$ are ``generalised forces'' in the thermodynamical sense.


\subsection{The unsheared state}

Elastic matter at a given density $n$ has an {\em unsheared} state
that minimises $\epsilon$ at fixed $n$, but one cannot assume that
there exists a {\em relaxed} state that minimises $\epsilon$
absolutely, including under variation of $n$. This is because at
sufficiently low pressure, and hence $n$, the matter may be in a fluid
rather than solid state \cite{KS}.

It is intuitively clear that the unsheared state corresponds to
${\eta^A}_B={\delta^A}_B$. In fact, we see from
(\ref{pi1def},\ref{pi2def}) that $\pi_{ab}$ vanishes for all values of
$\psi^A{}_a$ if and only if ${\eta^A}_B={\delta^A}_B$. This means that
$\eta_{AB}$ is the matrix inverse of $g^{AB}$, or
\begin{equation}
\eta_{AB}=g_{AB}. 
\end{equation}
Hence
\begin{equation}
\label{unsheared}
k_{AB}=n^{2/3}\,g_{AB}
\end{equation}
in the unsheared state. It is natural to assume that matter freezes in
the unsheared state. Hence we set $k_{AB}$ to (\ref{unsheared}) at the
moment of freezing, and advect it via (\ref{kadvection})
afterwards. Note that $g_{AB}$ is the pull-back of $h_{ab}$, which
even in special relativity is not flat, so in general $k_{AB}$ will
not be flat, except in the Newtonian limit where $h_{ab}=\gamma_{ab}$
is flat and even then only if $n$ takes a constant value at freezing. 


\section{Hyperbolicity}
\label{section:hyperbolicity}


\subsection{Overview}

For smooth solutions, it is natural to consider the relativistic
elasticity equations as a system of second-order PDEs in the variables
$\chi^A(x^a)$. In order to show existence and uniqueness of solutions,
Beig and Schmidt \cite{BS} have introduced an explicit reduction to
first order of these equations, and have shown that the reduction is a
first-order symmetric hyperbolic system, at least in the unsheared
state.

The reduction of any second-order system to first-order hyperbolic
form is complicated by the fact that the reduction creates definition
constraints on the auxiliary variables (here, $\psi^A{}_{[i,j]}=0$),
which can be added to the evolution equations to change their
principal part and hence their hyperbolicity properties. (We note in
this context that in \cite{CGJMM} a definition of symmetric
hyperbolicity for a second-order system has been given as the {\em
  existence} of a symmetric hyperbolic reduction to first, together
with a {\em necessary and sufficient} criterion for this reduction to
exist, which is purely algebraic in terms of the principal symbol of
the second-order system. Hence if well-posedness of the second-order
system is the only concern, constructing an explicit first-order
reduction is unnecessary.)

We have a different reason for constructing an explicit first-order
reduction: we want to construct a numerical scheme that can accurately
reproduce weak solutions of the relativistic elasticity equations. As
for weak solutions of fluid mechanics, the standard way of doing
this is to construct HRSC numerical schemes for the equations in an
appropriate first-order balance law form. 

In this section we will show that the kinematic evolution equations
(\ref{EAitres}), together with the dynamical evolution equations
$\nabla_bT^{ab}=(\hbox{constraints})$, form a symmetric hyperbolic
system of evolution equations for $\psi^A{}_a$, or equivalently
$\psi^A{}_i$ and $\hat v^i$, if the constraints (\ref{CAij}) are
obeyed or not. We also show that (\ref{EAitres}), together with
just $\nabla_bT^{ab}=0$, as used in the Newtonian formalisms
\cite{BDRT,TRT,MillerColella2001,MillerColella2002}, is strongly
hyperbolic but not symmetric hyperbolic.

For completeness, relevant standard definitions of
hyperbolicity are summarised in Appendix~\ref{appendix:hyperbolicity}.


\subsection{The second-order system}

Roughly speaking, the first-order equations for $\psi^A{}_a$ must be
the second-order equations for $\chi^A$, replacing $\chi^A{}_{,ab}$ by
$\psi^A{}_{a,b}$ and adding multiples of the constraint
$\psi^A{}_{[a,b]}$ to the right-hand sides. We therefore derive the
second-order equations first, following \cite{BS}. In particular, this
will allow us to establish the standard connection between the matter
evolution equations and stress-energy conservation.

Hence, in this subsection we consider $g^{ab}$ and $\chi^A$
as the independent variables. We consider $\psi^A{}_a=\chi^A{}_{,a}$
as a derived object, and we consider $k_{AB}$, any other matter
space tensors, and $s$, as fixed tensor
fields on matter space that are not varied in the following. The
action is
\begin{equation}
S:=\int e(g^{ab},\chi^A,\chi^A{}_{,a},k_{AB},\dots,s)\,\sqrt{g_x}\, d^4x.
\end{equation}
After integration by parts, and neglecting the boundary terms, its
variation is
\begin{equation}
\delta S=\int \left( {1\over 2} T_{ab}\  \delta g^{ab}
+{\cal E}_A \ \delta\chi^A \right)\sqrt{g_x}\, d^4x,
\end{equation}
where the stress-energy tensor is given as before by (\ref{Tabfrome}),
and the Euler-Lagrange equations are
\begin{equation}
{\cal E}_A:={\partial e\over \partial\chi_A}-{1\over \sqrt{g_x}}
\left(\sqrt{g_x}{\partial e\over\partial (\chi^A{}_{,a})}\right)_{,a}.
\end{equation}
Note that these are second-order differential equations for $\chi^A$.

Variations generated by an infinitesimal change of coordinates $x^a\to
x^a+\zeta^a$ on the spacetime take the form
\begin{eqnarray}
\delta \chi^A&=&{\cal L}_\zeta\chi^A=\zeta^c \psi^A{}_c, \\
\delta g^{ab} &=&{\cal L}_{\zeta}g^{ab}=2\nabla^{(a}\zeta^{b)}.
\end{eqnarray}
The action must be invariant under such changes, and hence after
another integration by parts
\begin{equation}
\label{TconsEL}
\nabla^bT_{ab}=\psi^A{}_a{\cal E}_A.
\end{equation}
Hence stress-energy conservation holds if and only if the elastic
matter field equations hold. 

${\cal E}_A=0$ has only three independent components, while
$\nabla^bT_{ab}=0$ has four. However, $u^a\nabla^bT_{ab}=0$ is equivalent to
\begin{equation}
n\left(\dot \epsilon-{p\over n^2}\dot n\right)+\pi^{ab}\nabla_au_b=0,
\end{equation}
where a dot denotes $u^a\nabla_a$.  But this is just the first law
(\ref{firstlaw}), evaluated along a particle worldline, for smooth
solutions, so that $\dot s=0$. Hence it is an identity if the
stress-energy tensor is thermodynamically consistent with the equation
of state.

The matter equations in their second-order form can be written as
\begin{equation}
\label{calEAdef}
{\cal E}_A=M^{ab}{}_{AB}\chi^B{}_{,ba}-G_A=0
\end{equation}
where 
\begin{equation}
\label{Mdef}
M^{ab}{}_{AB}:={\partial^2 e\over\partial
  \psi^A{}_a\partial\psi^B{}_b}
\end{equation}
are the coefficients of the principal part and $G_A$ comprises
all lower-order terms. From its definition,
\begin{equation}
\label{Msymmetry}
M^{ab}{}_{AB}=M^{ba}{}_{BA}.
\end{equation}
We shall see that $M^{ab}{}_{AB}$ as defined by (\ref{Mdef}) is not
symmetric in $ab$ alone, even though $M^{[ab]}{}_{AB}$ does not
contribute to (\ref{calEAdef}).


\subsection{The principal symbol}

We shall write the principal symbol more explicitly in terms of the
shear and the equation of state. From (\ref{gABonly}),
\begin{equation}
M^{ab}{}_{AB} = 
4{\partial^2 e\over\partial g^{AC} \partial g^{BD}} \psi^{Ca} \psi^{Db}
+2{\partial e\over\partial g^{AB}}g^{ab},
\end{equation}
where we have defined
\begin{equation}
\psi^{Aa}:=\psi^A{}_b g^{ab}.
\end{equation}
With 
\begin{equation}
\label{gupsi}
g^{ab}=-u^au^b+\psi^{Aa}\psi^{Bb}g_{AB},
\end{equation}
this can be split into parts parallel and normal to the 4-velocity as
\begin{equation}
\label{Mdef3}
M^{ab}{}_{AB}= -\mu_{AB}u^au^b+U_{ACBD}\psi^{Ca} \psi^{Db},
\end{equation}
where
\begin{eqnarray}
\mu_{AB}&:=& 2{\partial e\over\partial g^{AB}}, \\
U_{ACBD}&:=&4{\partial^2 e\over\partial g^{AC} \partial g^{BD}} 
+2{\partial e\over\partial g^{AB}}g_{CD}.
\end{eqnarray}
Note that there are no cross terms, that is $u_ah_{bc}M^{ab}{}_{AB}=0$.

We now evaluate the symbols $\mu_{AB}$ and $U_{ACBD}$ further. With
(\ref{epsilondef}), using (\ref{dndgAB}), 
we can rewrite
\begin{eqnarray}
\mu_{AB}&=& n\tau_{AB}+ e g_{AB}, \\
U_{ACBD}&=&n\bigl(g_{AC}\tau_{BD}+g_{BD}\tau_{AC}+\tau_{AB}g_{CD}
\nonumber \\ &&
\label{UACBD}
+\tau_{ACBD}\bigr)+2eg_{A[C}g_{D]B}, 
\end{eqnarray}
where $\tau_{AB}$ was defined above in (\ref{tauABdef}), and analogously we
have defined
\begin{equation}
\tau_{ABCD}:=4{\partial^2 \epsilon\over\partial g^{AB} \partial g^{CD}}.
\end{equation}

Using the chain rule, we now express $\tau_{AB}$ and $\tau_{ABCD}$ as
a sum of terms, each of which is a product of a matter scalar (such as
$e$, $p$, $c_s^2$ etc.) and a tensor that depends only on the
deformation. We can rewite the expression (\ref{tauABexpression}) for
$\tau_{AB}$ more compactly as
\begin{equation}
\tau_{AB}={p\over n}g_{AB}+2f_\alpha \pi^\alpha_{AB},
\end{equation}
where $\alpha=1,2$ labels the shear scalars, and we use a summation
convention over $\alpha$. With the same notation, we can write
\begin{eqnarray}
\tau_{ABCD}&=&-2{p\over n}g_{A(C}g_{D)B} + \left(c_s^2-{p\over
  n}\right)g_{AB} g_{CD} 
\nonumber \\ &&
+2n \left( g_{AB} f_{n\alpha}\pi^\alpha_{CD}+g_{CD}
f_{n\alpha}\pi^\alpha_{AB}\right)
\nonumber \\ &&
+4f_{\alpha\beta}\pi^\alpha_{AB}\pi^\beta_{CD}
+4f_{\alpha}\pi^\alpha_{ABCD},
\end{eqnarray}
where 
\begin{equation}
c_s^2:={\partial p\over\partial n}, \qquad
f_{n\alpha}:={\partial^2\epsilon\over\partial n\partial I^\alpha}, \qquad
f_{\alpha\beta}:={\partial^2\epsilon\over\partial I^\alpha\partial
  I^\beta}, 
\end{equation}
and 
\begin{eqnarray}
\pi^1_{ABCD}&:=&{\partial^2 I^1\over \partial g^{AB}\partial g^{CD}}
\nonumber \\
&=&\left({1\over 3}g_{A(C}g_{D)B}+{1\over 9}g_{AB}g_{CD}\right)I^1
\nonumber \\ &&
-{1\over 3}(\eta_{AB}g_{CD}+\eta_{CD}g_{AB}), \\
\pi^2_{ABCD}&:=&{\partial^2 I^2\over \partial g^{AB}\partial g^{CD}} 
\nonumber \\
&=&\left({2\over 3}g_{A(C}g_{D)B}+{4\over 9}g_{AB}g_{CD}\right)I^2
\nonumber \\ &&
-{4\over 3}(\eta_{AE}\eta^E{}_Bg_{CD}+\eta_{CE}\eta^E{}_Dg_{AB})
\nonumber \\ &&
+2\eta_{A(C}\eta_{D)B}.
\end{eqnarray}


\subsection{The unsheared state}

The principal symbol simplifies considerably in the unsheared
  state, denoted by a circle, where
\begin{eqnarray}
\qquad \mathring I^\alpha&=&3, \\
\mathring \pi^\alpha_{AB}&=&0, \\
\mathring \pi^1_{ABCD}&=&g_{A(C}g_{D)B}-{1\over 3}g_{AB}g_{CD}, \\
\mathring \pi^2_{ABCD}&=&4\pi^1_{ABCD}, 
\end{eqnarray}
and therefore
\begin{eqnarray}
n\mathring \tau_{AB}&=&pg_{AB}, \\
n\mathring \tau_{ABCD}&=&2{r}\,g_{A(C}g_{D)B}
+ {q}\,g_{AB}g_{CD},
\end{eqnarray}
where
\begin{eqnarray}
{r} &:=& -p+2n(f_1+4f_2), \\
{q} &:=& nc_s^2-p-{4\over 3}n(f_1+4f_2).
\end{eqnarray}
Note that in the unsheared state only the combination $f_1+4f_2$
appears. 

We finally obtain 
\begin{eqnarray}
\label{muABunsheared}
\mathring \mu_{AB}&=&(p+e)g_{AB}, \\
\mathring U_{ACBD}&=&
(2p+q+e) g_{AC}g_{BD} +(p+r)g_{AB}g_{CD}
\nonumber \\ &&
+(r-e) g_{AD}g_{BC}.
\label{UACBDunsheared}
\end{eqnarray}
(This reduces to Eq.~(4.16) of \cite{BS} in the special case $p=\epsilon=0$.)


\subsection{First-order systems}

Any first-order reduction of the second-order system must have the
form
\begin{equation}
\label{barcalEAdef} 
\bar {\cal E}_A:=\bar M^{ab}{}_{AB}\psi^B{}_{b,a}-G_A=0 
\end{equation}
with
\begin{equation}
\bar M^{ab}{}_{AB}:=M^{ab}{}_{AB} + D^{ab}{}_{AB},
\end{equation}
where
\begin{equation}
D^{ab}{}_{AB}=D^{[ab]}{}_{AB}
\end{equation}
governs constraint addition. 

In particular, $\nabla_bT^{ab}=0$ should give us the dynamical part of
the equations, but as we shall see, in order to achieve symmetric
hyperbolicity of the entire system, we will have to add constraints to
these equations. It turns out that adding constraints only to the
``spatial'' part of $M^{ab}{}_{AB}$ is sufficient. Hence, we consider
the evolution equations
\begin{equation}
\bar
E^a:=\nabla_bT^{ab}+\bar\Lambda_{ACBD}\psi^{Aa}\psi^{Cc}\psi^{Db}\psi^B{}_{[b,c]}=0,
\end{equation}
where
\begin{equation}
\bar\Lambda_{ACBD}=-\bar\Lambda_{ADBC}
\end{equation}
parameterises a family of constraint additions. 
To write $\nabla_bT^{ab}$ in terms of $\psi^A{}_a$, we start
from
\begin{equation}
T^{ab}=2{\partial e\over\partial g^{AB}}\psi^{Aa}\psi^{Bb}-eg^{ab}.
\end{equation}
Keeping only the principal part in the matter variables, that is terms
of the form $\psi^B{}_{b,c}$, we find after some calculation that
\begin{equation}
\label{intermediate}
\nabla_bT^{ab}=\left(\psi^{Aa}M^{cb}{}_{AB}-4{\partial e\over\partial
  g^{AB}}\psi^{A[b}g^{c]a}\right)\psi^B{}_{b,c}+\hbox{l.o.}
\end{equation}

Substituting (\ref{gupsi}) into
(\ref{intermediate}), we find that
\begin{equation}
\label{Ebar}
\bar E^a=\Bigl(\psi^{Aa} \bar M^{cb}{}_{AB}
+2u^a\mu_{AB}\psi^{A[b}u^{c]}\Bigr)\psi^B{}_{b,c}+\hbox{l.o.},
\end{equation}
where
\begin{equation}
\label{Mbardef}
\bar M^{ab}{}_{AB}:=M^{ab}{}_{AB}+(\bar\Lambda_{ACDB}-2g_{A[C}
  \mu_{D]B})\psi^{Ca}\psi^{Db}.
\end{equation}
The modification of $M^{ab}{}_{AB}$ can be pulled back to a
modification of $U_{ACBD}$, namely
\begin{equation}
\bar U_{ACBD}:=U_{ACBD}+\bar\Lambda_{ACDB}-2g_{A[C}\mu_{D]B}
\end{equation}

Splitting $\bar E^a$ into its parts parallel and normal to the
4-velocity, we have
\begin{eqnarray}
\label{uEbar}
u_a\bar E^a&=&-2\mu_{AB}\psi^{Ab}u^{c}\psi^B{}_{[b,c]}, \\
\label{psiEbar}
\psi^A{}_a\bar E^a&=& \bar M^{cb}{}_{AB}\psi^B{}_{b,c} -G^A.
\end{eqnarray}
But if $\mu_{AB}$ is invertible, $u_a\bar E^a=0$ is equivalent to the
kinematic evolution equations {\em with} hyperbolicity fix
(\ref{EAitres}).


\subsection{Symmetric hyperbolicity}

The definition of symmetric hyperbolicity for a general system of
first-order evolution equations is reviewed in
Appendix~\ref{appendix:hyperbolicity}. Roughly speaking, the principal
symbol must be symmetric, and its time component must be positive
definite. We begin with the first condition. 

\paragraph*{Symmetry}

The principal symbol of neither (\ref{barcalEAdef}) nor of
(\ref{Ebar}) has the correct index structure $P_{\alpha\beta}{}^c$ in
the composite index defined by $w^\alpha:=\psi^A{}_a$. We follow the
approach of Beig and Schmidt \cite{BS}. Define
\begin{eqnarray}
\label{WabABcdef}
W^{ab}{}_{AB}{}^c&:=&u^a\bar M^{cb}{}_{AB}-2u^{[c}\bar M^{b]a}{}_{BA}
\end{eqnarray}
and consider the system of first-order equations 
\begin{eqnarray}
\label{Weqn}
{\cal E}^a{}_A&:=&W^{ab}{}_{AB}{}^c \psi^B{}_{b,c}-G_A u^a = 0, \\
\label{uchi}
{\cal A}_A&:=&-u^c g_{AB}\chi^B{}_{,c}=0,
\end{eqnarray}
where $\chi^A$ and $\psi^A{}_a$ are now considered as independent
variables, and $u^A$ is defined by the $\psi^A{}_a$ through
(\ref{nabcdef},\ref{jdef},\ref{ufromj}). It is clear that each
solution $\chi^A$ of the second-order system generates a solution
$\psi^A{}_a=\chi^A{}_{,a}$ of this first-order system. Beig and
Schmidt \cite{BS} prove the converse, that a solution $\psi^A{}_i$ of
the first-order system obeying $\psi^A{}_{[i,j]}=0$ gives rise to a
second-order solution $\chi^A$.

The principal symbol of this system has now the correct index
structure.  It is easy to see that it is symmetric, in the sense that
\begin{equation}
W^{ab}{}_{AB}{}^c=W^{ba}{}_{BA}{}^c, \qquad -u^c g_{AB}=-u^c g_{BA},
\end{equation}
if and only if $\bar M^{ab}{}_{AB}$ has the symmetry
(\ref{Msymmetry}). To achieve this, we set
\begin{equation}
\bar\Lambda_{ACBD}=\Lambda_{ADBC}+2g_{A[C}\mu_{D]B},
\end{equation}
where $\Lambda_{ADBC}$ has the symmetries
\begin{equation}
\Lambda_{ACBD}=\Lambda_{BDAC}=-\Lambda_{ADBC},
\end{equation}
and will be determined when we consider positivity of the principal
symbol below.

We now verify that the system (\ref{Weqn},\ref{uchi}) is equivalent to
our evolution equations. From (\ref{Mdef3}) and (\ref{Mbardef}), we
find
\begin{equation}
u_a \bar M^{ab}{}_{AB}=u^b\mu_{AB}
\end{equation}
and hence
\begin{eqnarray}
\label{uW}
u_aW^{ab}{}_{AB}{}^c&=&-\bar M^{cb}{}_{AB}, \\
\label{psiW}
\psi_{Ca}W^{ab}{}_{AB}{}^c&=&-2 \psi^{D[b} u^{c]} U_{ACBD}.
\end{eqnarray}
If $U_{ACBD}$ is invertible as a matrix with composite indices
$AC$ and $BD$, we finally have the decomposition
\begin{eqnarray}
\label{uE}
u_a {\cal E}^a{}_A=0 \quad & \Leftrightarrow & 
\quad \bar M^{cb}{}_{AB}\psi^B{}_{b,c}=G_A, \\ 
\label{psiE}
\psi^E{}_a {\cal E}^a{}_A=0 \quad &
\Leftrightarrow & \quad u^b\psi^A{}_{[a,b]}=0,\\
\label{chiAadvection}
{\cal A}_A=0 \quad & \Leftrightarrow & \quad u^a \chi^C{}_{,a}=0, 
\end{eqnarray}
We see that (\ref{uE}) and (\ref{psiE}) are the same as (\ref{uEbar}),
(\ref{psiEbar}) in our formalism.  Finally, (\ref{chiAadvection}) is
equivalent to (\ref{matteradvection}) (plus similar equations for any
other matter tensors), as
\begin{equation}
u^a k_{AB,a}=k_{AB,C}\, u^a\chi^C{}_{,a}
\end{equation}
by the chain rule.

\paragraph*{Positive definiteness}

The second condition for symmetric hyperbolicity is the existence of a
timelike covector $t_a$ which makes the quadratic form (energy norm)
\begin{equation}
E:=t_c\left(W^{ab}{}_{AB}{}^c m^A{}_a m^B{}_b -u^cg_{AB}l^A l^B\right)
\end{equation}
positive definite, called a subcharacteristic vector. (The formal
arguments $m^A{}_a$ and $l^A$ of this quadratic form are in the
tangent bundle of the phase space, and can be thought of as
perturbations of $\psi^A{}_a$ and $\chi^A$ about a background
solution.) Decomposing $m^A{}_a$ uniquely as
\begin{equation}
\label{mdef}
m^A{}_a=:\alpha^A u_a+\alpha^{AC} \psi_{Ca},
\end{equation}
and choosing $t_a=u_a$, we have
\begin{equation}
E=\mu_{AB}{}\alpha^A\alpha^B+\bar U_{ACBD}\alpha^{AC}\alpha^{BD}
+g_{AB}l^A l^B.
\label{alphasquared}
\end{equation}
Hence $u_a$ is a subcharacteristic vector if $\mu_{AB}$ and $U_{ACBD}$
are positive definite. (Note that then they are also invertible, as we
assumed earlier.) From (\ref{muABunsheared}), $\mu_{AB}$ is positive
definite in the unsheared state if $p+e>0$. 

It remains to look at the positive definiteness of $\bar U_{ACBD}$. We choose
\begin{equation}
\Lambda_{ACBD}=2(d-e-p)g_{A[C}g_{D]B},
\end{equation}
or equivalently 
\begin{equation}
\bar\Lambda_{ACBD}=4nf_\alpha g_{A[C}\pi^\alpha_{D]B}+2dg_{A[C}g_{D]B}
\end{equation}
for the total constraint addition, where $d$ is a parameter to be
determined now. For simplicity, we look again at the unstrained
case. Uniquely decomposing $\alpha^{AB}$ as
\begin{equation} 
\alpha^{AB}=\omega^{AB}+\kappa^{AB}+{\kappa\over 3}g^{AB},
\end{equation}
where the first term is antisymmetric and the second symmetric and
tracefree, we find
\begin{eqnarray}
\mathring{\bar U}_{ACBD}\alpha^{AC}\alpha^{BD}
&=&\left(nc_s^2+{2d\over 3}\right)\kappa^2+d\,\omega^{AB}\omega_{AB}
\nonumber \\ &&
+\left[4n(f_1+4f_s)-d\right]\kappa^{AB}\kappa_{AB}.
\nonumber \\
\end{eqnarray}
We now see that this quadratic form is positive definite, and hence
our evolution equations are symmetric hyperbolic, for
$0<d<4n(f_1+4f_2)$ (assuming that $c_s^2\ge 0$). Hence, adding some
constraints to $\nabla_b T^{ab}=0$ is necessary for symmetric
hyperbolicity, for example with the mid-range value of
$d=2n(f_1+4f_2)$.

Tracing all the definitions back, we can write this particular
constraint addition as
\begin{equation}
\bar E^a=\nabla_bT^{ab}
+h^{ac}\psi^{Db}\left[2nf_\alpha\pi^\alpha_{DB}+4n(f_1+4f_2)g_{DB}\right]
\psi^B_{[b,c]}.
\end{equation}
Looking back, the first term in the square brackets
makes the principal part of the second-order system symmetric, and the
second makes it positive definite. 


\subsection{Characteristics of the first-order system}
\label{section:characteristics_1}

As reviewed in Appendix~\ref{appendix:hyperbolicity}, $k_a$ is a
characteristic covector of the first-order equations with
characteristic variable $w^\beta=m^B{}_b$ if
\begin{equation}
\label{firstchar}
W^{ab}{}_{AB}{}^cm^B{}_bk_c=0.
\end{equation}
Once again we decompose $m^B{}_b$ in the form (\ref{mdef}). Fixing an
irrelevant overall factor, we parameterise the wave number $k_a$ as
\begin{equation}
k_a=\lambda u_a-e_a,
\end{equation}
where $e_a=\psi^A{}_ae_A$ is a unit covector on spacetime normal to
$u^a$ and $e_A$ the corresponding unit (with respect to $g^{AB}$)
covector on matterspace.  As reviewed in
Appendix~\ref{appendix:hyperbolicity}, $\lambda$ is then the physical
velocity of the mode relative to the matter. Using the decomposition
(\ref{uW},\ref{psiW}), (\ref{firstchar}) is equivalent to the pair
\begin{eqnarray}
\label{firstchar1}
\bar U_{ACBD}\left(\alpha^B e^D+\lambda \alpha^{BD}\right) &=&0, \\
\label{firstchar2}
\lambda \mu_{AB}\alpha^B+\bar U_{ACBD}e^C\alpha^{BD}&=&0.
\end{eqnarray}
Moreover, symmetric hyperbolicity implies that $\bar U_{ACBD}$ is
invertible and so (\ref{firstchar1}) is equivalent to
\begin{equation}
\label{firstchar1bis}
\alpha^B e^D+\lambda \alpha^{BD}=0.
\end{equation}

Eq.~(\ref{firstchar1bis}) has two classes of solutions. One class obeys
\begin{equation}
\label{unphysicalmodes}
\lambda=0, \quad \alpha^B=0, 
\end{equation}
with $\alpha^{BD}$ restricted by (\ref{firstchar2}) to obey
\begin{equation}
\label{class1}
\bar U_{ACBD}e^C\alpha^{BD}=0.
\end{equation}
These modes travel at zero speed relative to the matter. As
(\ref{class1}) represents 3 equations for 9 components of
$\alpha^{BD}$, there are 6 such modes. They can be parameterised
explicitly as
\begin{equation}
\label{class1parameterised}
\alpha^{BD}=(\bar U^{-1})^{ACBD}v_Aw_C, \quad w_Ce^C=0.
\end{equation}

The other class obeys
\begin{equation}
\lambda\ne 0, \quad \alpha^{BD}=-\lambda^{-1}\alpha^B e^D,
\end{equation}
or equivalently
\begin{equation}
\label{charfirstorder1}
m^B{}_b=\lambda^{-1}\alpha^Bk_b.
\end{equation}
Hence these modes are physical, obeying the constraints. Furthermore,
all physical modes are of this form, which indicates that the modes in
the class (\ref{unphysicalmodes}) are all unphysical. Substituting
(\ref{charfirstorder1}) into (\ref{firstchar2}), we find
\begin{equation}
\label{charfirstorder2}
(-\lambda^2 \mu_{AB}+\bar U_{ACBD}e^Ce^D)\alpha^B=0, 
\end{equation}
or equivalently 
\begin{equation}
\label{charfirstorder2bis}
(-\lambda^2 \mu_{AB}+U_{ACBD}e^Ce^D)\alpha^B=0
\end{equation}
(constraint addition drops out). This can be written as
\begin{equation}
\label{charvar}
\Delta_{AB}\alpha^B=0, \qquad \Delta_{AB}:=M^{ab}{}_{AB}k_ak_b.
\end{equation}
But, as reviewed in Appendix~\ref{appendix:hyperbolicity}, this is
precisely the condition for $k_a$ to be a characteristic covector of
the second-order system with characteristic variable
$\alpha^A$. Hence the physical modes of the first-order system
correspond one-to-one to the modes of the second-order system. There
are 6 of these, forming 3 pairs with speeds $\pm\lambda$ relative to
the matter.


\subsection{Characteristics of the second-order system}
\label{section:characteristics_2}

We now look at the solutions of (\ref{charvar}) in more detail. The
general expression for $\Delta_{AB}$ is quite long, and so we begin
our analysis with the unsheared state. We find
\begin{equation}
\mathring\Delta_{AB}=(-A\lambda^2+B)g_{AB}+Ce_Ae_B,
\end{equation}
where
\begin{equation}
A=e+p, \quad B=p+{r}, \quad C=2p+{r}+{q}.
\end{equation}
We can now read off the characteristic covectors and characteristic
variables by inspection. Transversal waves have eigenvectors obeying
$\alpha^Be_B=0$ (and so have two polarisations travelling with the
same velocity), and
$\Delta_{AB}\alpha^B=0$ then gives
\begin{equation}
\lambda^2={B\over A}={2f_1+8f_2\over1+\epsilon+{p\over n}}=:\lambda_T^2.
\end{equation}
Longitudinal waves have eigenvectors $\alpha^B\propto e^B$ and
$\Delta_{AB}\alpha^B=0$ gives
\begin{equation}
\lambda^2={B+C\over A}={c_s^2\over1+\epsilon+{p\over n}}+{4\over
  3}\lambda_T^2=:\lambda_L^2.
\end{equation}

Taking the Newtonian limit of these characteristic speeds, we can
identify the shear modulus $\mu$ and the bulk modulus $K$ as
\begin{eqnarray}
\label{shearmodulus}
\mu&=&n(2f_1+8f_2), \\
\label{bulkmodulus}
K&=&nc_s^2.
\end{eqnarray}
(These expressions hold in units where the speed of light $c$ is one,
and where $n$ is the rest mass density, rather than the particle
number density. Otherwise they have to be multiplied by
$c^2$ and the particle mass.)

In the general, sheared, case the matter space tensor $\Delta_{AB}$ is
constructed from $g_{AB}$, $g^{AB}$, $\eta_A{}^B$ and $e_A$. It would
therefore be natural to decompose $e_A$ (and $\alpha^B$) into
eigenvectors of $\eta_A{}^B$, which are automatically also
eigenvectors of $g_A{}^B=\delta_A{}^B$. This can be done trivially by
assuming that the index $A$ labels that basis, so that $\eta_A{}^B$ is
diagonal. The result is of the form
$\Delta(\lambda)=\lambda^2\Delta_2+\Delta_0$ (dropping the indices on
$\Delta_{AB}$). We have solved the resulting cubic equation for
$\lambda^2$ by computer algebra but the result is too complicated to
be illuminating.

Furthermore, for numerical purposes we are interested in the {\em
  coordinate} speeds $\lambda=dx^i/dt$ in the $x^i$, $i=1,2,3$
direction of some Eulerian coordinate system, that is, in
characteristic covectors of the form
\begin{equation}
k_a=\lambda (dt)_a-(dx^i)_a.
\end{equation}
This gives a characteristic equation of the form
\begin{equation}
\label{coordinate_delta}
\Delta\alpha=(\lambda^2\Delta_2+\lambda\Delta_1+\Delta_0)\alpha=0. 
\end{equation}
The resulting $\lambda$ are not related to the characteristic speeds
relative to the matter 4-velocity in a simple way, because of the
appearance of the Lorentz factor $W$ in the relativistic velocity
addition. To solve (\ref{coordinate_delta}) numerically, we use a
standard linear algebra package to find the right eigenvectors
$(\lambda\alpha,\alpha)^T$ and eigenvalues $\lambda$ of the matrix
\begin{equation}
\left(\begin{array}{cc}
-(\Delta_2)^{-1}\Delta_1 & -(\Delta_2)^{-1}\Delta_1 \\
I & 0 \\
\end{array}\right).
\end{equation}


\subsection{Strong hyperbolicity}

In the Newtonian literature, the evolution equations for $\psi^A{}_i$
are taken to be (\ref{EAitres}) (with constraint addition), but no
constraints are added to $\nabla_bT^{ab}=0$. Our results above show
that the first-order system is then definitely not symmetric
hyperbolic, as the term $\bar U_{ACBD}$ in the principal symbol is
then not positive definite even in the unsheared state, and is not
symmetric in general (although it is symmetric in the unsheared
state). However, the first-order system is still strongly hyperbolic
if it admits a complete set of characteristic variables. We have just
done the calculation in Sec.~\ref{section:characteristics_1}, and need
to see only what changes if we cannot assume that $\bar U_{ACBD}$ is
symmetric and positive definite.

(\ref{firstchar1}) is no longer equivalent to (\ref{firstchar1bis}) 
because $\bar U_{ACBD}$ may not have an inverse, but solutions of
(\ref{firstchar1bis}) are solutions of (\ref{firstchar1}). Furthermore,
(\ref{class1}) only admits a larger solution space if $\bar U_{ACBD}$
does not have maximal rank, so it will still have 6 solutions, even if
they can no longer be explicitly parameterised by
(\ref{class1parameterised}). Hence the 6 unphysical modes still
exist. The 3 physical modes are completely unaffected by constraint
addition (as one would expect) because we solve
(\ref{charfirstorder2bis}) to find them. 

Hence we have proved that the first-order system consisting of
(\ref{EAitres}) and $\nabla_aT^{ab}=0$ is strongly hyperbolic (but not
symmetric hyperbolic) as long as the second-order system is
strongly hyperbolic. 


\section{Stress-energy conservation in 3+1 form}
\label{section:valencia}


\subsection{Conservation laws}

The energy and momentum conservation laws in general relativity are
the spacelike and timelike components of stress-energy conservation
\begin{equation}
\nabla_a T^{ab}=0. 
\end{equation}
In \cite{Font}, this is decomposed into four balance laws as
\begin{eqnarray}
\label{Sjdot}
\nabla_a\left[T^{ab} (\partial_j)_b\right] &=&
  T^{ab}\nabla_{(a}(\partial_j)_{b)}, \\
\label{taudot}
\nabla_a\left(-T^{ab} n_b\right) &=&
  -T^{ab}\nabla_{(a} n_{b)},
\end{eqnarray}
where $n^a$ is the unit normal on the $t=\rm const$ surfaces.  The
right-hand sides can be seen as a failure of stress-energy
conservation to split into separate proper conservation laws for the
energy and each momentum component, due to the failure of the thee
spatial coordinate basis vectors $(\partial_j)^a$ and the timelike
unit normal vector $n^a$ to be Killing. (The choice of the four basis vectors
is merely conventional).  In a 3+1 split, (\ref{Sjdot}) and
(\ref{taudot}) become
\begin{eqnarray}
\label{cons1}
\left(\alpha\sqrt{\gamma_x}{T^0}_j\right)_{,t}
+\left(\alpha\sqrt{\gamma_x}{T^i}_j\right)_{,i} &=& \dots, \\
\label{cons2}
\left(\alpha^2\sqrt{\gamma_x}T^{00}\right)_{,t}
+\left(\alpha^2\sqrt{\gamma_x}T^{0i}\right)_{,i} &=& \dots.
\end{eqnarray}

We now restrict to the elastic matter stress-energy tensor 
\begin{equation}
T^{ab}=(e+p) u^a u^b + p g^{ab}+\pi^{ab}.
\end{equation}
To insert this into (\ref{cons1}) and (\ref{cons2}), we need 
certain components of $\pi^{ab}$.  From (\ref{piab}) and
(\ref{psiAt}) we have
\begin{equation}
\pi_{00}=\pi_{ij}\hat v^i\hat v^j, \qquad
\pi_{0i}=-\pi_{ij}\hat v^j.
\end{equation}
Using the 3+1 decomposition of the metric, the components that we need
in (\ref{cons1},\ref{cons2}) are
\begin{eqnarray}
\pi^{00}&=&\alpha^{-2}\pi_{ij}v^iv^j, \\
{\pi^0}_i&=&\alpha^{-1}\pi_{ij}v^j, \\
{\pi^i}_j&=&(\gamma^{ik}-\alpha^{-1}\beta^iv^k)\pi_{kj}, \\
\pi^{0i}&=&(\alpha^{-1}\gamma^{ij}-\alpha^{-2}\beta^iv^j)v^k\pi_{jk}.
\end{eqnarray}
From $g^{ab}\pi_{ab}=0$, we have
\begin{equation}
v^iv^j\pi_{ij}=\gamma^{ij}\pi_{ij}=:\pi.
\end{equation}

In numerical hydrodynamics, the conservation laws for the
stress-energy are closed by the equation of state together with the
explicit particle number conservation law
\begin{equation}
\label{rhoconservation}
\nabla_a(n u^a)=0.
\end{equation}
As we have shown in Sec.~\ref{section:particlecurrent}, this evolution equation for $n$ is equivalent
to that for $\psi^A{}_i$ (with the hyperbolicity fix) even when the
constraints are not obeyed. Therefore, where we write $n$ below,
either value could be used without changing the
hyperbolicity. However, we shall test this numerically by obtaining a
value $n_\psi$ from $\psi^A{}_i$ and a value $n_D$ from $D$, and using
either the one or the other. 

The conservation laws (\ref{Sjdot},\ref{taudot},\ref{rhoconservation})
can be written in the form
\begin{equation}
\label{eq:consform1}
\left(\sqrt{\gamma_x}{\cal U}\right)_{,t}+\left(\alpha\sqrt{\gamma_x}{\cal
  F}^i\right)_{,i}=\hbox{source terms}.
\end{equation}
(Note the explicit insertion of $\sqrt{\gamma_x}$ and $\alpha$, which
is only a convention -- we follow \cite{Font}). 
The conserved variables ${\cal U}=(D,S_i,\tau)$ are
related to the primitive variables via
\begin{eqnarray}
\label{Ddef}
D&=&\alpha n u^0=n W, \\
\label{Sidef}
S_i&=& \alpha {{T}^0}_i = n  hW^2v_i+\alpha\pi^0{}_i
\nonumber \\
&=&n  hW^2v_i+\pi_{ij}v^j, \\ 
\label{taudef}
\tau &=& \alpha^2 {T}^{00} - D = n  h W^2-p + \alpha^2 \pi^{00} - D
\nonumber \\ &=&n  (h W^2-W)-(p-\pi),
\end{eqnarray}
where we have defined the standard specific enthalpy
\begin{equation}
\label{hdef}
h:= 1+\epsilon+{p\over n }.
\end{equation}
The corresponding fluxes are 
\begin{eqnarray}
\label{FDdef}
{\cal F}(D)^i&=& n u^i=n \alpha^{-1}W \hat v^i, \\
{\cal F}(S_j)^i&=& T^i{}_j=n  h W^2\alpha^{-1}\hat v^i
v_j+p\delta^i{}_j+\pi^i{}_j, \\
{\cal F}(\tau)^i&=& \alpha T^{0i}-{\cal F}(D) \nonumber \\
&=& n  (h W^2-W)\alpha^{-1}\hat v^i
+p\alpha^{-1}\beta^i+\alpha \pi^{0i} \nonumber \\
&=&n  (h W^2-W)\alpha^{-1}\hat v^i \nonumber \\ 
&& +(p-\pi)\alpha^{-1}\beta^i+\gamma^{ij}\pi_{jk}v^k.
\end{eqnarray}
Following \cite{Font}, we have subtracted the rest energy from the
total energy density in order to obtain the usual Newtonian energy
conservation law in the Newtonian limit.


\subsection{Conversion of conserved to primitive variables}
\label{section:conversion}

In any numerical scheme, we frequently need to calculate the primitive
variables $n$, $v^i$ and $\epsilon$ from the related conserved
variables $D$, $S_i$ and $\tau$, and hence further variables such as
$p$ and $\pi_{ij}$ that appear in the fluxes. We assume that we have
an equation of state that relates $\epsilon$, $n$, $s$ and $I^\alpha$,
and that we can use this to find the generalised forces $p$ and
$f_\alpha$. We also assume that $g_{ab}$ is evolved using the Einstein
equations, that $k_{AB}$ is advected using (\ref{kadvection}) and that
$\psi^A{}_i$ is evolved using the balance law (\ref{EAitres}). (Note
that for our purposes $\psi^A_i$ is both a conserved and a primitive
variable.)

The obvious difficulty is that to calculate $\pi_{ij}$ from
$\pi_{AB}$, we need $v^i$ while $\pi_{ij}$ is needed to extract $v^i$
from $S_j$. We therefore need to proceed iteratively. We {\em guess}
the primitive variables
\begin{equation}
\label{eq:p2cguesses}
p-\pi, \quad \pi_{ij}v^j.
\end{equation}
From
\begin{eqnarray}
\tau+D+(p-\pi) =n  h W^2&=:&Z, \\
(S_i-\pi_{ik}v^k)(S_j-\pi_{jl}v^l)\gamma^{kl}&=&Z^2 v^2
\end{eqnarray}
we obtain $Z$ and $v^2$ and hence $W$ from (\ref{Wdef}), followed by
$n_D$ from (\ref{Ddef}) and $n_\psi$ from (\ref{rhoGR}) (we compute
both, but choose one value to use as $n$ in the remainder of the
calculation), $v^i$ from (\ref{Sidef}) and $h$ from (\ref{taudef}). We
now have a complete set of primitive matter variables, but these will
not be consistent with the equation of state. We therefore now
recompute $p$ from the equation of state, compute $g^{AB}$ from
(\ref{gABdef}), and $\eta^A{}_B$ from $k_{AB}$ and $g^{AB}$, and hence
obtain $\pi^{1,2}_{AB}$ and $I^{1,2}$. We then compute $\pi_{ab}$ from
(\ref{piab},\ref{piAB}), using $v^i$ and the equation of state. Hence
we finally recompute $\pi_{ij}v^j$ and $p-\pi$. We then have four
residuals giving the discrepancy (four numbers) between our original
guesses~\eqref{eq:p2cguesses} and the recomputed values, as a function
of the four initial guesses. We can now use any standard root-finding
method, such as a Newton solver, to find the solution (and hence the
correct primitive values) to desired accuracy. (Note: this will
converge only with a good initial guess, and a solution may not exist
or be unique in general.)

Note that in the fluid limit we only need to guess $p$, and our scheme
then reduces to the standard conserved-to-primitive conversion,
requiring a root find in one variable \cite{Font}.


\section{Numerical tests}
\label{section:numerics}


\subsection{Description of the code}
\label{section:code}

The computer code used to produce the following results uses planar
symmetry; all of the variables in the code are 3-dimensional, but the
system is only evolved in one or two dimensions. The numerical methods
employed are those in~\cite{Millmore}. Briefly, the code uses a HRSC
method with a third-order Runge-Kutta time evolution. In the
reconstruction, standard slope limiting techniques, applied to the
primitive variables are used -- all results shown used van Leer's MC
limiter (\cite{vanLeerMC}). The HLL approximate Riemann solver
(\cite{HLL}) is used to calculate the fluxes. The code can be run
using either the relativistic or Newtonian set of governing
equations.

The HLL flux is
\begin{equation}
{\bf f}_{i-\frac{1}{2}} = \frac{{\bf f}({\bf q}^R_{i-1}) + {\bf f}({\bf q}^L_{i}) + \bar{\lambda}_{\rm
    HLL} \left({\bf q}^R_{i-1} - {\bf q}^L_{i}\right)}{2},
\end{equation}
where ${\bf q}^R_{i-1}$ and ${\bf q}^L_{i}$ are the right and left
reconstructed vectors of conserved variables for the
$(i-1)^{\text{th}}$ and $i^{\text{th}}$ cells, respectively, and
$\bar{\lambda}_{\rm HLL}$ is an estimate of the absolute value of the
largest coordinate characteristic speed. We set this either to
$\max|\lambda|$ at one point (using the numerical calculation outlined
in Sec.~\ref{section:characteristics_2}), to $\max|\lambda|$ over the
whole grid, or to a constant (for example, $\bar{\lambda}_{\rm HLL}=1$
in highly relativistic situations and assuming the matter evolution is
causal).

In two dimensions standard directional splitting techniques are
used. Specifically, on our (logically) Cartesian grid we compute the
appropriate one dimensional fluxes ${\cal F}^i$ required by
equation~\eqref{eq:consform1} by sweeping through the grid lines one
dimension at a time. The update terms are accumulated and applied
simultaneously to minimize symmetry errors caused by the splitting.

We briefly note that the performance of the code has been compared to
a relativistic hydrodynamics code by reducing the elasticity code
explicitly to the hydrodynamic limit. As none of our codes have been
optimized for performance, any comparisons will be
approximate. Nevertheless, as the elasticity code is approximately 8
times slower than the hydrodynamic code on the same problem
(equivalently, the run-time of the hydrodynamic code is approximately
12\% of the run-time of the elastic code) we see that performance will
likely be an issue in realistic simulations.


\subsection{$n_D$ versus $n_\psi$}\label{rhovn}

As discussed in Sec.~\ref{section:conversion}, the particle number
density $n$ can either be obtained from the conserved variable
$\psi^A{}_i$ (or its inverse $F^i{}_A$ in the mixed framework), or
from the conserved variable $D$. If we evolve $D$ as a dynamical
variable and use $n_D$ to represent the primitive variable, we have
one more variable than if we use $n_\psi$. However, we have shown in
Sec.~\ref{section:particlecurrent} the evolution equations for $n_D$
and $n_\psi$ are equivalent even if the constraints are not obeyed,
and so we expect that both formulations have identical stability
properties.

In fact, when these two evolutions are compared, the RMS relative
error in the resulting data is small; we expect that this is
finite-differencing error, as it converges away between first and
second order. Hoewever, when $n_\psi$ and $n_D$ are compared for a single
evolution where $n_D$ is dynamical, the difference is of the order of
round-off error rather than finite-differencing error; we suspect that
this is an artifact of planar symmetry.


\subsection{$\psi$ versus $F$}\label{psivF}

A mixed framework using the inverse $F^A{}_i$ of the configuration
gradient $\psi^i{}_A$ is outlined in
Appendix~\ref{appendix:mixed}. For constraint satisfying initial data
the results of the two frameworks should be the same. We have
implemented both frameworks numerically and compared them. We find that the
difference is on the order of the finite differencing error for
constraint satisfying initial data, as expected. 

Some of the tests given in \cite{BDRT} and \cite{TRT} do not satisfy
the constraints -- namely the second test in \cite{BDRT} and the fifth
test in \cite{TRT}. 
As the evolution of such data depends on the choice of constraint
addition to the equations, we would expect it to depend on the
framework used. Our numerical results obtained in the ``mixed''
framework (presented in Appendix~\ref{appendix:mixed}), which is that
used in \cite{BDRT, TRT}, matches their numerical results for all
tests. Our results using the Eulerian framework (presented in
Sec.~\ref{section:kinematics} and also used by
\cite{MillerColella2001, MillerColella2002}) match only for the
physical tests, where the initial data obey all constraints.


\subsection{Newtonian Riemann tests}
\label{compare}

To validate our Newtonian code, and the Newtonian limit of our
relativistic code, we have compared our results
to two previously published studies (\cite{BDRT} and
\cite{TRT}). These results use the Newtonian theory and the mixed
framework outlined in Appendix~\ref{appendix:mixed}. 

\begin{figure}
  \includegraphics[width=0.49\textwidth]{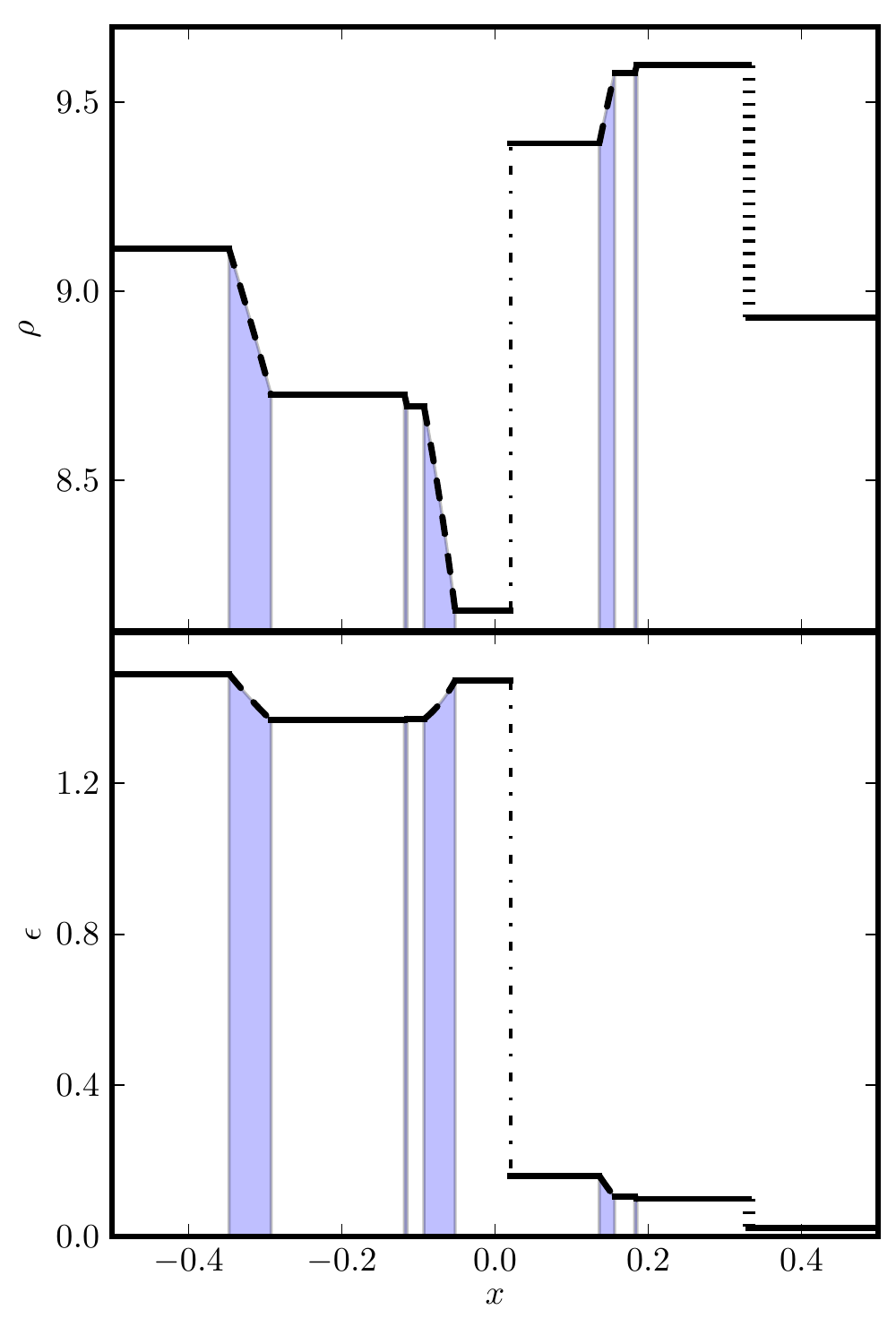}
  \caption{The density and specific internal energy in the first
    Newtonian Riemann test from \cite{BDRT} (from now on BDRT1),
    illustrating the seven waves possible in elastic matter. This is
    the exact solution, illustrating the wave structure in detail. The
    rarefactions -- the 1, 2, 3, 5 and 6-waves -- are given by the
    blue dashed lines and are shaded beneath to show the width of the
    fan. The contact -- the linear 4-wave -- is given by the dotted
    black line. The shock -- the 7-wave -- is given by the thick
    dash-dotted black line. It is clear that resolving some of the
    rarefaction waves will be difficult at moderate resolution. }
  \label{fig:BDRT1_Structure}
\end{figure}

\begin{figure}
  \includegraphics[width=0.49\textwidth]{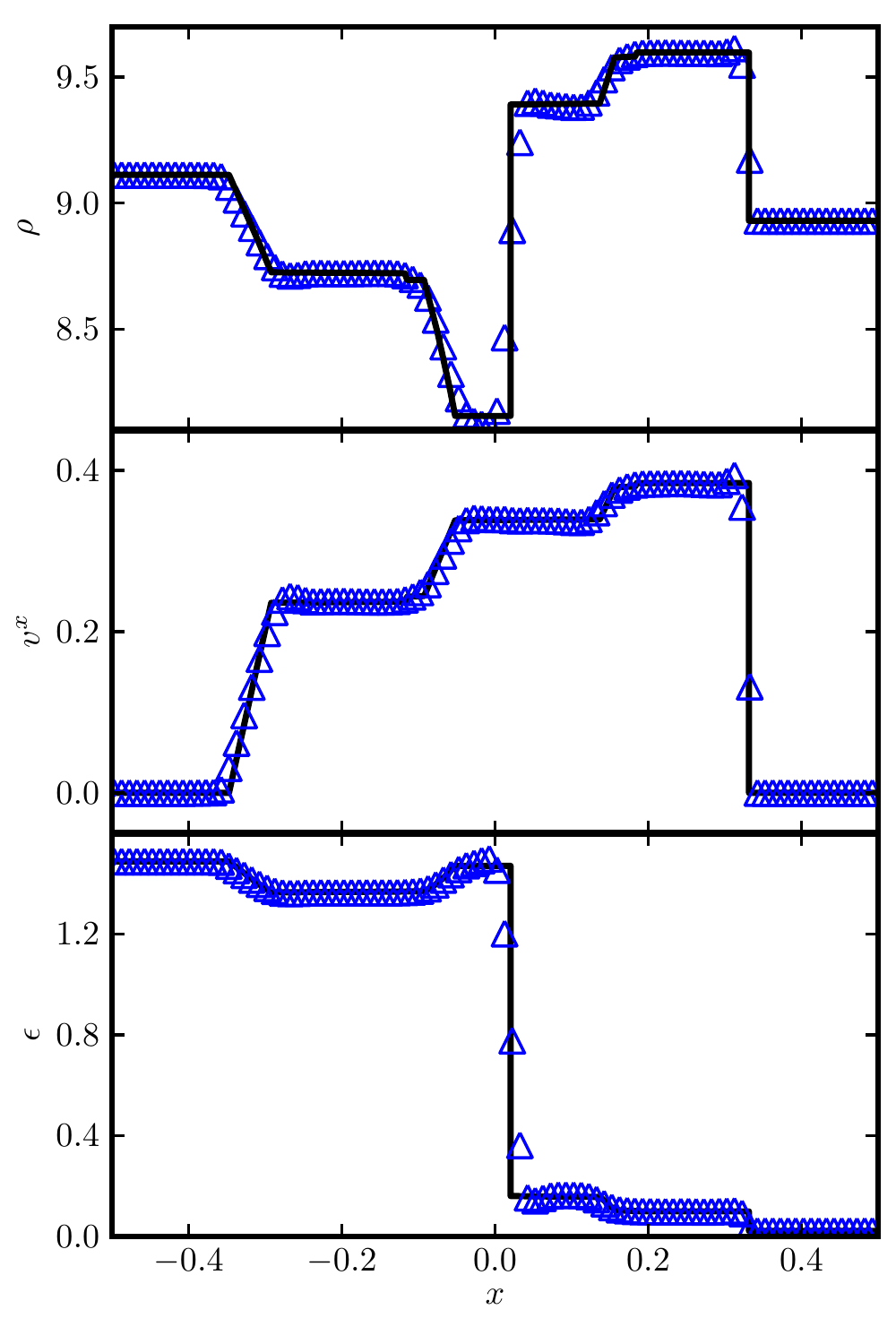}
  \caption{Numerical solution of the BDRT1 test. We show the results
    of our Newtonian code using 200 points (only 100 are plotted for
    clarity), with the exact solution given by the solid
    line. Density, specific internal energy and normal velocity are
    shown. Only minor under/over shoots are visible.}
  \label{fig:BDRT_Scalars_200}
\end{figure}

Broadly the results obtained from our codes matched those shown in
\cite{BDRT} and \cite{TRT}. As an example, we show the results for the
first test of \cite{BDRT} in
Figs.~\ref{fig:BDRT_Scalars_200}--\ref{fig:BDRT_Psi_1000}, using the
results of the Newtonian code. The precise initial data used is
outlined in Appendix~\ref{appendix:initialdata}. We see the seven
waves expected for this solution; three left travelling rarefactions
(the second is very small), a contact, two right travelling
rarefactions (again the second is very small), and a fast shock.  For
clarity, the wave structure of the exact solution is shown in
Fig.~\ref{fig:BDRT1_Structure}.  All waves are captured with only
minor under/over shoots, and the numerical solutions converge to the
exact solution \cite{PB} with resolution, as seen by comparing
Figs.~\ref{fig:BDRT_Scalars_200} and \ref{fig:BDRT_Psi_200} with
Figs.~\ref{fig:BDRT_Scalars_1000} and \ref{fig:BDRT_Psi_1000}.

Similar results are seen for all comparison tests.  However, not all
of the tests run robustly for all numerical methods possible within
our code. An example is the sonic point test problem outlined in
\cite{TRT} (see in particular Figs.~5-7 there). At the contact
discontinuity there is an unphysical ``dip'' in the density and a
corresponding ``jump'' in the internal energy. This is the classical
``wall-heating'' effect seen by most numerical methods when strong
rarefactions separate (e.g., on reflections from walls or the origin
in spherical symmetry -- see \cite{Noh} for the classical case and
\cite{MM} for a brief discussion of the relativistic case). In our
case these artefacts lead, for certain choices of numerical
parameters, to numerical results that are unphysical. This typically
manifests itself by an imaginary characteristic speed, usually as the
squared sound speed becomes negative.  Variants of the code which rely
on calculations of the characteristic information immediately
fail. This problem will only affect some numerical methods at low
accuracy in certain, somewhat artificial, situations, so is unlikely
to cause problems in realistic situations.

Finally, we note that a direct and comprehensive comparison to the
results of \cite{TRT} is complicated by two issues. Firstly the units
for the entropy appear inconsistent there, as detailed in
Appendix~\ref{appendix:initialdata}. Secondly we do not find agreement
in the comparison of the pressure tensor $p_{ij}$ (denoted
$\sigma$ there). As all other values and wave speeds match up well,
and we have comprehensive quantitative agreement with the results of
\cite{BDRT}, we believe our results to be correct.

%
\begin{figure}
  \includegraphics[width=0.49\textwidth]{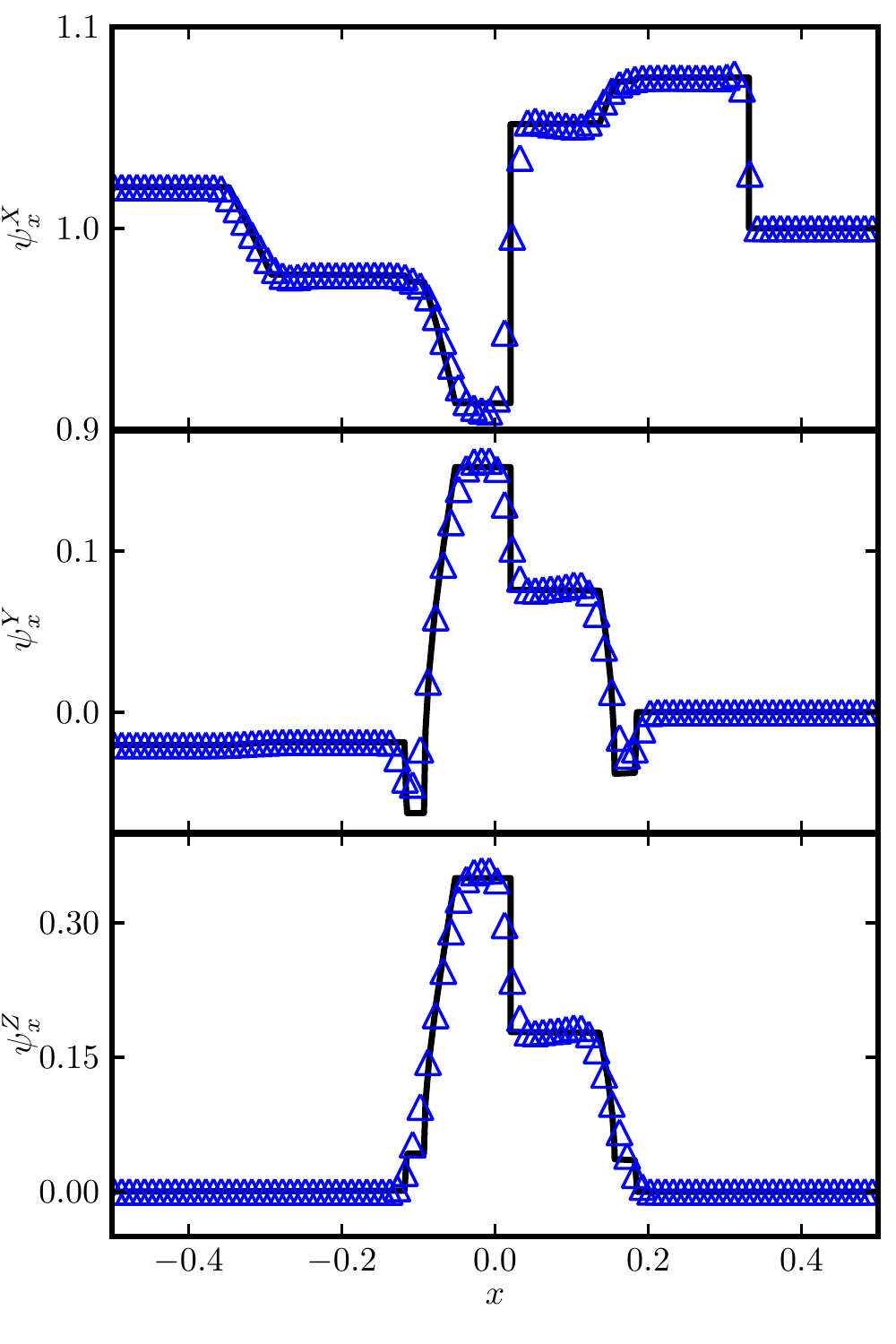}
  \caption{Numerical solution of the BDRT1 test. We show the results
    of our Newtonian code using 200 points (only 100 are plotted for
    clarity), with the exact solution given by the solid
    line. Components of the configuration gradient are shown. Only
    minor under/over shoots are visible.}
  \label{fig:BDRT_Psi_200}
\end{figure}
%
\begin{figure}
  \includegraphics[width=0.49\textwidth]{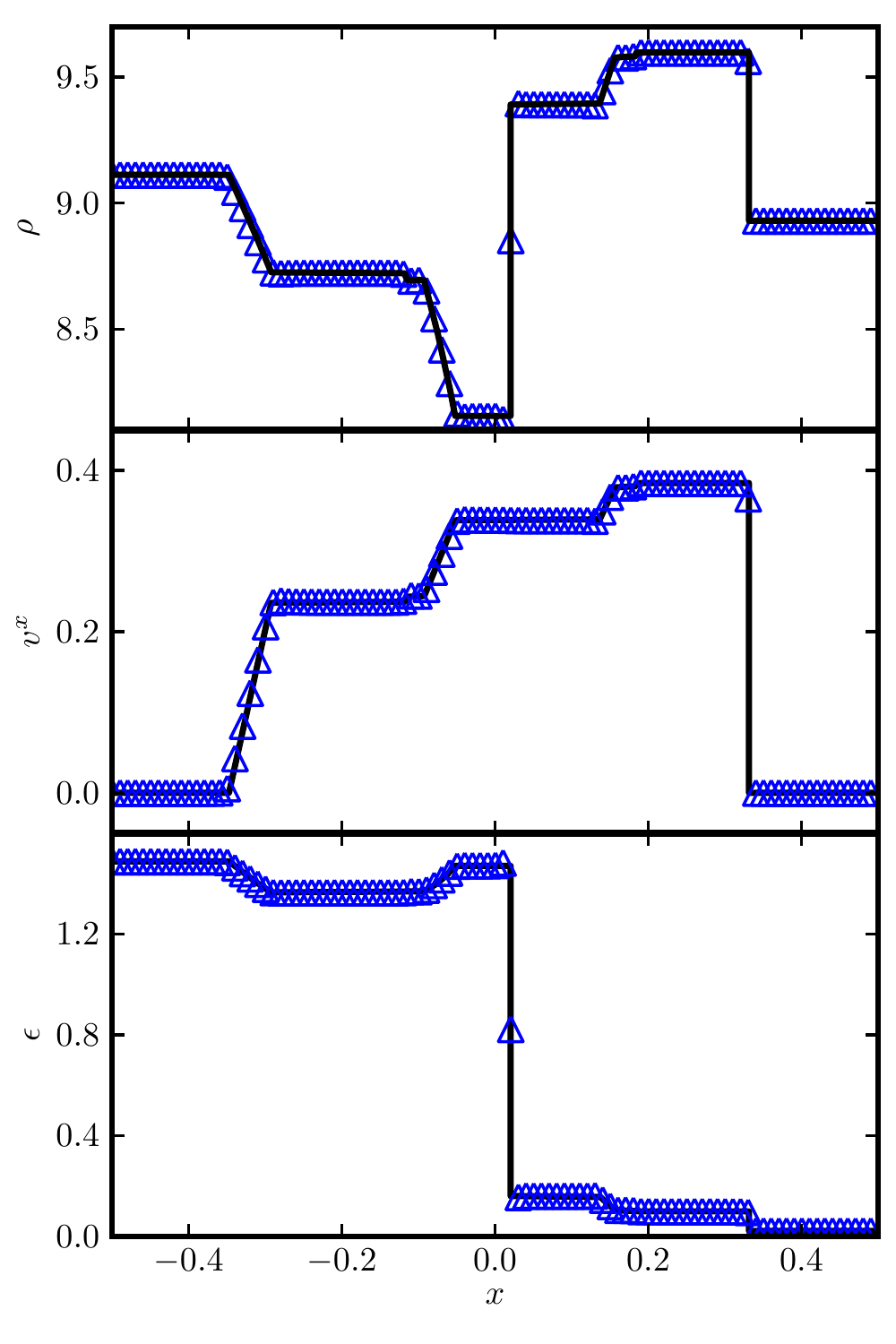}
  \caption{Numerical solution of the BDRT1 test. We show the results
    of our Newtonian code using 1000 points (only 100 are plotted for
    clarity), with the exact solution given by the solid
    line. Density, specific internal energy and normal velocity are
    shown. Only minor under/over shoots are visible. Comparing against
    the results in Fig.~\ref{fig:BDRT_Scalars_200} we see convergence
    to the correct weak solution.}
  \label{fig:BDRT_Scalars_1000}
\end{figure}
\begin{figure}
  \includegraphics[width=0.49\textwidth]{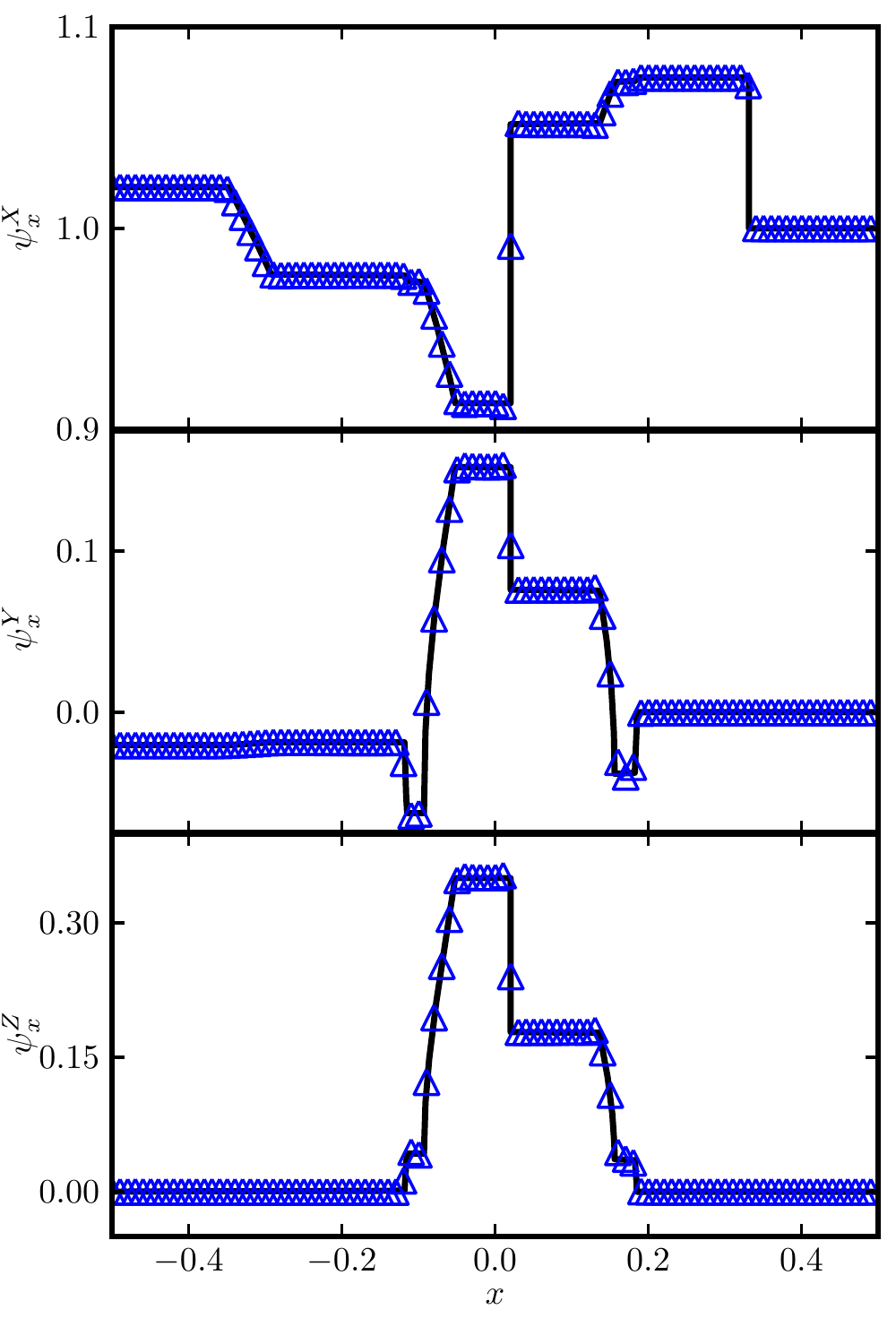}
  \caption{Numerical solution of the BDRT1 test. We show the results
    of our Newtonian code using 1000 points (only 100 are plotted for
    clarity), with the exact solution given by the solid
    line. Components of the configuration gradient are shown. Only
    minor under/over shoots are visible. Comparing against the results
    in Fig.~\ref{fig:BDRT_Psi_200} we see convergence to the
    correct weak solution.}
  \label{fig:BDRT_Psi_1000}
\end{figure}
\begin{figure}
  \includegraphics[width=0.49\textwidth]{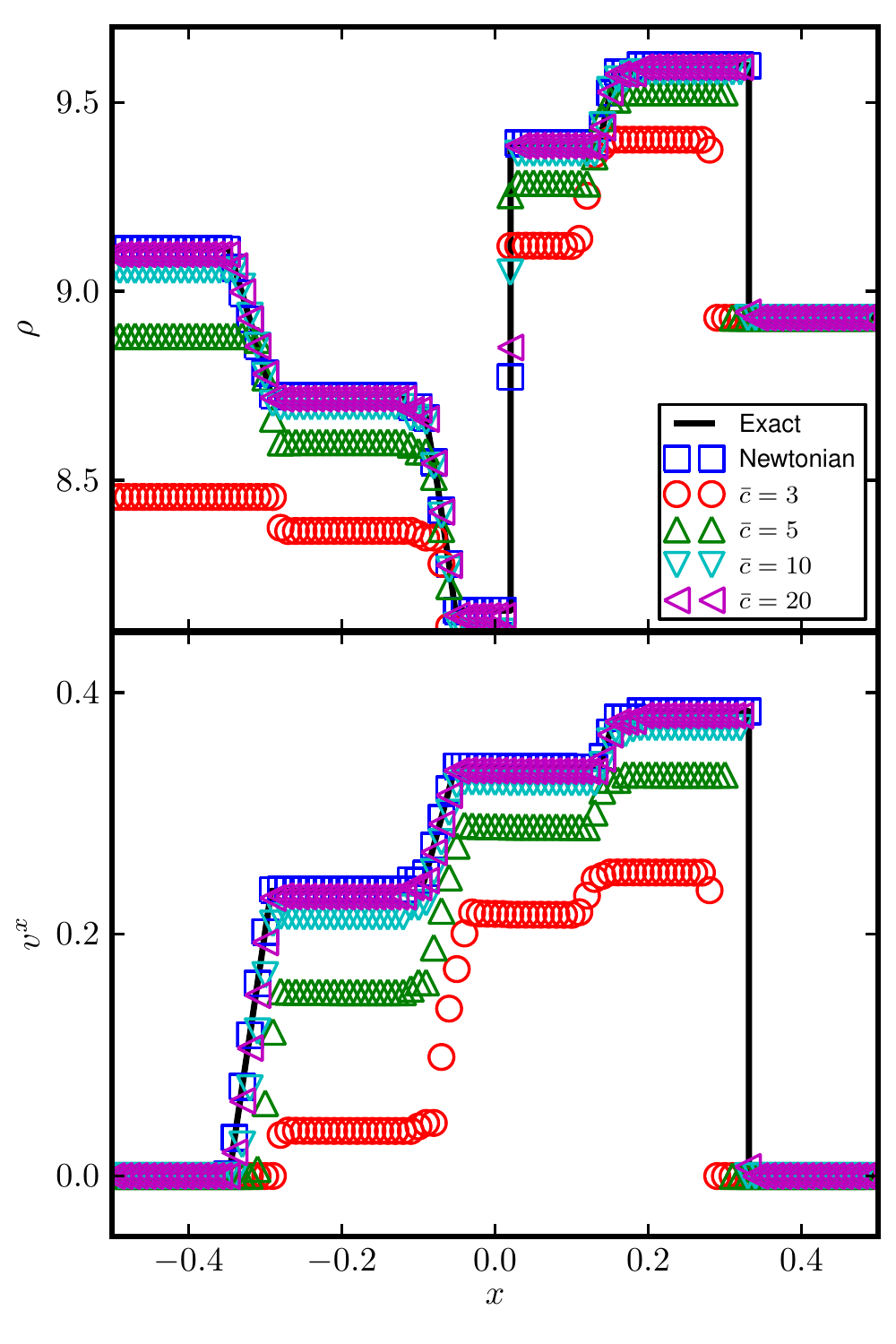}
  \caption{BDRT1 again, but now run using the \emph{relativistic} code
    with various values of $\bar{c}$, and compared against the
    Newtonian results. The results for $\rho, v^x$ (appropriately
    scaled by $\bar{c}$) are representative of the behaviour of all
    quantities. We see that as $\bar{c}$ increases the Newtonian limit
    is approached. 10000 points were used in each case and only 100
    plotted for clarity.}
  \label{fig:BDRT_NewtonianLimit}
\end{figure}

\subsection{Newtonian limit vs Newtonian code}
\label{compare_rel}

The code (both relativistic and Newtonian) uses geometric units where
the speed of light is one. In particular, all velocities are of the
form $v=\bar v/\bar c$ where $v$ is a dimensionless velocity, $\bar v$
its value in conventional units and $\bar c$ the value of the speed of
light in the same units. All parameters in the equation of state, such
as $\epsilon$ and $c_s^2$, are treated analogously. There is no need
to rescale rest mass and length, as long as units are used
consistently.

Changing $\bar c$ while keeping $\bar v$ etc. fixed is a trivial scale
invariance of the Newtonian equations and their solution, but in the
relativistic equations decreasing $\bar c$ with $\bar v$ etc. fixed
makes the same test problem more relativistic. We can use this to
obtain an insight into the effects of (special) relativity, and to
verify that our relativistic code has the correct Newtonian limit as
$\bar c\to\infty$. 

In Fig.~\ref{fig:BDRT_NewtonianLimit} we show the results from the
relativistic code run with a small range of values for $\bar{c}$. Only
relatively small values of $\bar{c}$ are shown -- $3, 5, 10$ and $20\,
\text{km s}^{-1}$, compared to a typical velocity in the (Newtonian)
Riemann problem of $1\, \text{km s}^{-1}$ -- as for sufficiently large
values of $\bar{c}$ the results are visually indistinguishable. We see
that the results from the relativistic code are qualitatively similar,
in terms of wave structure and accuracy, and approach the Newtonian
results in the limit $\bar{c} \rightarrow \infty$.


\subsection{Relativistic Riemann tests}
\label{exact}

In the genuinely relativistic limit we have tested our code against
exact solutions constructed by solving a pre-determined wave
structure. The explicit procedure is detailed in
Appendix~\ref{sec:exactsolns} and follows the method used in the
Newtonian case outlined in \cite{BDRT}, without constructing a full
Riemann problem solver.

\begin{figure}
  \includegraphics[width=0.49\textwidth]{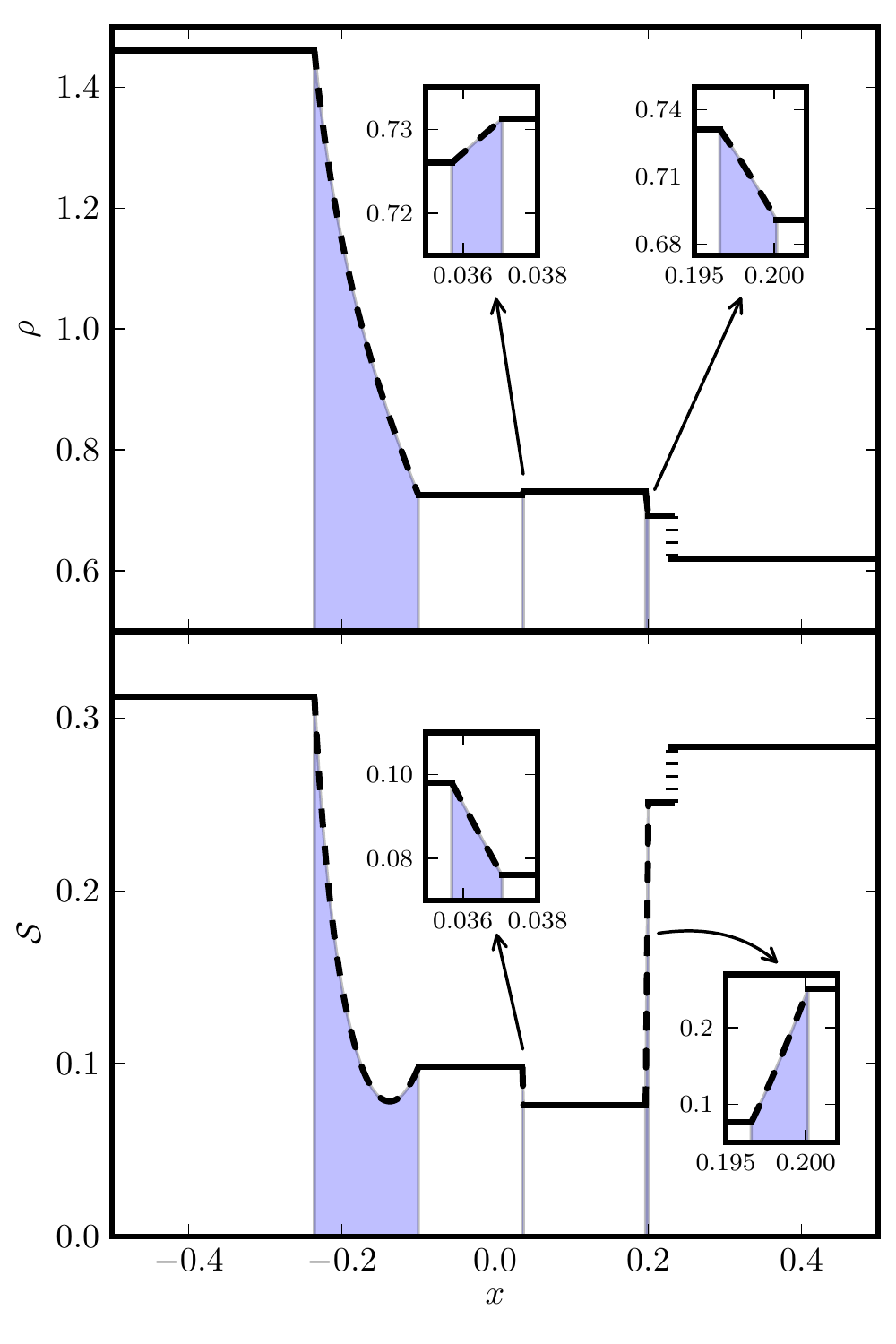}
  \caption{The density and shear scalar for the relativistic 4-wave
    test. This is the exact solution, illustrating the wave structure
    in detail. The rarefactions -- the 1, 2, and 6-waves -- are given
    by the blue dashed lines and are shaded beneath to show the width
    of the fan. The very narrow 2 and 6-waves are shown in detail in
    the insets. The contact -- the linear 4-wave -- is trivial. The
    shock -- the 7-wave -- is given by the thick dash-dotted black
    line. It is clear that resolving some of the rarefaction waves
    will be difficult at moderate resolution. }
  \label{fig:SR_4wave_Structure}
\end{figure}

\begin{figure}
  \includegraphics[width=0.49\textwidth]{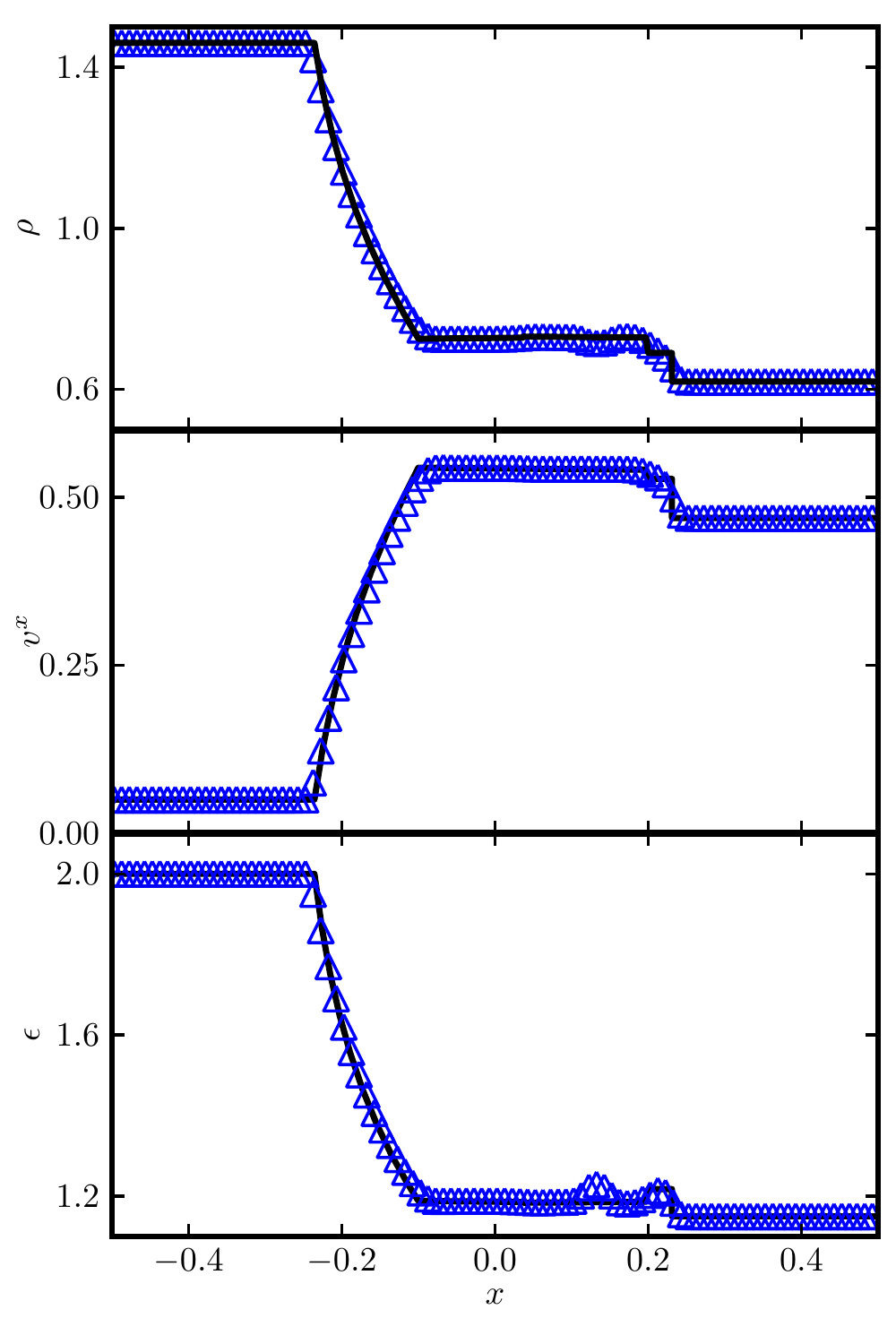}
  \caption{Numerical solution of the relativistic 4-wave test.
    Density, specific internal energy and normal velocity are
    shown. The 4 wave structure (two left-going rarefactions, one
    right-going rarefaction and one right going shock) is difficult to
    see in these variables. The solution is computed using 200 points
    but only 100 are plotted for clarity. We see that all waves are
    captured well and with only minor under/over shoots, most visible
    for the second rarefaction wave in quantities such as $\epsilon$.}
  \label{fig:SR_4wave_Scalars_200}
\end{figure}

\begin{figure}
  \includegraphics[width=0.49\textwidth]{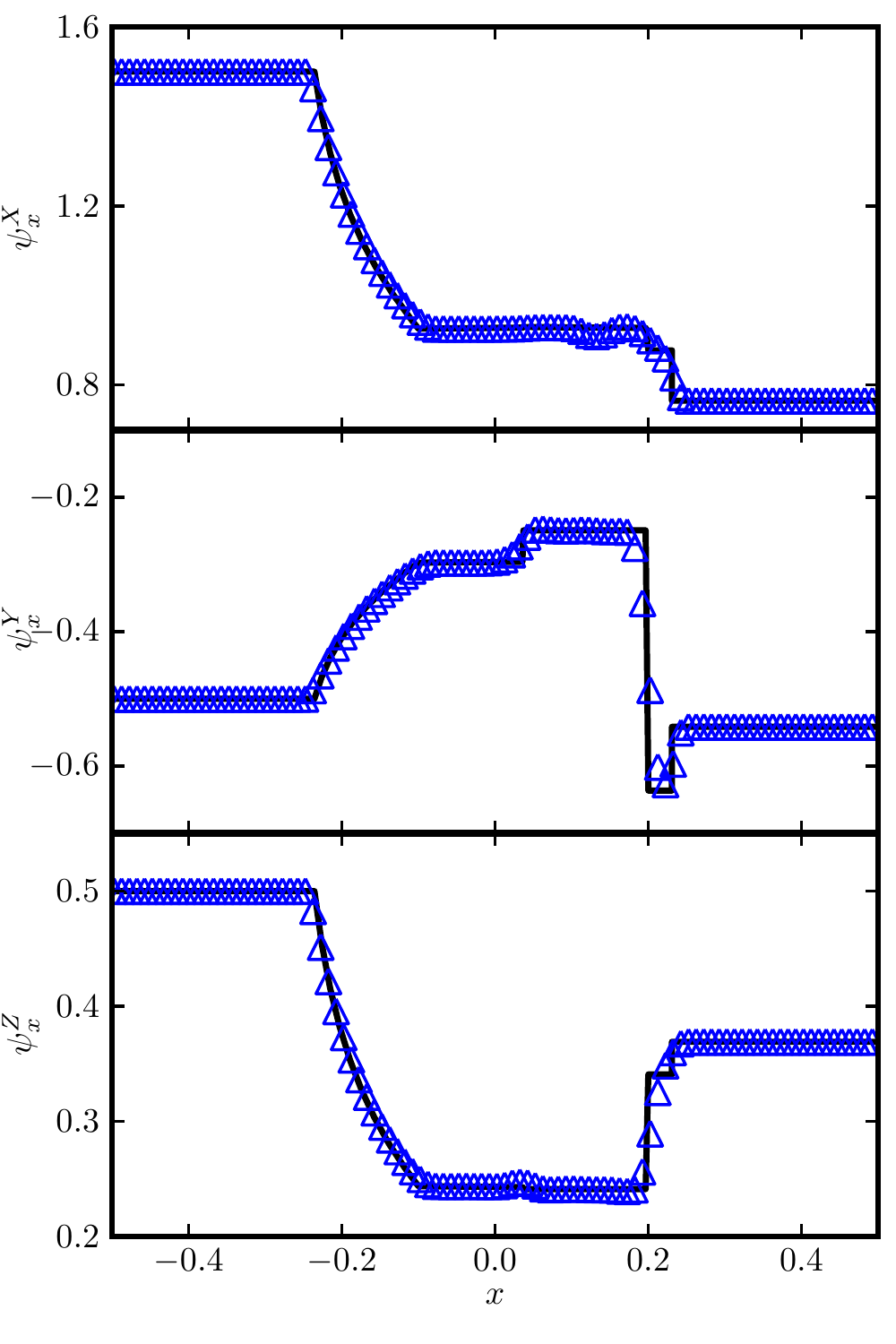}
  \caption{Numerical solution of the relativistic 4-wave
    test. Components of the configuration gradient are shown. The 4
    wave structure (two left-going rarefactions, one right-going
    rarefaction and one right going shock) is most clearly seen in the
    plot of $\psi^Y{}_x$. The solution is computed using 200 points
    but only 100 are plotted for clarity. We see that all waves are
    captured well and with only minor under/over shoots.}
  \label{fig:SR_4wave_Psi_200}
\end{figure}

\begin{figure}
  \includegraphics[width=0.49\textwidth]{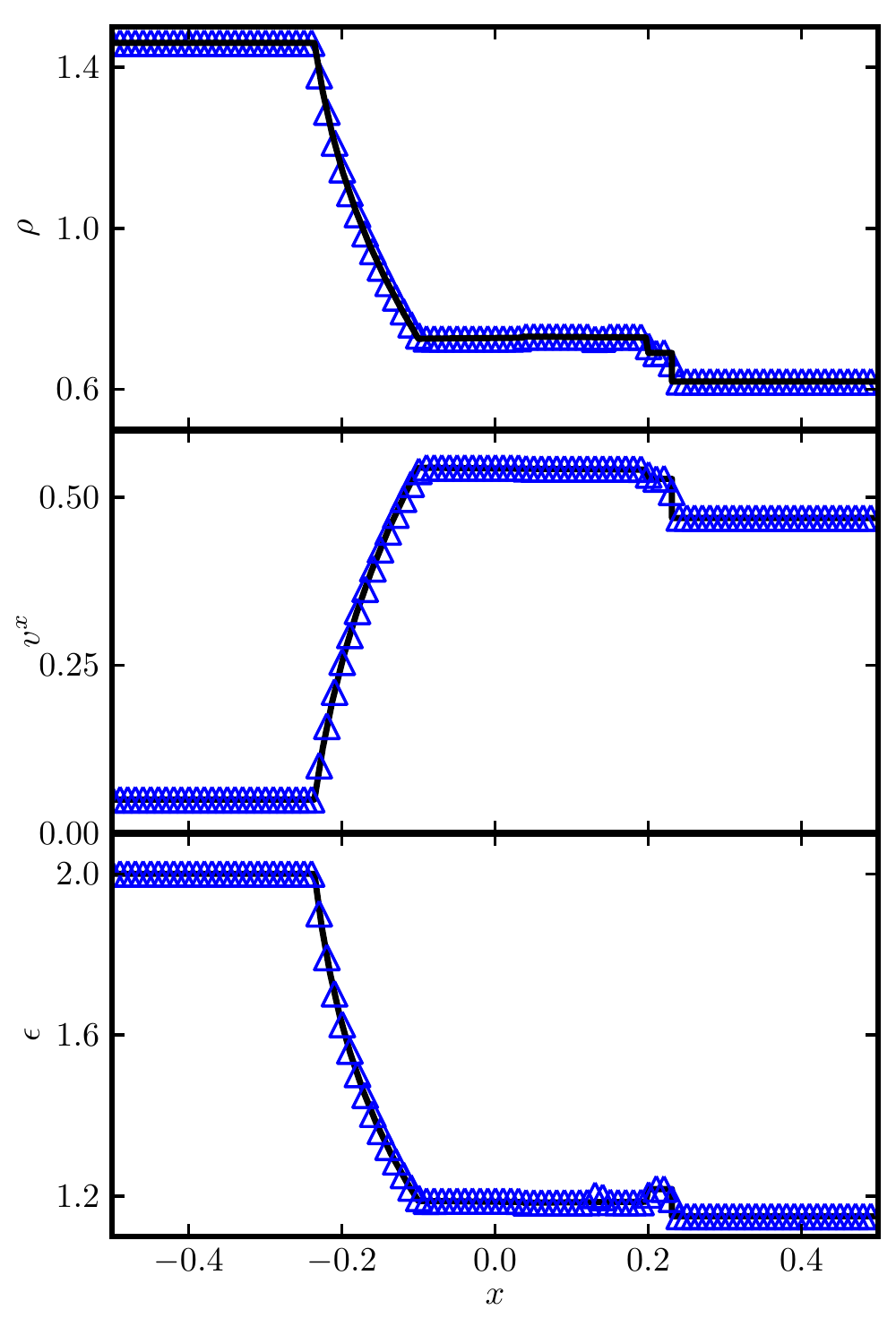}
  \caption{The relativistic 4-wave test again, but now using 1000
    points (only 100 are plotted for clarity). We see that all waves
    are captured well and with only minor under/over shoots, and
    comparing to Figure~\ref{fig:SR_4wave_Scalars_200} we see the
    expected convergence.}
  \label{fig:SR_4wave_Scalars_1000}
\end{figure}

\begin{figure}
  \includegraphics[width=0.49\textwidth]{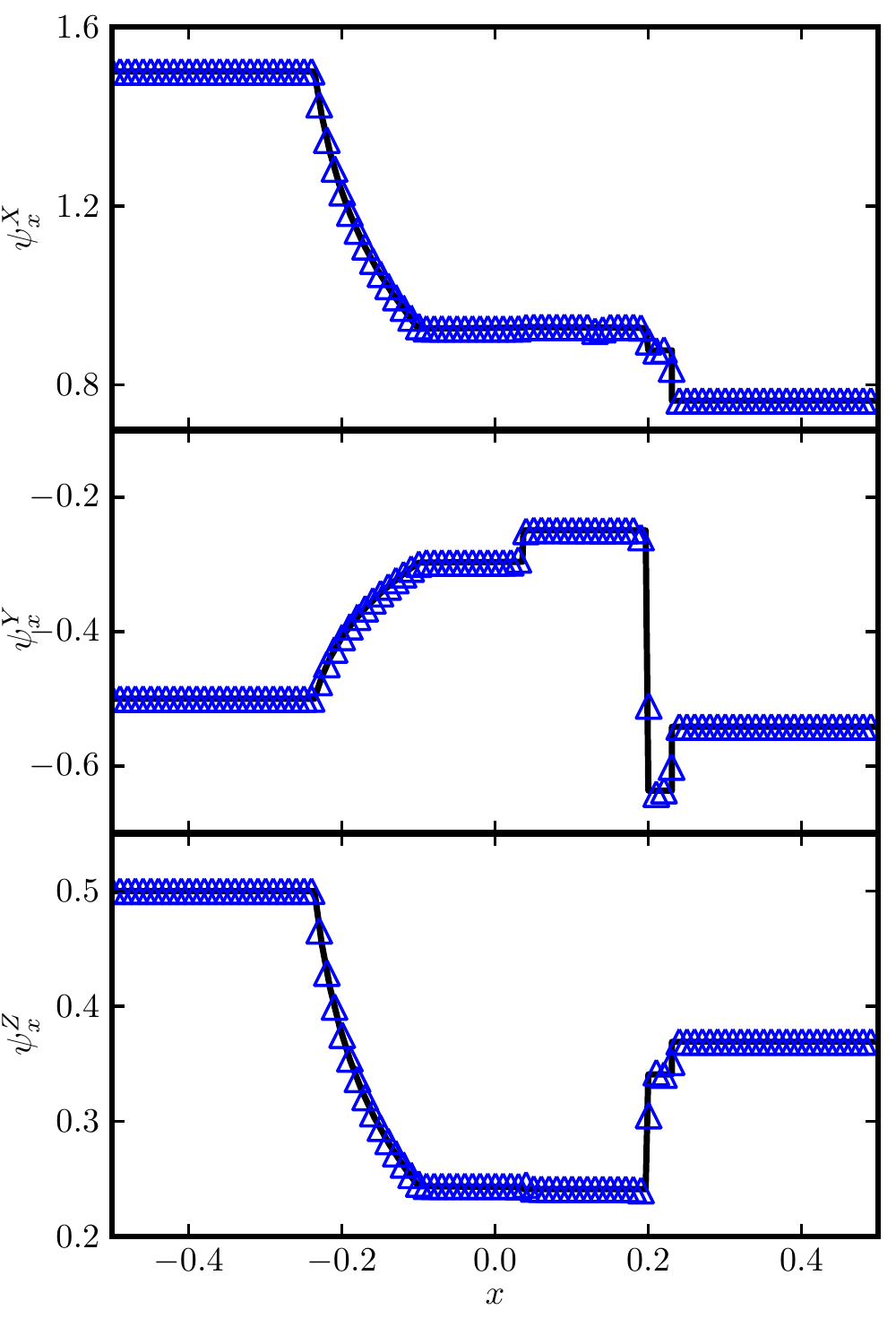}
  \caption{The relativistic 4-wave test again, but now using 1000
    points (only 100 are plotted for clarity). We see that all waves
    are captured well and with only minor under/over shoots, and
    comparing to Figure~\ref{fig:SR_4wave_Scalars_200} we see the
    expected convergence.}
  \label{fig:SR_4wave_Psi_1000}
\end{figure}

We have verified that the code behaves correctly for single shocks and
rarefactions in the relativistic limit, and for some invented initial
data sets that test a range of wave structures. As an example, we show
in Figs.~\ref{fig:SR_4wave_Scalars_200}--\ref{fig:SR_4wave_Psi_1000}
the results for a four wave problem.  For clarity, the wave structure
of the exact solution is shown in Fig.~\ref{fig:SR_4wave_Structure}.
There are two left-going rarefactions (1 and 2-waves), one right-going
rarefaction (a 6-wave) and a right going shock (7-wave). The central
three waves -- the nonlinear 3 and 5-waves and the contact -- are all
trivial. We note that some of the quantities change so rapidly across
some rarefaction waves that they are only visually distinguishable
from shocks at high magnification.

%

Even with the violent behaviour displayed across some waves in this
four wave test, we find our code matching the exact solution well,
with no unphysical oscillations and only minor under and overshoots
that converge away with resolution. There are the expected minor
oscillations near the trivial waves, most noticeable near the contact,
but again these converge with resolution.


\subsection{Two-dimensional Riemann tests}



\begin{figure}
\includegraphics[width=0.49\textwidth]{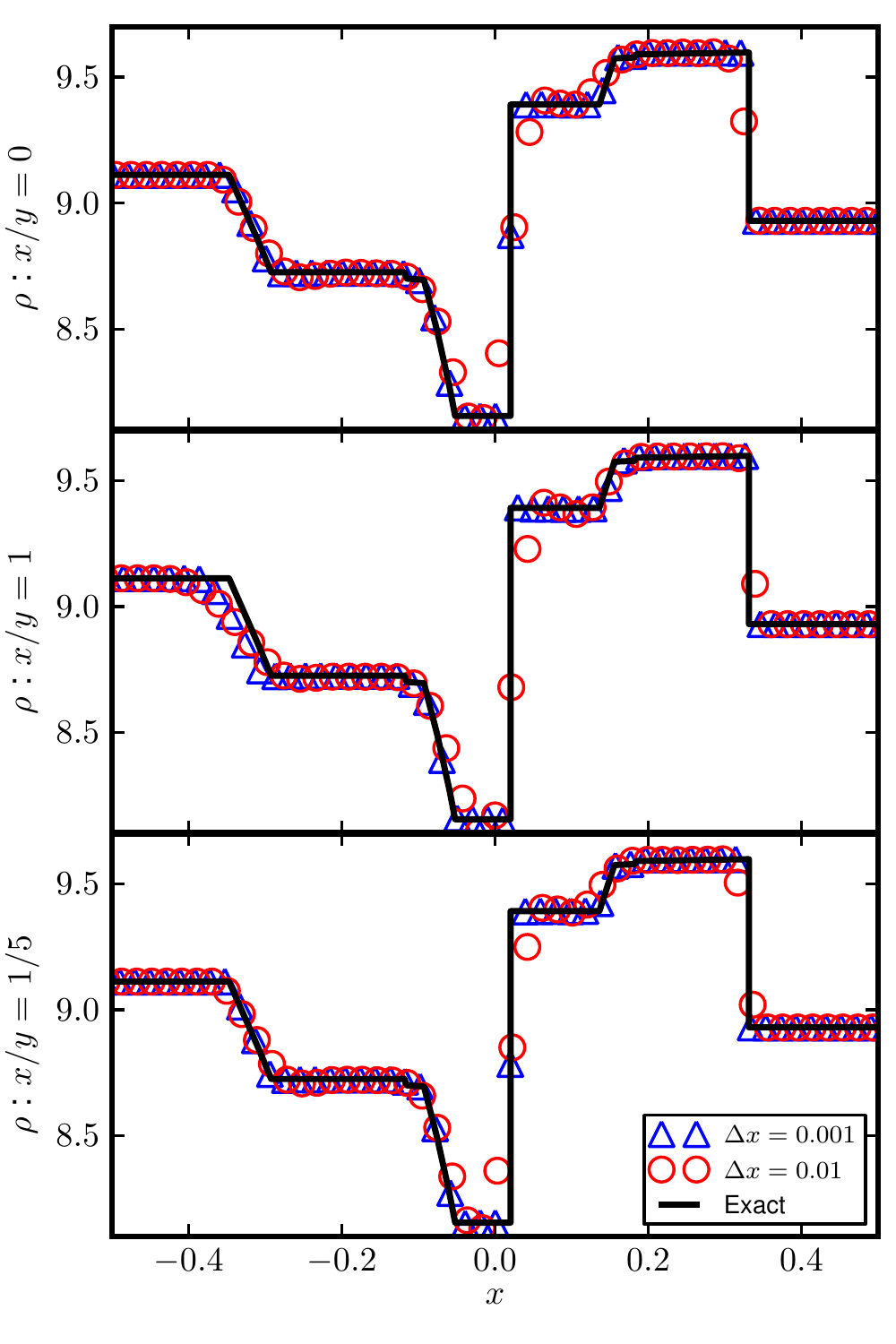}
\caption{
\label{fig:2dwithexact}
Results for the BDRT1 Riemann test, calculated on a two-dimensional
grid for three different angles between the initial discontinuity and
the grid, each at two resolutions. The initial discontinuity was
placed on the line given above each plot. In order to compare the
results to the exact Riemann solver presented in \cite{BDRT}, a slice
through the two-dimensional grid is taken along the $x$ axis (as an
approximation to a line normal to the waves), and $x$ is scaled to
correspond to distance perpendicular to the initial discontinuity. The
spatial resolution is independent of the angle of the initial
discontinuity, and the snapshot is always taken at the same time, for
all angles. The relativistic code is used in the Newtonian limit (as
the exact solution is Newtonian). The high-resolution results were
produced using $\Delta x = \Delta y = 0.001$. The low-resolution
version was produced using $\Delta x =
\Delta y = 0.01$. (For clarity, only 1 in 2 or 1 in 20 points are
plotted for the $x/y = 0$ and $x/y = 1/5$ cases, while 1 in 3 or 1 in
28 points are plotted for $x/y = 1$.)  All three evolutions look
similar, with the results approaching the exact solution as the
resolution is increased; the only notable feature is that the
left-most rarefaction wave does not appear to be well captured by the
evolution with a slope of $1$. }
\end{figure}


The constraints are trivial if all variables depend only
on one coordinate, for example when a Riemann problem is aligned with
the numerical grid. As a first test of the behaviour in three
dimensions and the role of the constraints, we have solved Riemann
problems also at an arbitrary angle to a two-dimensional Cartesian
grid.  A method for carrying out such 2D simulations efficiently is
described in Appendix~\ref{appendix:2d}. 

We put the initial discontinuity along lines $x/y=0$ (our 1D tests),
$1$, $1/2$ and $1/5$, and use a cut along the $x$ axis as an
approximation to a line normal to the initial discontinuity. We
compare this cut, suitably foreshortened, against the exact solution.

We have not implemented the ``hyperbolicity fix'' constraint additions
for either the kinematic or dynamical evolution equations. In 1D the
equations are symmetric hyperbolic anyway, as there are no constraints
then, but in 2D our equations are not even strongly hyperbolic. The
error at the same time is somewhat larger in 2D than in 1D, see
Fig.~\ref{fig:2dwithexact}, but there is no sign of numerical
instability in 2D. We have no explanation for this, but expect that
constraint addition will be necessary in other tests.


\subsection{Two-dimensional Rotor tests}

To study a genuinely two-dimensional problem we consider a test
suggested by~\cite{Dumbser2008}. The initial data, detailed in
Appendix~\ref{appendix:initialdata}, represents an elastic rotor
problem, where an inner rotating bearing is instantaneously welded to
the non-rotating exterior, causing the rotor to slow and propagating
elastic waves through the material. In all cases the rotor has
coordinate radius $0.1$, whilst the exterior is at rest. In all
numerical experiments shown here $400^2$ points were used.

\begin{figure}
  \centering
  \includegraphics[width=0.49\textwidth]{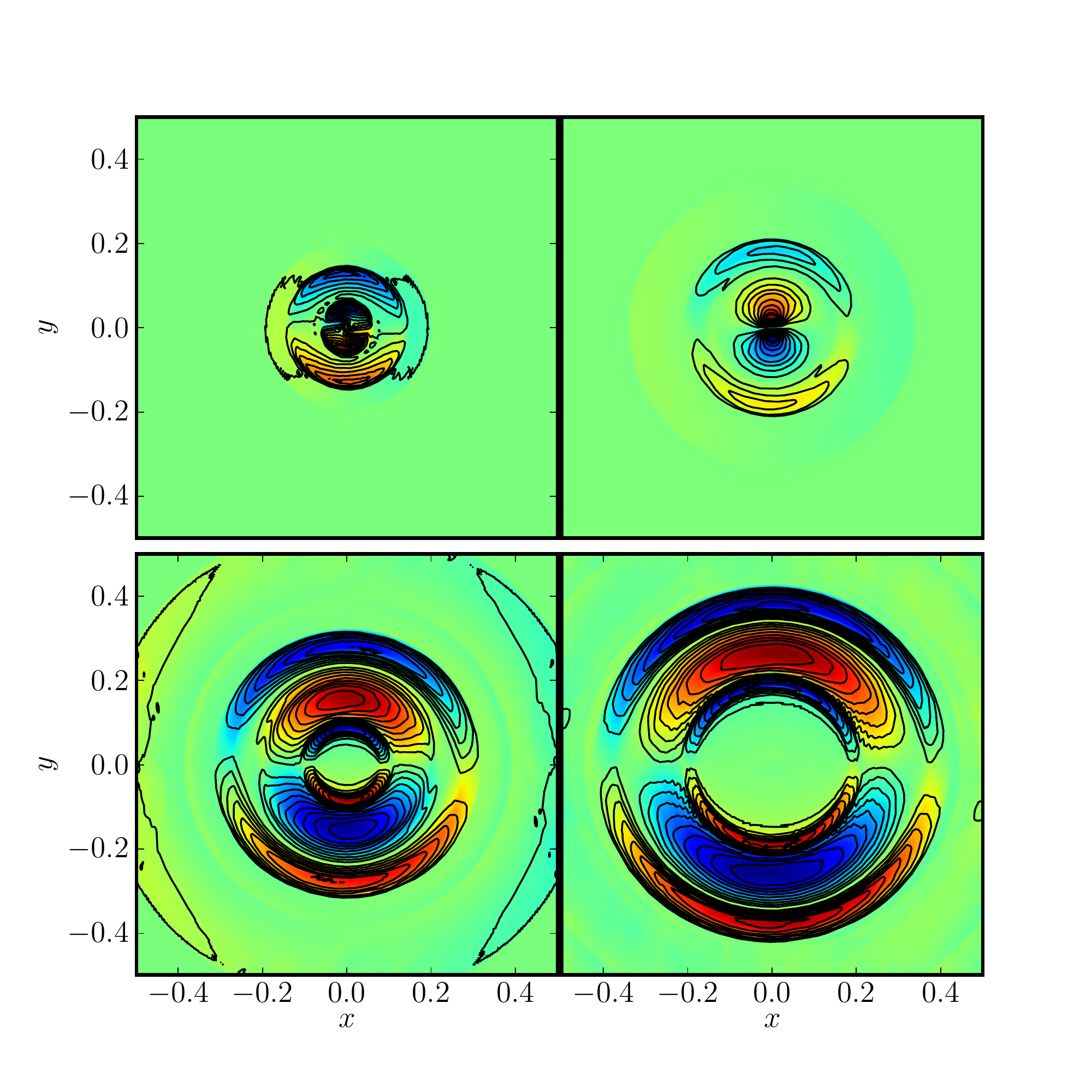}
  \caption{Newtonian rotor test, following~\cite{Dumbser2008}. An
    initially rotating central cylinder is slowed by the interaction
    with the exterior, which is initially at rest. These figures show
    $\rho v^x$ at coordinate times $t=0.02, 0.05, 0.1$ and $0.15$. The
    results qualitatively match those in Figure 24
    of~\cite{Dumbser2008}.}
  \label{fig:RotorNewtonian_v}
\end{figure}

\begin{figure}
  \centering
  \includegraphics[width=0.49\textwidth]{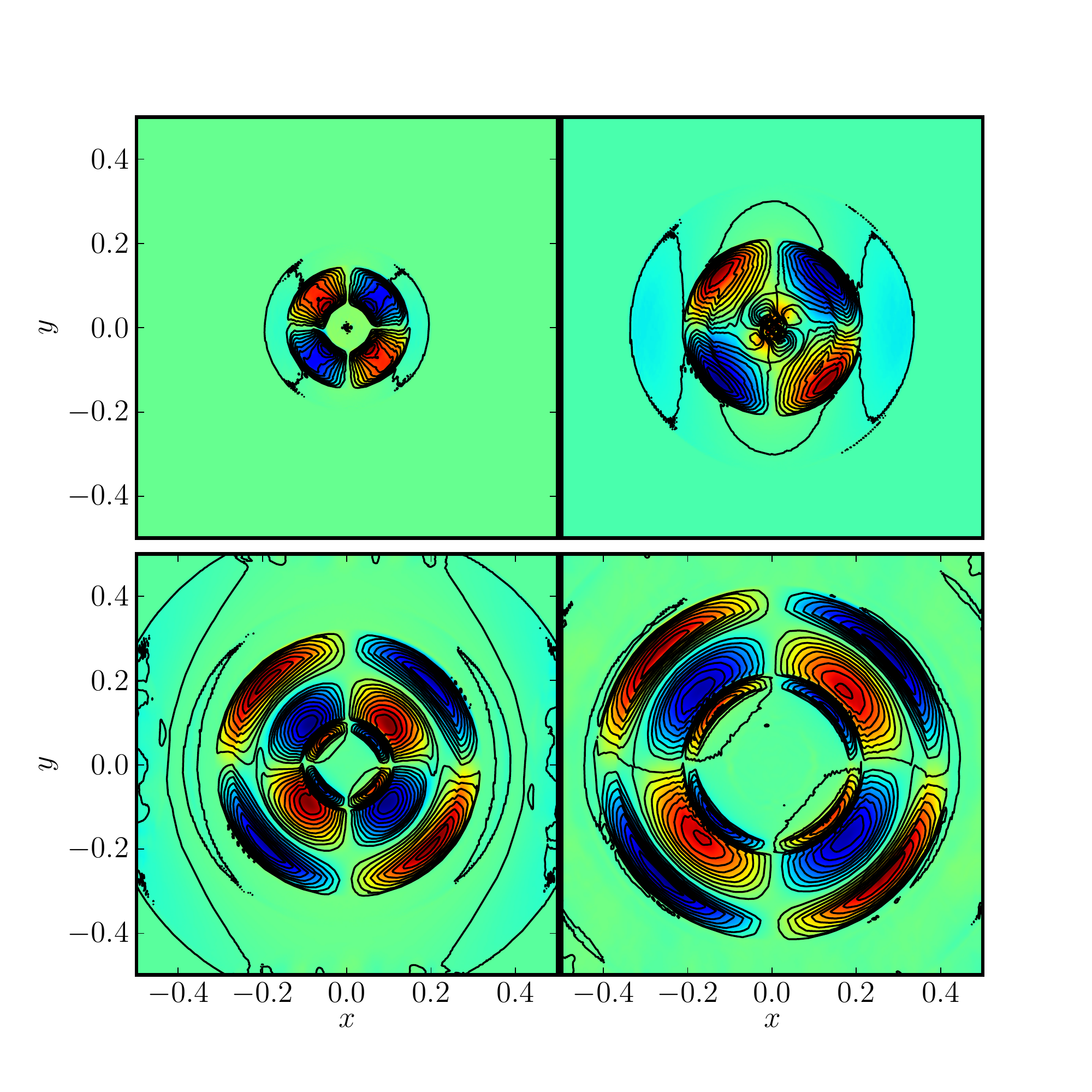}
  \caption{Newtonian rotor test, following~\cite{Dumbser2008}. An
    initially rotating central cylinder is slowed by the interaction
    with the exterior, which is initially at rest. These figures show
    $\rho F^y{}_Y$ at coordinate times $t=0.02, 0.05, 0.1$ and
    $0.15$. The results qualitatively match those in Figure 25
    of~\cite{Dumbser2008}.}
  \label{fig:RotorNewtonian_F}
\end{figure}
Results for the Newtonian case are shown are representative coordinate
times are shown in figures~\ref{fig:RotorNewtonian_v}
and~\ref{fig:RotorNewtonian_F}. These should be compared to the
results shown by Dumbser et al.\ in \cite{Dumbser2008}. The results in
the literature use a considerably more accurate numerical method,
which is both higher order and uses finite elements better adapted to
the symmetry of the problem. Despite this, we see qualitative
agreement in the waves emitted during the evolution of the problem.

\begin{figure}
  \centering
  \includegraphics[width=0.49\textwidth]{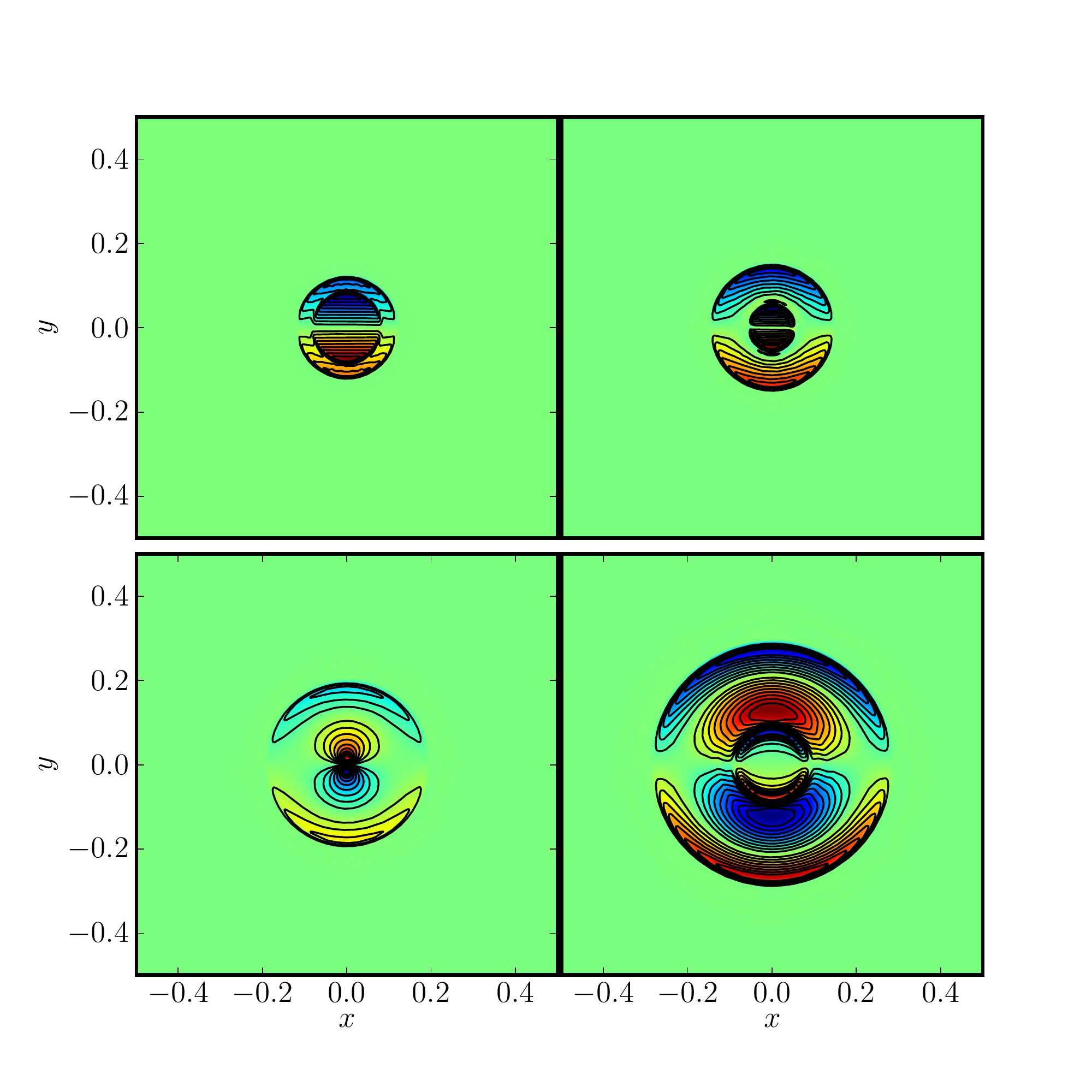}
  \caption{Relativistic rotor test, to be compared with the Newtonian
    results in figure~\ref{fig:RotorNewtonian_v}. An initially
    rotating central cylinder is slowed by the interaction with the
    exterior, which is initially at rest. These figures show $\rho
    v^x$ at coordinate times $t=0.02, 0.05, 0.1$ and $0.2$. The
    emitted waves are qualitatively similar to the Newtonian results.}
  \label{fig:RotorRelativistic_v}
\end{figure}

\begin{figure}
  \centering
  \includegraphics[width=0.49\textwidth]{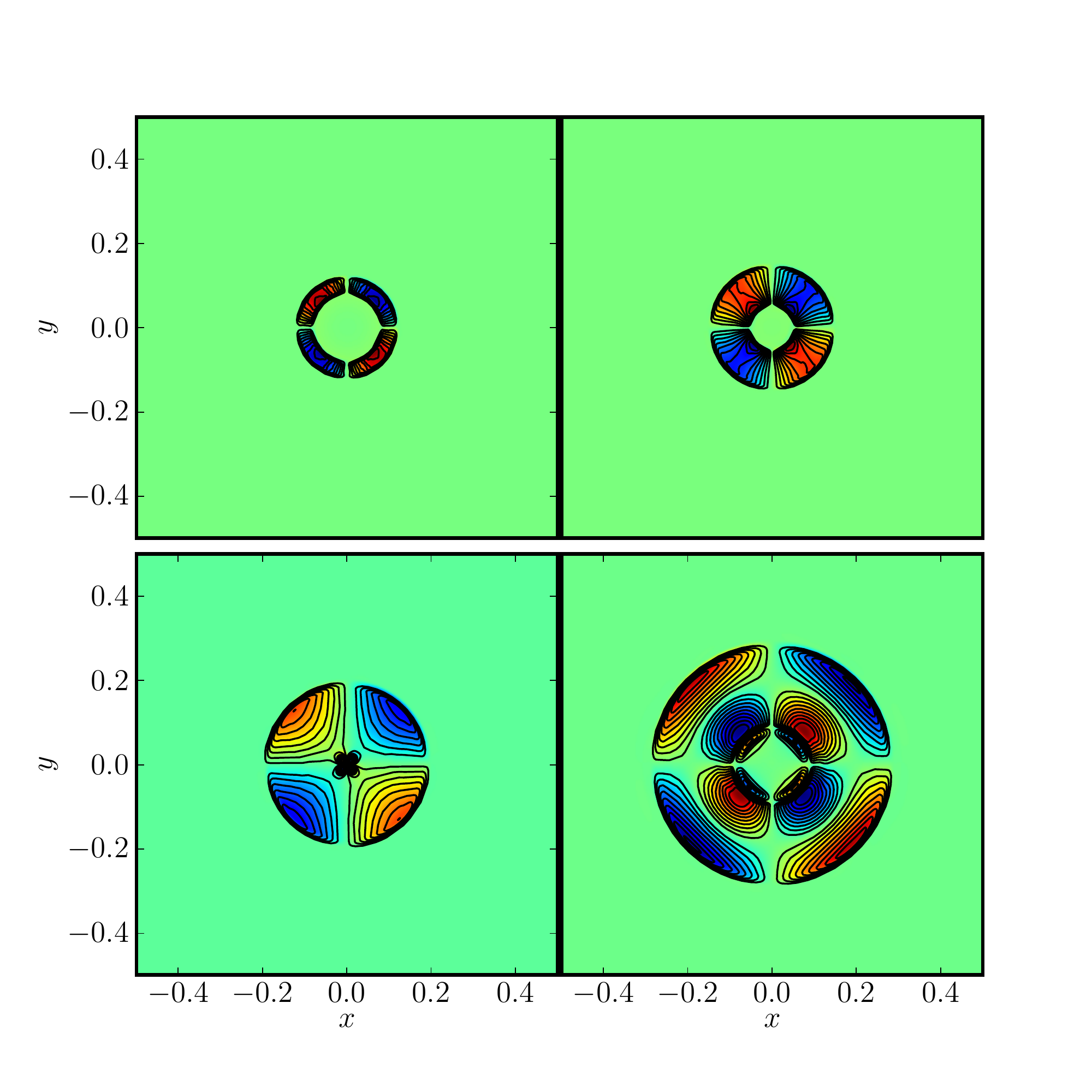}
  \caption{Relativistic rotor test, to be compared with the Newtonian
    results in figure~\ref{fig:RotorNewtonian_F}. An initially
    rotating central cylinder is slowed by the interaction with the
    exterior, which is initially at rest. These figures show $\rho
    F^y{}_Y$ at coordinate times $t=0.02, 0.05, 0.1$ and $0.2$. The
    emitted waves are qualitatively similar to the Newtonian results.}
  \label{fig:RotorRelativistic_F}
\end{figure}
Results for the relativistic case are shown are representative
coordinate times are shown in figures~\ref{fig:RotorRelativistic_v}
and~\ref{fig:RotorRelativistic_F}. Again we see qualitative agreement
in the emitted wave structure, despite the differences in the models.


\section{Conclusions}
\label{section:conclusions}

We have presented a framework that can be used for simulating
nonlinear elasticity in numerical relativity, and checked its
viability in Riemann tests. The framework can be directly related
to existing approaches and is a first step towards the simulation of
neutron star crusts. 

Our numerical simulations show that the results from the Newtonian
limit of the equations match those in the literature, and that the
Newtonian limit of the \emph{relativistic} code also match the results
from the Newtonian literature.

The equations in first order form consist of three groups: evolution
equations for a \emph{configuration gradient} $\psi^A{}_i$, auxiliary
constraints $\psi^A{}_{[i,j]}=0$ for this variable due to the fact that
$\psi^A{}_i=\partial\chi^A/\partial x^i$ for an implicit underlying
\emph{configuration} $\chi^A$, and energy-momentum conservation laws.

The first two groups are purely kinematical in the sense that they are
independent of the geometry of both spacetime and matter space, and
hence are the same in Newtonian and relativistic elasticity. However,
from a spacetime point both the evolution equations and constraints
naturally arise as components of a single spacetime constraint
$\psi^A{}_{[a,b]}=0$. (In fact, without the benefit of this point of
view, some of the constraints seem to have been systematically
overlooked in the Newtonian literature, giving rise to the numerical
solution of unphysical Riemann problems in \cite{TRT, BDRT}.)

Energy-momentum conservation is due to time and space translation
invariance, and this fixes their correct weak form \cite{Font}. The
weak form of the kinematical equations appears to have been assumed
{\it ad hoc} in the Newtonian literature. Here we have rigorously
derived it from the absence of dislocations in the elastic matter.

There are two rather different frameworks in the Newtonian
literature. One of these \cite {TrangensteinColella,
  MillerColella2001, MillerColella2002} is fully Eulerian and arises
naturally as the Newtonian limit of our relativistic framework. The
other \cite{GodunovRomenski,GodunovPeshkov,PS} mixes Eulerian and
Lagrangian points of view and gives rise to more complicated evolution
equations. For completeness, we have proved that the two frameworks
are equivalent in their \emph{weak} form, and hence that the weak form
of the second framework is also correct. This is borne out by our
numerical tests, which agree for both frameworks (if the initial data
obey the constraints).

The {\em dynamical} equations of our framework can be related to the
standard Valencia formalism for relativistic hydrodynamics. Although,
as noted in section~\ref{section:kinematics}, the fluid limit is
singular, the system presented takes the form of the Valencia
equations with additional terms. We also note that steps within the
numerical code, such as the conversion to primitive variables outlined
in Sec.~\ref{section:conversion}, tend towards standard algorithms
in the fluid limit.

Using the methods of \cite{BS}, we have shown that our framework can
be made symmetric hyperbolic, at least in a neighbourhood of the
unsheared state of the matter, if certain linear combinations of the
auxiliary constraints are added as source terms to the conservation
laws. We have also shown that if constraints are added only to the
kinematic evolution equations, bringing them into the form
(\ref{EAitres}) (the ``hyperbolicity fix''), but not to
$\nabla_aT^{ab}=0$, the resulting first-order system is strongly
hyperbolic but not symmetric hyperbolic. 

The latter is precisely the situation in the Newtonian literature, and
so the Newtonian limit of our result shows that the equations given
there \cite{PS,MillerColella2001} are only strongly hyperbolic but
could be made symmetric hyperbolic by a simple constraint
addition.

There remain two outstanding issues before this framework can be used
in a fully nonlinear GR simulation of a neutron star. The first is the
issue of the integrability constraints in higher dimensional
simulations. In the Newtonian literature it is clear that the
hyperbolicity fix included here is required to obtain stable
evolutions. However, there is no agreement as to the impact of
constraint violations on the \emph{accuracy} of the simulations. In
analogy with MHD simulations where the $\nabla \cdot {\bf B} = 0$
constraint is crucial for accuracy, we might expect that methods for
reducing constraint violations (such as the parabolic damping term
used by \cite{Miller} -- similar to the Powell method for MHD), or
alternatively a discretisation that maintains a discrete version of
the constraints along the lines of
Appendix~\ref{section:discreteconstraints}, will be important.

Secondly, to be useful for simulating a neutron star, we must couple
the elastic crust to the fluid interior. A framework for the nonlinear
simulation of multiple matter models separated by sharp interfaces in
GR was studied in \cite{Millmore}, but only for fluid-fluid
interactions. This model built on standard Newtonian methods which
have themselves been extended to deal with solid-fluid interactions;
we expect that these methods will extend to relativity as well.


\begin{acknowledgments}

We are grateful to Philip Barton for discussions and for the numerical
data corresponding to the exact solutions of the problems in
\cite{BDRT}. We are also grateful to Lars Samuelsson, Bobby Beig, Lars
Andersson and members of the Southampton General Relativity Group for
discussions relating to this work.

\end{acknowledgments}


\appendix


\section{3+1 split of spacetime}
\label{appendix:3+1}

For reference, we assemble some standard formulas.
In 3+1 numerical relativity, the spacetime metric $g_{ab}$ is split
into a spatial metric $\gamma_{ij}$ with inverse $\gamma^{ij}$, a
lapse $\alpha$ and shift $\beta^i$, as
\begin{equation}
g_{00}=-\alpha^2+\beta_i\beta^i, \quad g_{0i}=\beta_i, \quad
g_{ij}=\gamma_{ij},
\end{equation}
where we define indices on $\beta^i$ to be moved implicitly with
$\gamma_{ij}$. The (absolute value of the) determinant of the 4-metric
is given by
\begin{equation}
\label{gtogamma}
g_x=\alpha^2\gamma_x.
\end{equation}
and hence the volume forms on $M^3$ and $M^4$ are related by 
\begin{equation}
\label{3eps4eps}
\epsilon^{0ijk}=\alpha\epsilon^{ijk}.
\end{equation}

The inverse 4-metric is
\begin{equation}
g^{00}=-\alpha^{-2}, \quad 
g^{0i}=\alpha^{-2}\beta^i, \quad 
g^{ij}=\gamma^{ij}-\alpha^{-2}\beta^i\beta^j.
\end{equation}
The covector normal to the surfaces of constant $t$ has components
\begin{equation}
n_0=-\alpha, \qquad n_i=0,
\end{equation}
and hence
\begin{equation}
n^0=\alpha^{-1}, \qquad n^i=-\alpha^{-1}\beta^{i}.
\end{equation}
Hence the projector into the surfaces of constant $t$
\begin{equation}
\label{gammaab}
\gamma_{ab}:=g_{ab}+n_an_b
\end{equation}
has components
\begin{equation}
\gamma_{00}=\beta_i\beta^i, \quad 
\gamma_{0i}=\beta_i, \quad 
\gamma_{ij}=\gamma_{ij}, 
\end{equation}
and
\begin{equation}
\gamma^{00}=0, \quad 
\gamma^{0i}=0, \quad 
\gamma^{ij}=\gamma^{ij}.
\end{equation}

We define the convective derivative to be the derivative along the
4-velocity,
\begin{equation}
u^a{\partial\over\partial x^a} \propto{\partial\over \partial t}
+\hat v^i{\partial\over\partial x^i}.
\end{equation}
The factor of proportionality is given by the normalisation condition
\begin{equation}
u^a u^b g_{ab}=-1.
\end{equation}
We find
\begin{eqnarray}
\label{Valenciau}
u^a&=&(u^t,u^i)=\alpha^{-1}W(1,\hat v^i),\\
\label{Valenciaudown}
u_a&=&(u_t,u_i)=W(-\alpha+v_j\beta^j,v_i),
\end{eqnarray}
where
\begin{eqnarray}
\label{vhatdef}
\hat v^i&:=&\alpha v^i - \beta^i, \\ 
\label{Wdef}
W&:=& \left(1-v_iv^i\right)^{-1/2},
\end{eqnarray}
and where we define the indices on $v^i$ to be moved implicitly with
$\gamma_{ij}$.
The scalar
\begin{equation}
-u^a n_a=W
\end{equation}
gives the Lorentz factor of the relative velocity between the matter
and the time slices.  


\section{Definitions of hyperbolicity}
\label{appendix:hyperbolicity}

We summarise some standard definitions \cite{BS,Anile} in
our notation. Let $w^\alpha$ be a vector of variables obeying the
system of first-order partial differential equations
\begin{equation}
\label{Palphabeta}
P_{\alpha\beta}{}^c w^\beta_{,c}+{\rm l.o.}=0,
\end{equation}
where l.o. stands for lower order terms. Obviously the index $\alpha$
labelling the equations needs to take as many values as the index
$\beta$ labelling the variables.

Assume, however, that $\alpha$ is an index of the same type as $\beta$
and that $P_{\alpha\beta}{}^c=P_{\beta\alpha}{}^c$. Then we have a
conserved current (up to lower order terms) in the sense that
\begin{equation}
J^c{}_{,c}={\rm l.o.}, \qquad J^c:=P_{\alpha\beta}{}^c w^\alpha w^\beta.
\end{equation}
If furthermore there exists a covector $t_c$ with the property that
\begin{equation}
E(w,w):=t_cJ^c=t_cP_{\alpha\beta}{}^c w^\alpha w^\beta
\end{equation}
is positive definite, called a subcharacterisic vector, then the
system is called symmetric hyperbolic. (In a relativistic context we
expect $t_c$ to be timelike.) $E$ allows us to estimate an $L^2$ norm
called an energy norm of the solution in terms of the initial data and
boundary data.

A characteristic direction is a covector $k_c$ such that 
\begin{equation}
\label{charvarfirstorder}
\det k_cP_{\alpha\beta}{}^c=0
\end{equation}
and the corresponding characteristic variable $w^\alpha$ is the
non-zero vector obeying
\begin{equation}
k_cP_{\alpha\beta}{}^c w^\beta=0.
\end{equation}
This means that a plane wave with amplitude $w^\beta$ and wave number
$k_c$ is a solution of the principal part. For a causal system in
relativity, influence cannot travel faster than
light, and so $k_c$ must be spacelike or null.

For a second-order system
\begin{equation}
P_{\alpha\beta}{}^{cd}w^\beta_{,cd}+{\rm l.o.}=0
\end{equation}
the equivalent definition of a characteristic direction and variable is 
\begin{equation}
k_ck_dP_{\alpha\beta}{}^{cd}w^\beta=0,
\end{equation}
and it has the same interpretation as a plane wave solution of the
principal part.

It is often useful to decompose the characteristic equation with
respect to a preferred hypersurface. Let
\begin{equation}
k_a=\lambda n_a-e_a
\end{equation}
where $n_a$ is a unit timelike covector and $e_a$ a unit spacelike
covector normal to $n_a$. $\lambda$ is called the characteristic
velocity (relative to $n_a$) of the characteristic variable
$w^\alpha$. $k_a$ is normal to the characteristic plane spanned
by the vectors
\begin{equation}
v^a=n^a+\lambda e^a+s^a
\end{equation}
where $s^a$ is any vector normal to both $n_a$ and $e_a$ (so that
$k_av^a=0$). The relative speed between $n^a$ and $v^a$ (calculated
from $n^av_a/|n||v|$) is $\sqrt{\lambda^2+s^as_a}\ge\lambda$.  The
disturbance itself moves along $n^a+\lambda e^a$, that is in the
direction $e^a$ with speed $\lambda$ as measured by $n^a$ observers.
One natural choice of $n_a$ is the unit normal to the surfaces of
constant time $t$, and the resulting values of $\lambda$ are used in
the numerical scheme.  By contrast, choosing
$n_a=u_a$ gives the speed of the disturbances relative to the matter,
which are simpler to compute.

To make contact with non-relativistic concepts of hyperbolicity, we
rewrite the first order characteristic equation
(\ref{charvarfirstorder}) as
\begin{equation}
{\cal P}_e w=\lambda w, \qquad {\cal P}_e:=(n_aP^a)^{-1}(e_bP^b)
\end{equation}
where we have not written the Greek indices for simplicity. (If $n_a$
is subcharacteristic, $n_aP^a$ is positive definite and so has an
inverse.) The system is then called weakly hyperbolic with respect to
the time direction $n_a$ if ${\cal P}_e$ has real eigenvalues
$\lambda$ for all unit vectors $e_a$ normal to $n_a$. It is called
strongly hyperbolic if furthermore ${\cal P}_e$ has a basis of real
eigenvectors that depends continuously on $e_a$. It is called
symmetric hyperbolic if ${\cal P}_e$ is symmetric. As a real symmetric
matrix is always diagonalisable with real eigenvalues, symmetric
hyperbolicity implies strong hyperbolicity. More generally, the system
is called symmetric hyperbolic, or symmetrisable, if there exists a
symmetriser, a positive definite symmetric matrix $\cal H$ independent
of $e^a$ such that ${\cal P}_e{\cal H}$ is symmetric. In this case
$E={\cal H}_{\alpha\beta}w^\alpha w^\beta$.


\section{The Newtonian limit} 
\label{appendix:newtonianlimit}


We obtain the limit of Newtonian motion in the absence of gravity in
two steps. In the first step, we let the spacetime go to Minkowski
spacetime in adapted coordinates,
\begin{equation}
ds^2=-dt^2+\gamma_{ij}\,dx^i\,dx^j,
\end{equation}
where $\gamma_{ij}$ is flat and independent of $t$, but $x^i$ could
still be curvilinear coordinates. Hence
\begin{equation}
\hat v^i=v^i,
\end{equation}
and the advection equation~\eqref{kadvection} becomes
\begin{equation}
(\partial_t+v^i\partial_i)k_{AB}=0.
\end{equation}

In the second step, we use dimensional analysis of the special
relativistic equations of motion to insert a parameter
$c$ representing the speed of light, as follows:
\begin{eqnarray}
&& n, \\
&&c^{-1}v^i, \\
&&c^{-2}\epsilon, \quad c^{-2}p, \quad c^{-2} \pi_{ij}, \\
&&c^{-3}\pi_{0i}, \quad c^{-4}\pi_{00},
\end{eqnarray}
for the primitive variables, and
\begin{eqnarray}
&&D, \quad c^{-1} S_i, \quad c^{-2}\tau, \\
&&c^{-1}{\cal F}(D)^i, \quad c^{-2}{\cal F}(S_j)^i, \quad c^{-3}{\cal
  F}(\tau)^i,
\end{eqnarray}
for the conserved variables. We then take the limit $c\to \infty$ of
the relevant equations for Minkowski spacetime. In this limit,
\begin{eqnarray}
W &=& 1, \\
u^a &= & n^a, \\
h_{ab} &=& \gamma_{ab}, \\
\psi^A{}_t &=& 0, \\
\pi &=& \gamma^{ij}\pi_{ij} = 0, \\
D &=& n, \\
S_i &=& n v_i, \\
\tau &=& n(v^2/2+\epsilon), \\
{\cal F}(D)^i&=&n v^i, \\
{\cal F}(S_j)^i&=&n v_jv^i+p\delta^i{}_j+\pi^i{}_j, \\
{\cal F}(\tau)^i&=&n(v^2/2+\epsilon) v^i+p v^i+\pi^i{}_jv^j,
\end{eqnarray}
where $v^i$ and $\pi_{ij}$ are now the Newtonian velocity and stress
tensor, and their indices are moved implicitly with the metric
$\gamma_{ij}$ of Euclidean space.
Instead of requiring $p$, $f_1$ and $f_2$ as functions of $h$ (the
relativistic enthalphy, which includes the rest mass energy) and
$n$, we need them as functions of $\epsilon$ and $n$. The
reconstruction of $n$, $v_i$ and $\epsilon$ from $D$, $S_i$ and
$\tau$ becomes explicit for the equations of state we consider. 


\section{The mixed framework}
\label{appendix:mixed}


\paragraph*{Variables}

In the alternative Newtonian framework of \cite{PS,GodunovPeshkov},
the deformation is given by a map from a 3-dimensional matter space
and time to 3-dimensional space
\begin{eqnarray}
F:\quad R\times X_3 &\to& M_3, \\
(t,\xi^A)&\mapsto & x^i
\end{eqnarray}
with derivatives
\begin{equation}
{F^i}_A:={\partial x^i\over\partial \xi^A}, 
\qquad \hat v^i:=\left.{\partial x^i\over\partial
  t}\right|_{\xi^A={\rm const}},
\end{equation}
where $F^i{}_A$ is the $3\times 3$ matrix inverse of $\psi^A{}_i$.
We shall call this the mixed framework, as the dependent variables are
Lagrangian, but the independent ones are still Eulerian. (A purely
Lagrangian framework also exists, but is not relevant for us because
we are interested in finite volume methods for weak solutions.)

For the purpose of a systematic derivation of the kinematic
equations, and a comparison with the Eulerian framework, we add a time
coordinate $\tau$ to matter space, which now has coordinates
$\xi^\alpha=(\tau,\xi^A)$. To make this extension trivial, 
we then fix $\tau=t$. Note that
\begin{equation}
\label{convectivederivative}
{\partial\over\partial\tau} = {\partial\over\partial t}
+ \hat v^i {\partial\over\partial x^i}
\end{equation}
is then the usual convective derivative. This extension gives us the
extended derivatives
\begin{eqnarray}
\label{tildeFdef}
{\tilde F}^a{}_\alpha&=& \left(
\begin{array}{cc}
1 & 0_A \\
\hat v^i & {F^i}_A \\
\end{array}
\right), \\ 
\label{tildepsidef}
{\tilde\psi}^\alpha{}_a&=& \left(
\begin{array}{cc}
1 & 0_i \\
\psi^A{}_t & {\psi^A}_i \\
\end{array}
\right), 
\end{eqnarray}
which are now $4\times 4$ matrix inverses of one another, assuming
(\ref{psiAt}).


\paragraph*{Kinematic equations}

We derive the evolution equations and constraints
in the mixed framework by working in the 4-dimensional
notation at first. The integrability condition
\begin{equation}
\label{calCaalphabeta1}
{}^{(1)}\tilde{\cal C}^a{}_{\alpha\beta}:={\tilde F}^a{}_{[\alpha,\beta]}=0
\end{equation}
can be written as the commutator of the vector fields
$\partial_\alpha$ and $\partial_\beta$ pushed forward to spacetime,
\begin{equation}
\label{calCaalphabeta2}
{}^{(2)}\tilde{\cal C}^a{}_{\alpha\beta}:=
{\tilde F}^a{}_{[\alpha} {\tilde F}^b{}_{\beta],a}=0.
\end{equation}

We define the determinant
\begin{equation}
{\tilde F_{x\xi}}:={1\over 4!}\delta_{abcd}\delta^{\alpha\beta\gamma\delta}
{\tilde F}^a{}_\alpha {\tilde F}^b{}_\beta {\tilde F}^c{}_\gamma
{\tilde F}^d{}_\delta,
\end{equation}
where the suffixes indicate that this depends explicitly on the
coordinates $x^a$ and $\xi^\alpha$.  With $\tilde\psi$ the inverse of
$\tilde F$, we have the
variation-of-determinant rule
\begin{equation}
\label{variationofdeterminant}
{\delta{\tilde F_{x\xi}}\over \delta\tilde F^a{}_\alpha}={\tilde
  F_{x\xi}}\ \tilde\psi^\alpha{}_a .
\end{equation}

As $\delta_{abcd}$ and $\delta_{\alpha\beta\gamma\delta}$
are constant, we therefore have
\begin{equation}
\label{detderiv}
{\tilde F}_{x\xi,b}={\tilde F}_{x\xi}\ \tilde\psi^\alpha{}_a \ \tilde
F^a{}_{\alpha,b}.
\end{equation}

From (\ref{detderiv}) and (\ref{calCaalphabeta1}), we find that
\begin{equation}
\label{calCalpha}
\tilde{\cal C}_\alpha:=
\left({\tilde F}_{x\xi}^{-1}{\tilde F{}^a}_\alpha\right)_{,a}=0,  
\end{equation}
while combining (\ref{calCaalphabeta2}) and (\ref{calCalpha}) we 
obtain
\begin{equation}
\label{calCalalphabeta3}
{}^{(3)}\tilde{\cal C}^b{}_{\alpha\beta}:=
2\left({\tilde F}_{x\xi}^{-1}{\tilde F}^a{}_{[\alpha}{\tilde F}^b{}_{\beta]}\right)_{,a}=0. 
\end{equation}

Developing (\ref{tildeFdef}) into its first row, we find
\begin{equation}
{\tilde F_{x\xi}}={1\over 3!}\delta_{ijk} \, \delta^{ABC}
F^i{}_A F^j{}_B F^k{}_C=:{F_{x\xi}}
\end{equation}
Hence we obtain the 3+1 split of the 4-dimensional constraints into
evolution equations and constraints:
\begin{eqnarray}
\label{calCA}
\tilde{\cal C}_A&=&\left({F}_{x\xi}^{-1}{F^i}_A\right)_{,i}=:{\cal C}_A=0, \\
\label{calE}
\tilde{\cal C}_\tau&=&
\left({F}_{x\xi}^{-1}\right)_{,t}+\left(\hat
v^i{F}_{x\xi}^{-1}\right)_{,i}=:{\cal E}=0,\\
\label{calCiAB}
-{}^{(3)}\tilde{\cal C}^i{}_{AB}&=&\left({F}_{x\xi}^{-1}{F^i}_{[A}{F^j}_{B]}\right)_{,j}
=:{\cal C}^i{}_{AB}=0, \\
\label{calEiA}
-{}^{(3)}\tilde {\cal C}^i{}_{A\tau}&=&\left({F}_{x\xi}^{-1}{F^i}_A\right)_{,t}
+\left[{F}_{x\xi}^{-1}\left(
\hat v^j{F^i}_A-\hat v^i{F^j}_A\right)\right]_{,j}\nonumber \\ 
&=:&{\cal E}^i{}_A=0.
\end{eqnarray}
The remaining components
\begin{eqnarray}
{}^{(3)}\tilde {\cal C}^t{}_{A\tau}&=&{\cal C}_A, \\
{}^{(3)}\tilde{\cal C}^t{}_{AB}&=&0
\end{eqnarray}
are redundant. Note that everything is now expressed in terms of
$F^i{}_A$ and $\hat v^i$, and we no longer need $\tilde F^a{}_A$. 

The corresponding jump conditions are 
\begin{eqnarray}
\label{calCAjump}
\left[{F}_{x\xi}^{-1}F^{n}{}_A\right] &=& 0, \\
\label{calEjump}
\left[\hat F_{x\xi}^{-1}(\hat v^{n}-s)\right] &=& 0, \\
\label{calCiABjump}
\left[{F}_{x\xi}^{-1}F^{n}{}_{[A}F^{\parallel i}{}_{B]}\right] &=& 0, \\
\label{calEiAjump}
\left[F_{x\xi}^{-1}F^{\parallel i}{}_A(\hat
v^{n}-s)\right]-\left[{F}_{x\xi}^{-1}F^{n}{}_A\hat v^{\parallel i}\right] &=& 0,
\end{eqnarray}
where the $n$, $\parallel$ notation is as in
Sec.~\ref{section:kinematicjump}.  Note that these jump conditions are
both more numerous and more complicated than the jump conditions
(\ref{psijump1},\ref{psijump2}) of the Eulerian framework.


\paragraph*{Equivalence with the Eulerian framework}

Note that we have used $\cal E$ and $\cal C$ to denote the evolution
equations and constraints in the mixed framework, and $E$ and $C$ for
the Eulerian framework. 

We have the following relations between the full and contracted
equations for $F^i{}_A$,
\begin{eqnarray}
{\cal C}_A&=&2\psi^B{}_i {\cal C}^i{}_{AB}, \\
{\cal E}&=&{1\over 2}\psi^B{}_i\left(\hat v^i{\cal C}_B-{\cal
  E}^i{}_B\right),
\end{eqnarray}
and the following relations between these equations and those for
$\psi^A{}_i$:
\begin{eqnarray}
{\cal E}^i{}_A&=&\delta_{ABC}\delta^{ijk}
 \left( \psi^C{}_l \hat v^l C^B{}_{jk}+\psi^C{}_j E^B{}_k \right), \\
{\cal C}^i{}_{AB}&=&\delta_{ABC}\delta^{ijk} C^C{}_{jk}.
\end{eqnarray}

As these relations between differential equations involve
multiplication by one or more factors of $\psi^A_i$, which in general
is not continuous, the corresponding jump conditions may be
inequivalent. In particular, 
it is not clear if (\ref{calEjump})
follows from (\ref{calEiAjump}), if (\ref{calCAjump}) follows from
(\ref{calCiABjump}), if (\ref{calCiABjump}) is equivalent to
(\ref{psijump1}) or if (\ref{calEiAjump}) is equivalent to
(\ref{psijump2}). However, a detailed calculation shows that all these
relations hold.

As an example of these calculations, consider
\begin{eqnarray}
\left[F_{x\xi}^{-1}F^{n}{}_A\right]&=& \left[\delta^{njk}\delta_{ABC}
\psi^B{}_j \psi^C{}_k\right] \nonumber \\
\label{calCAjumpbis}
&=&2\delta^{nij}\delta_{ABC}
\left[\psi^B{}_{\parallel i} \psi^C{}_{\parallel j}\right],
\end{eqnarray}
where in the first equality we have used the cofactor rule and the
assumption that $F$ is the inverse of $\psi$, and in the second
equality we have used that $\delta_{ABC}$ and
$\delta^{ijk}$ are continuous. 
From (\ref{calCAjumpbis}) we see that (\ref{psijump1}) implies
(\ref{calCAjump}) (as claimed above), but the reverse is not true. In
fact, the right-hand side of (\ref{calCAjumpbis}) vanishes if and only
if
\begin{equation}
\label{wrongpsiAijump}
\left[\psi^A{}_{\parallel i}\right]=\alpha^Ak_{\parallel i}
\end{equation}
for some matter space vector $\alpha^A$ and spatial covector
$k_i$. That is why the jump condition (\ref{calCiABjump}) also needs
to be imposed. 

In the papers
\cite{PS,GodunovPeshkov,GodunovRomenski,Garaizar,BDRT,TRT} only
(\ref{calEjump}), (\ref{calEiAjump}) and (\ref{calCAjump}) are
explicitly given, but (\ref{calCiABjump}) appear to have been
overlooked. In particular, the initial data for the second Riemann
numerical test of \cite{BDRT} (BDRT2) and the initial data for the
fifth Riemann numerical test of \cite{TRT} (TRT5) explicitly violate
(\ref{calCiABjump}). As noted in \cite{Miller}, not imposing the constraints
(\ref{CAijjump}), or equivalently (\ref{calCiABjump}), in full
corresponds to performing surgery (of the type illustrated in
Fig.~\ref{fig:nosurgery2}) at the discontinuity. Moreover, once the
initial data violate the constraints, the subsequent evolution depends
on how constraints have been added to the evolution equations.


\paragraph*{Equations written in terms of the density}

We have already noted that with (\ref{kadvection}) and (\ref{rhoGR}),
(\ref{calE}) is just particle number conservation
(\ref{restmassconservationGR}). Note that in weak solutions, we must
demand that $\sqrt{k_\xi}$ is everywhere continuous, a property that
is conserved under advection.

$F_{x\xi}$ can also be replaced by $n$ in the other equations of
the mixed framework. Defining
\begin{equation}
f^i{}_A:={F^i{}_A\over \sqrt{k_\xi}},
\end{equation}
we can write (\ref{calCA}) and (\ref{calEiA}) as 
\begin{eqnarray}
\left(W\sqrt{\gamma_\xi}\,n f^i{}_A\right)_{,i} &=& 0, 
\nonumber \\ \label{fdiv} \\
\left(W\sqrt{\gamma_\xi}\,n f^i{}_A\right)_{,t} 
+\left[W\sqrt{\gamma_\xi}\,n \left(f^i{}_A\hat v^j-f^j{}_A\hat
  v^i\right)\right]_{,j} &=& 0. 
\nonumber \\ \label{finduction}
\end{eqnarray}
For fixed matter space index ${}_A$, these happen to be identical with
the divergence constraint and the induction equation for the magnetic
field in the formulation \cite{ValenciaMHD} of ideal
magnetohydrodyamics (MHD) in general relativity.

Taking the Newtonian limit $W=1$, $\gamma_{ij}$ flat and assuming
Cartesian coordinates so that $\sqrt{\gamma_x}=1$,
(\ref{restmassconservationGR}), (\ref{fdiv}) and (\ref{finduction})
reduce to Eqs.~(3.12), (3.26) and (3.22) of \cite{PS}, where
$\sqrt{k_\xi}$ is called $\rho_{\rm ref}$. Further assuming
$\sqrt{k_\xi}=1$, they reduce to Eqs.~(6), (9) and (1b) of \cite{BDRT}
and Eqs.~(3), (1) and (2) of \cite{TRT}.

In terms of $f^i{}_A$, the remaining equation, Eq.~(\ref{calCiAB}) can
be written as
\begin{equation}
\left(\sqrt{k_\xi}\,W\sqrt{\gamma_\xi}\,n
f^i{}_{[A}f^j_{B]}\right)_{,j}=0,
\end{equation}
where $\sqrt{k_\xi}$ reappears. As we have already noted, this
constraint is not mentioned in \cite{PS,BDRT,TRT,GodunovPeshkov}. It
also does not have an equivalent in MHD.


\section{Discrete constraint preservation}
\label{section:discreteconstraints}

The following class of conservative numerical schemes preserves a
discrete version of the integrability constraints. With all other
numerical variables defined as cell averages with, by convention,
integer grid index values, define $\psi^A{}_i$ on relevant cell
faces. To initialise them consistently, assign values to $\chi^A$ at
cell centres at the initial time (in liquid as well as solid
regions). Then initialize
\begin{eqnarray}
\label{eq:psi_chi_rep1}
\psi^A{}_{x,i+1/2,j,k}={1\over\Delta x}
\left(\chi^A_{i+1,j,k}-\chi^A_{i,j,k}\right), \\
\psi^A{}_{y,i,j+1/2,k}={1\over\Delta y}
\left(\chi^A_{i,j+1,k}-\chi^A_{i,j,k}\right), \\
\label{eq:psi_chi_rep3}
\psi^A{}_{z,i,j,k+1/2}={1\over\Delta z}
\left(\chi^A_{i,j,k+1}-\chi^A_{i,j,k}\right).
\end{eqnarray}
The $\chi^A_{i,j,k}$ are used only for initialisation, and are not
required afterwards.
We then evolve using the conservative equations
\begin{equation}
{d\over dt}\psi^A{}_{x,i+1/2,j,k}={1\over \Delta
  x}\left({\cal F}^A_{i+1,j,k}-{\cal F}^A_{i,j,k}\right)
\end{equation}
and similarly for $\psi^A{}_y$ and $\psi^A{}_z$, where the numerical
flux ${\cal F}^A_{i,j,k}$ is some approximation to $\psi^A{}_j\hat
v^j$ at cell centres, suitably limited to enforce the TVD
property. Then a discrete version of
$\psi^A{}_{x,y}-\psi^A{}_{y,x}=0$, evaluated at relevant cell edges:
\begin{eqnarray}
&&{1\over\Delta y}(\psi^A{}_{x,i+1/2,j+1,k}-\psi^A{}_{x,i+1/2,j,k})
\nonumber \\
&-&{1\over\Delta x}(\psi^A{}_{y,i+1,j+1/2,k}-\psi^A{}_{y,i,j+1/2,k})=0,
\end{eqnarray}
and similarly for the other two commutators, is obeyed at all times if
it is obeyed initially. The fundamental idea is that the numerical
fluxes ${\cal F}^A$ are the time derivatives of the underlying
$\chi^A$, and hence are the same for the $\psi^A{}_x$, $\psi^A{}_y$
and $\psi^A{}_z$. The discrete constraints act as discrete
integrability conditions that allow us to reconstruct the
$\chi^A_{i,j,k}$ by summation if desired.


\section{Riemann tests on a 2-dimensional grid}
\label{appendix:2d}


\begin{figure}
\includegraphics[width=9cm]{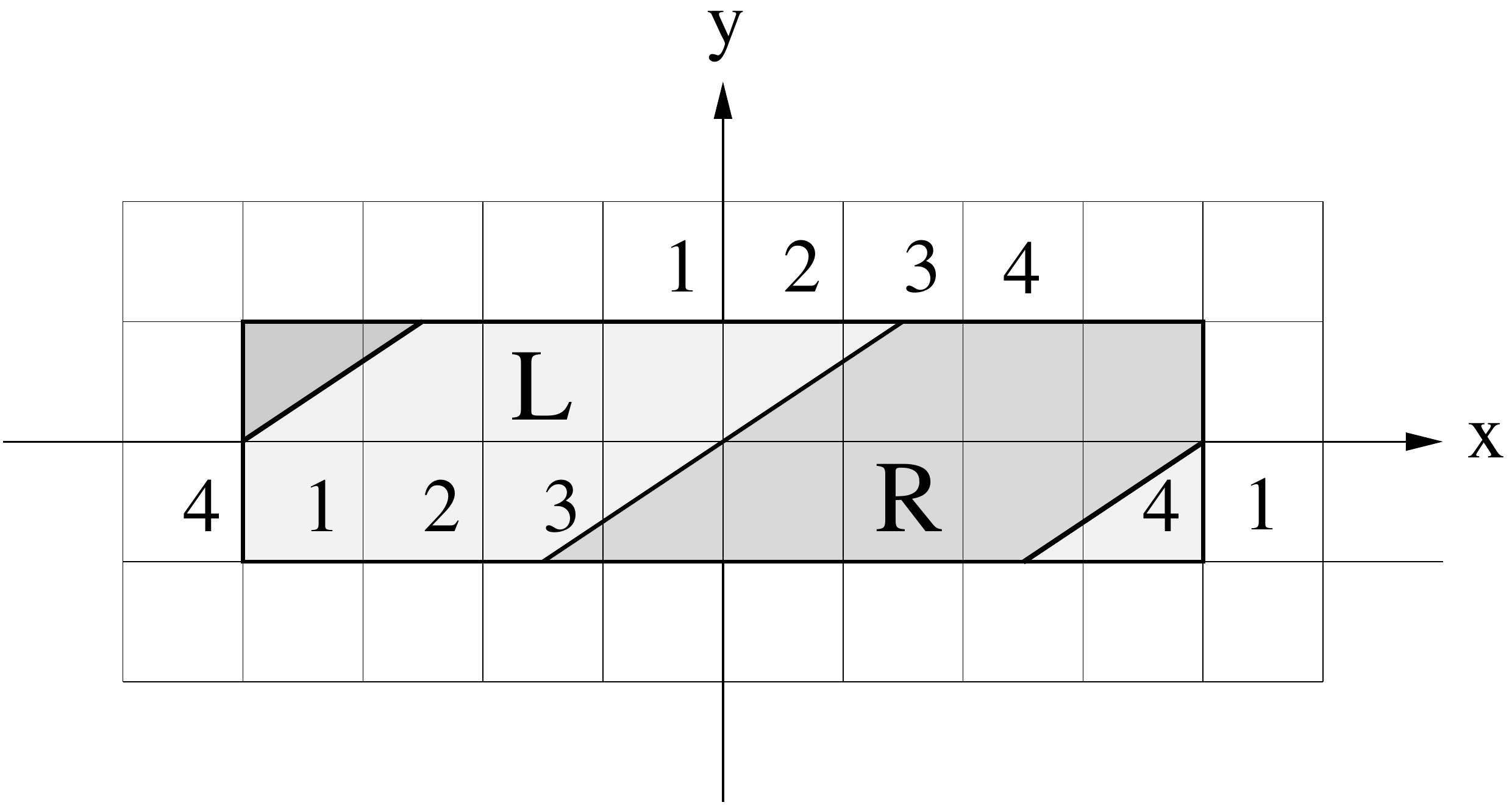}
\caption{
\label{fig:twodimgrid}
Example of a two-dimensional grid with shifted periodic boundary
conditions, with $n_x=8$, $n_y=2$ and $\delta_x=3$. The physical grid
is surrounded by one ghost cell on each side. In reality, $n_x$ would
be much larger, while $n_y$ ranges from 1 to a few, and
$\delta_x$ from 0 to a few, with no common
factor. The placement of the left and right state of a Riemann problem
is shown by the letters L and R and shading. Cells are initialised
with the left or right state depending on the position of the cell
{\em center}. The numbers 1, 2, 3, 4 identify four physical cells and
the ghost cells they donate values to. The initial
discontinuity is at an angle $\alpha$ from the $y$ axis, with
$\tan\alpha=\delta_x/n_y$ ($=3/2$ in this example), and goes through
the point $x=y=0$.}
\end{figure}


As a first test of the role of the constraints in hyperbolicity, we
numerically solve Riemann problems on a 2-dimensional grid, with the
initial discontinuity at an angle to the grid. Assume the grid
consists of $n_x\times n_y$ cells, surrounded by the necessary number
of ghost cells. After each time update, the ghost points are filled
using periodic boundary conditions, identifying cell $(i,j)$ with
$(i+n_x,j)$ in the $x$ direction, but $(i,j)$ with
$(i+\delta_x,j+n_y)$ in the $y$ direction, where $\delta_x$ is an
offset. Consistently with these boundary conditions, the initial
discontinuity is then placed on a line of $x/y=\delta_x/n_y$ (assuming
that the grid spacing is the same in the $x$ and $y$ directions). This
is illustrated in Fig.~\ref{fig:twodimgrid}.

As the $x$ and $y$ directions are interchangeable, the slope
$\delta_x/n_y$ and its inverse pose the same Riemann test. (Less
obviously, in our implementation those two tests also have
roughly equal computational cost.) We choose $n_y\ge\delta_x$ (and
typically $\delta_x=1$) so that the initial discontinuity is always
closer to the $y$ axis (where it is in the 1D tests), and use the $x$
axis as an approximation to a line normal to the initial discontinuity
when taking a cut through the solution.


\section{Shear scalars}

The three eigenvalues of $\eta^A{}_B$ can be parameterised as
$\{a,b,1/(ab)\}$. We then find that in the unsheared state $a=b=1$,
\begin{eqnarray}
 I^1=I^2&=&3, \\
 I^1_{,a}=I^1_{,b}=I^2_{,a}=I^2_{,b}&=&0, \\
 I^1_{,aa}=I^1_{,bb}=2, \quad I^1_{,ab}&=&1, \\
 I^2_{,aa}=I^2_{,bb}=8, \quad I^2_{,ab}&=&4.
\end{eqnarray}
Hence $4(I^1-3)$ and $I^2-3$ are the same function of the shear up to
quadratic order. This is not a bad choice of $I^\alpha$ but a
property of any shear invariant. It is related to the fact that the
characteristic speeds in the unsheared state depend on $f_1$ and $f_2$
only through the one combination $f_1+4f_2$ that appears in the shear
modulus (\ref{shearmodulus}). 

Therefore, to model linear elasticity correctly, it is sufficiently
general to make the ansatz
\begin{equation}
\label{EOSclass}
\epsilon(s,{n},I^\alpha)
=\check\epsilon({n},s)+{\check\mu({n},s)\over{n}}{\cal
  S}(I^\alpha),
\end{equation}
where the {\em shear scalar} ${\cal S}$ obeys
\begin{eqnarray}
{\cal S}&=&0, \\
2{\partial{\cal S}\over \partial I^1}
+8{\partial{\cal S}\over \partial I^2} &=& 1
\end{eqnarray}
in the unsheared state $I^1=I^2=3$, but is otherwise arbitrary. For
any such choice of $\cal S$, $\check\mu({n},s)$ evaluates to the usual
shear modulus (\ref{shearmodulus}) in the Newtonian limit, and the
equations of motion are the same when linearised about the unsheared state.

Clearly there are many possibilities of defining a shear scalar that
obeys these conditions, but we are not aware of any physical reason
given in the literature for why a specific choice should be preferred,
or of values given for $f_1$ and $f_2$ independently. 

An equation of state for copper in \cite{TRT} uses the shear scalar
\begin{equation}
\label{SCran}
{\cal S}_{\rm Cran}:= {3I^2-(I^1)^2\over 12},
\end{equation}
which is homogenously quadratic in the eigenvalues of $\eta^A{}_B$. In
\cite{KS} the shear scalar
\begin{equation}
\label{SKS}
{\cal S}_{\rm KS}:={(I^1)^3-I^1I^2-18\over 24}, 
\end{equation}
which is cubic, is suggested for what seem to be aesthetic
reasons. Yet another shear scalar is
\begin{equation}
\label{SVM}
{\cal S}_{\rm VM}:=s^{ab}s_{ab}={I^2-2I^1+3\over 4}, 
\end{equation}
where 
\begin{equation}
s_{ab}:={1\over 2}(h_{ab}-\eta_{ab})
\end{equation}
is the ``constant volume shear tensor'' defined in \cite{CQ}. In the
Newtonian limit, near the unsheared state, $\cal S_{\rm VM}$ is
related to the Von Mises stress scalar (assuming stress and strain are
related linearly). It gives the same values of $f_1$ and $f_2$ as
$\cal S_{\rm Cran}$.


\section{Equations of state}
\label{appendix:eos}

We now consider examples of equations of state of the form (\ref{EOSclass}).
The following general expressions will be useful:
\begin{eqnarray}
h&=&1+\check\epsilon+{\check\mu\over{n}}{\cal S}+{p\over {n}}, \\
p&=&{n}^2{\partial\check\epsilon\over\partial {n}}
+\left({n}{\partial\check\mu\over\partial{n}}-\check\mu\right){\cal
  S}, \\
f_\alpha&=&{\check\mu({n},s)\over{n}}{\partial {\cal
  S}\over\partial I^\alpha}.
\end{eqnarray}
In principle we can eliminate $s$ from these two equations to obtain
$p$, $f_1$ and $f_2$, as functions of $({n},h,I^1,I^2)$, which we
need in the recovery of the primitive from the conserved variables.


\paragraph*{A toy relativistic EOS}

As a toy model for a relativistic equation of state, we take
$\check\epsilon$ from the commonly used ``Gamma-law'' hot equation of
state, and make the shear modulus $\check\mu$ a power of the density only,
namely
\begin{eqnarray}
\check\epsilon({n},s)&=&{K(s)\over \Gamma-1}{n}^{\Gamma-1}, \\ 
\check\mu({n},s)&=& \kappa{n}^\lambda,
\end{eqnarray}
where $\Gamma$, $\kappa$ and $\lambda$ are constants.  This is
motivated by the fact that in neutron star crusts $\mu\propto
n^{4/3}$, with the factor of proportionality only weakly
temperature-dependent. The bulk modulus in neutron stars is given by
the nuclear interactions, while the shear modulus is provided by
Coulomb interactions, which makes it independent and much smaller.
Following \cite{KS}, we choose ${\cal S}$ as ${\cal S}_{\rm KS}$ given
by (\ref{SKS}).

The expressions we need for the conserved to primitive variables
conversion are then
\begin{eqnarray}
p(h,{n},I^\alpha)&=&{\Gamma-1\over\Gamma}{n}(h-1)+{\lambda-\Gamma\over\Gamma}
\kappa{n}^\lambda {\cal S},\\
p(\epsilon,{n},I^\alpha)&=&(\Gamma-1){n}\epsilon+(\lambda-\Gamma)\kappa{n}^\lambda
{\cal S}, \\
h(p,{n},I^\alpha)&=&1+{\Gamma\over\Gamma-1}{p\over
  {n}}
+{\Gamma-\lambda\over\Gamma-1}\kappa{n}^{\lambda-1}{\cal S},\\
f_1&=&\kappa{n}^{\lambda-1}{3(I^1)^2-I^2\over 24}, \\
f_2&=&-\kappa{n}^{\lambda-1}{I^1\over 24}.
\end{eqnarray}
The characteristic speeds in the unsheared state are
\begin{eqnarray}
\lambda_T^2&=&{\kappa n^{\lambda-1}\over 1+\Gamma\epsilon}, \\
\lambda_L^2&=&{\Gamma(\Gamma-1)\epsilon
+{4\over 3} \kappa n^{\lambda-1}
\over 1+\Gamma\epsilon}.
\end{eqnarray}


\paragraph*{Cranfield EOS}

The equation of state for copper used in \cite{TRT} for Newtonian shock tube
problems, translated into our notation, is
\begin{eqnarray}
\epsilon(s,{n},I^\alpha)&=& A({n})+B({n})K(s)+C({n}){\cal S}, \\
A&:=&{K_0\over 2\alpha^2}\left[\left({{n}\over{n}_0}\right)^\alpha-1\right]^2, \\
B&:=&c_VT_0\left({{n}\over{n}_0}\right)^\gamma, \\ 
K&:=& e^{s\over c_V}-1, \\
C&:=&B_0\left({{n}\over{n}_0}\right)^{\beta+4/3},
\end{eqnarray}
where ${\cal S}$ is ${\cal S}_{\rm Cran}$ given by (\ref{SCran}).  We
need the following forms of the equation of state:
\begin{eqnarray}
p(s,{n},I^\alpha)&=&{n}\left[{n} A'+\gamma BK+(\beta+4/3)C{\cal S}\right], \\
p(h,{n},I^\alpha)&=&{{n}\over\gamma+1}\Bigl[\gamma(h-1)
-\gamma A+{n} A' \nonumber \\ && +(\beta+4/3-\gamma)C{\cal S}\Bigr],
\\
p(\epsilon,{n},I^\alpha)&=&{n}\Bigl[\gamma\epsilon-\gamma A+{n} A'
 \nonumber \\ && +(\beta+4/3-\gamma)C{\cal S}\Bigr], \\
h(p,{n},I^\alpha)&=&1+{\gamma+1\over\gamma}{p\over{n}}+A-{1\over\gamma}{n}
A' \nonumber \\ && -{1\over\gamma}(\beta+4/3-\gamma)C{\cal S}, \\
f_1 &=&-{CI^1\over 6}, \\
f_2 &=&{C\over 4} 
\end{eqnarray}


\section{Constructing exact solutions}
\label{sec:exactsolns}

The exact solution of the Riemann problem is a standard test for HRSC
methods. For Newtonian elasticity exact solvers have been constructed
both by Miller~\cite{Miller} and by Barton et al.~\cite{BDRT}. In the
relativistic case here we have not constructed a generic solver to
compute the full Riemann problem solution. As noted by~\cite{BDRT},
this can be extremely sensitive to initial guesses used. Instead we
construct exact solutions by specifying the wave structure explicitly
in advance and solving across each wave.

As summarized in~\cite{BDRT}, with piecewise constant initial data the
generic solution will contain seven self-similar waves. The central
wave will be a contact discontinuity, and the other waves will be
genuinely nonlinear. We assume that the solutions are simple shocks or
rarefactions. We then solve across each wave in the following manner.

\paragraph*{Shock wave}

We assume that the primitive variables to the left of the wave, ${\bf
  w}_L$, are given. We then impose the value \emph{either} of the
shock speed $s^{(p)}$ or of one component of the variables to
the right of the wave, ${\bf w}_R$. The Rankine-Hugoniot conditions
\begin{equation}
  \label{eq:RH}
  {\bf f}({\bf w}_R) - {\bf f}({\bf w}_L) = s^{(p)} \left[ {\bf
      q}({\bf w}_R) - {\bf q}({\bf w}_L) \right] 
\end{equation}
then form a system of nonlinear equations for the remaining components
of ${\bf w}_R$ and, where necessary, for the shock speed
$s^{(p)}$. Here ${}^{(p)}$ denotes the wave number counting from the
left.

This problem is solved explicitly using the Matlab solver {\tt
  fsolve}. It is usually necessary to experiment with the imposed
value and initial guesses in order to construct a solution satisfying
the Lax entropy condition
\begin{equation}
  \label{eq:RH_Lax}
  \lambda^{(p)} ({\bf w}_L) > s^{(p)} > \lambda^{(p)} ({\bf w}_R).
\end{equation}
The construction of the eigenvalues $\lambda^{(p)}$ is discussed below.

\paragraph*{Contact discontinuity}

A contact must satisfy the Rankine-Hugoniot conditions~\eqref{eq:RH}
combined with the restriction that the wave speed $s$ matches the
normal velocity on either side of the wave. Hence we can use the same
techniques as for the shock with the value of the velocity imposed.

\paragraph*{Rarefaction wave}

As noted by~\cite{BDRT} the solution across a rarefaction wave is
given by
\begin{equation}
  \label{eq:rarefaction}
  \pda{{\bf w}}{\xi} = \frac{{\bf r}^{(p)} \left( {\bf w}
    \right)}{{\bf r}^{(p)} \left( {\bf w} \right) \cdot \nabla_{{\bf
        w}} \lambda^{(p)}  \left( {\bf w} \right)}.
\end{equation}
Here $\xi = x / t$ is the self-similarity variable. We have that
$\lambda^{(p)}({\bf w}_L) \le \xi$ where ${}^{(p)}$ labels the wave
number and ${\bf w}_L$ is given, as above. We impose that $\xi \le
\xi_R = \lambda^{(p)}({\bf w}_R)$ to stop the integration. In addition
${\bf r}^{(p)}$ are the right eigenvectors associated with the
$p^{\text{th}}$ eigenvalue $\lambda^{(p)}$, and $\nabla_{{\bf w}}$
denotes the gradient operator with respect to the vector of primitive
variables. 

All characteristic information ($\lambda^{(p)}, {\bf r}^{(p)}$) is
constructed from the Jacobian matrix
\begin{equation}
  \label{eq:jacobian}
  J = \pda{{\bf f}({\bf w})}{{\bf q}({\bf w})} = \left( \nabla_{{\bf w}}
    {\bf q} \right)^{-1} \nabla_{{\bf w}} {\bf f}.
\end{equation}
As in the Newtonian case discussed in \cite{BDRT} we need to
explicitly modify the calculated Jacobian to build in the
hyperbolicity corrections as in equation~\eqref{EAitres}.

Given an explicit left state ${\bf w}_L$ the numerical solution is
found by solving the ODE~\eqref{eq:rarefaction} for ${\bf w}$ with
initial data ${\bf w}_L$ in $\lambda^{(p)}({\bf w}_L) \le \xi \le
\xi_R$. Explicitly we use the {\tt ode45} routine with Matlab. The
Jacobian $J$ is constructed using explicit finite differencing by
varying each component of ${\bf w}$ by a small value $h$. Standard
Matlab routines were used to construct and sort the characteristic
information. The gradient $\nabla_{{\bf w}} \lambda^{(p)} \left( {\bf
    w} \right)$ was also constructed using explicit finite
differencing. In all cases $6^{\text{th}}$ order finite differencing
combined with Richardson extrapolation was used to ensure sufficient
accuracy. 

There are two potential problems with this construction. First, as
noted by~\cite{BDRT}, we have no guarantee that
equation~\eqref{eq:rarefaction} has a unique solution. This would
imply that the true solution is a compound wave, and breaks the
assumptions made here. Second, the numerical construction of the
characteristic information is extremely sensitive when the eigenvalues
are close to each other. This appears to be the case for the problems
and equations of state considered below, and means that for the slower
3 and 5 waves next to the contact we are forced to construct very small
rarefaction fans.

In principle there is no reason why the procedure above could not be
extended to construct a full Riemann solver. However, such a solver
would have little practical utility, even if it could be made generic
and robust. Numerical experiments have shown that it is faster to
compute an approximate solution using 800 grid cells than it is to
construct \emph{one} exact solution with a pre-specified wave
structure. Even allowing for the massive speed improvements possible
within our current exact solver, it is clearly impractical for use
within an evolution code.


\section{Initial data for numerical tests}
\label{appendix:initialdata}


We used several sets of initial data that were defined in published
papers; this was done to ensure that our code agreed with Newtonian
results produced previously \cite{BDRT} \cite{TRT}. Because both
papers chose entropy, $s$, as a primitive variable, instead of the
pressure, $p$, we list the initial entropy value here, and calculate
the pressure from the entropy when the system is initialized.

For the following sets of initial data, the spacetime metric is the
Minkowski metric, and the matter-space metric is the Euclidean metric
in Euclidean coordinates normalized with the initial density of the
elastic medium, $n_0$; we note that while we must convert units of
velocity to geometrized units, we do not need to convert units of
density or of length, as long as we are consistent throughout the
code. For this paper the value ${n}_0 = 8.93$~g/cm$^3$ was used for
the BDRT tests (from \cite{BDRT}) and ${n}_0 = 8.9$~g/cm$^3$ was used
for the TRT tests (from \cite{TRT}). In addition to this, for each of
these situations, the Cranfield equation of state, described in
Appendix~\ref{appendix:eos}, was used. For comparison purposes, the
velocities in this section are taken to be in km s$^{-1}$, while the
entropy is in kJ g$^{-1}$K$^{-1}$.

\paragraph*{BDRT1} 

This is the same as \textit{Testcase 1} in \cite{BDRT}. It allows us
to examine the entire seven-wave structure of the solution. Using the
Cranfield EOS above, the solution consists of three left-travelling
rarefaction waves, a right-travelling contact, two right-travelling
rarefactions, and a right-travelling shock wave. The initial data is
presented for the state vector ${\bf w} = ( v^i, F^i{}_A, s )$ in the
mixed framework given in Appendix~\ref{appendix:mixed}, and all other
quantities are derived from them:
\begin{align} 
  {\bf w}_L &= \left\{
    \begin{pmatrix}
      0 \\ 0.5 \\ 1
    \end{pmatrix}, 
    \begin{pmatrix}
      0.98 & 0 & 0 \\
      0.02 & 1 & 0.1 \\
      0 & 0 & 1
    \end{pmatrix}, 
    0.001
  \right\}, \nonumber \\
  {\bf w}_R &= \left\{
    \begin{pmatrix}
      0 \\ 0 \\ 0
    \end{pmatrix},
    \begin{pmatrix}
      1 & 0 & 0 \\
      0 & 1 & 0.1 \\
      0 & 0 & 1 
    \end{pmatrix},
    0
  \right\}.
\end{align}
Results are shown at coordinate time $t = 0.06$.

\paragraph*{4-wave relativistic solution}

We constructed a range of relativistic solutions, mostly consisting of
a single shock or rarefaction, using the technique outlined in
Appendix~\ref{sec:exactsolns}. The toy relativistic equation of state
given above is used, with parameters $\Gamma = 5/3$, $\lambda = 4/3$,
and $\kappa = 1/2$. In particular, we present a solution with four
nonlinear waves. The two left-going waves (1- and 2-waves) are
rarefactions. The contact is trivial, as are the central (3- and
5-waves) nonlinear waves. The slower right-going wave (a 6-wave) is a
rarefaction, and the fast right-going 7-wave is a shock. The initial
data is presented for the state vector ${\bf w} = ( v^i, \psi^A{}_i, p
)$, truncated to 6 significant figures, and all other quantities are
derived from them:
\begin{align} 
  {\bf w}_L &= \left\{
    \begin{pmatrix}
      0.05 \\ 0.1 \\ 0.2
    \end{pmatrix}, 
    \begin{pmatrix}
      \phantom{-}1.5 & 0 & 0 \\
      -0.5 & 1 & 0 \\
      \phantom{-}0.5 & 0 & 1 
    \end{pmatrix}, 
    1.86054 \right\}, \\
  {\bf w}_R &= \left\{
    \begin{pmatrix}
      \phantom{-}0.469381 \\ -0.0332532 \\ \phantom{-}0.349709
    \end{pmatrix},
    \begin{pmatrix}
      \phantom{-}0.764910 & 0 & 0 \\
     -0.541672 & 1 & 0 \\
      \phantom{-}0.369075 & 0 & 1
    \end{pmatrix},
    0.450123 \right\}.
\end{align}
Results are shown at coordinate time $t = 0.25$.


\vspace{1ex}

In addition to Riemann problem style tests we consider a genuinely
two-dimensional rotor test. The Newtonian rotor test was suggested
by~\cite{Dumbser2008}, where the evolution was shown using a
high-order finite element technique. The domain is cylindrical, of
total radius $0.5$. The material is initially at rest except in the
rotor, represented by a cylinder of radius $0.1$, within which it
rotates with angular velocity $\omega = 10$. The material is not
deformed (i.e., $F^i{}_A$ is the unit matrix) nor hot (i.e., $s =
0$). All other matter properties follow the Riemann tests above. That
is, the initial density is given by ${n}_0 = 8.93$~g/cm$^3$ and the
Cranfield equation of state, described in Appendix~\ref{appendix:eos},
was used. Here, as we have used a Cartesian grid, we have simulated
the full domain $x, y \in [-0.5, 0.5]$.

We suggest a relativistic rotor test as a direct comparison with the
Newtonian version. The domain remains the same as the Newtonian
case. The angular velocity is reduced to $\omega = 0.5$. The material
is initially set so that $\psi^A{}_i$ is the unit matrix and $p=1$. As
the shear also depends on the velocity through $\psi^A{}_t$, the
material is sheared within the rotor initially, in contrast to the
Newtonian case, but this is small. As in the Riemann tests above we
use the toy relativistic equation of state given in
Appendix~\ref{appendix:eos}, with parameters $\Gamma = 5/3$, $\lambda
= 4/3$, and $\kappa = 1/2$.


\end{document}